\documentclass[10pt]{article} 
\usepackage[preprint]{tmlr}

\usepackage{amsmath,amsfonts,bm}

\def\eqref#1{equation~\ref{#1}}

\def\1{\bm{1}}

\DeclareMathAlphabet{\mathsfit}{\encodingdefault}{\sfdefault}{m}{sl}
\SetMathAlphabet{\mathsfit}{bold}{\encodingdefault}{\sfdefault}{bx}{n}

\usepackage{hyperref}
\hypersetup{
  colorlinks=true,
  linkcolor=[HTML]{67008c},   
  citecolor=[HTML]{064e69}, 
  urlcolor=[HTML]{064e69},
  filecolor=[rgb]{0.55,0.1,0.1}
}
\usepackage{url}
\usepackage{graphicx}
\usepackage[most]{tcolorbox}
\usepackage{booktabs}
\usepackage{multirow}
\usepackage{adjustbox}
\usepackage{rotating}
\usepackage{pdflscape}
\usepackage{afterpage}
\usepackage{environ}
\usepackage{amssymb}
\usepackage{svg}
\usepackage[english]{babel}
\usepackage{amsthm}
\usepackage{subcaption}
\usepackage{enumitem}

\usepackage{tikz}
\usetikzlibrary{positioning,fit,calc,shapes,shadows,arrows,trees}
\usepackage[font=small,labelfont=bf]{caption}
\usepackage[dvipsnames, table]{xcolor}
\usepackage[edges]{forest}
\usepackage{glossaries}
\usepackage{soul}
\usepackage{pifont}
\usepackage{array}
\usepackage{fontawesome5}
\usepackage{tabularx}
\usepackage{longtable}

\usepackage[version=4,arrows=pgf-filled,
textfontname=sffamily,
mathfontname=mathsf]{mhchem}

\definecolor{red_0}{HTML}{CD1F1F}
\definecolor{yellow_0}{HTML}{D8D823}
\definecolor{green_0}{HTML}{16E716}
\definecolor{turqoise_0}{HTML}{1CCECE}
\definecolor{blue_0}{HTML}{2222D1}
\definecolor{purple_0}{HTML}{C427C4}

\definecolor{okabe_1}{HTML}{E69F00} 
\definecolor{okabe_2}{HTML}{56B4E9} 
\definecolor{okabe_3}{HTML}{009E73} 
\definecolor{okabe_4}{HTML}{F0E442} 
\definecolor{okabe_5}{HTML}{0072B2} 
\definecolor{okabe_6}{HTML}{D55E00} 
\definecolor{okabe_7}{HTML}{CC79A7} 

\definecolor{lightline}{gray}{0.9}

\newcommand{\out}{\mathrm{out}}%

\newcounter{takeaway}

\newtcolorbox{takeawaybox}[1]{%
    enhanced,
    breakable,
    colback=white,
    frame hidden,          
    boxrule=0pt,
    sharp corners,         
    boxsep=0pt,
    borderline west={1.1pt}{0pt}{gray!65!black},
    left=7pt, right=0pt, top=3pt, bottom=0pt,
    before skip=9pt, after skip=9pt,
    fontupper=\small,
    before upper={\textsf{\bfseries\footnotesize #1}\par\vspace{3pt}}
}

\theoremstyle{definition}
\newtheorem{definition}{Definition}

\newtcbox{\borderpill}[1][gray]{
  on line, 
  arc=3pt, 
  colback=white, 
  colframe=#1, 
  baseline=-0.4ex,
  before upper={\rule[-0.2ex]{0pt}{2ex}}, 
  boxrule=0.8pt, 
  boxsep=0pt, 
  left=2pt, right=2pt, top=1.5pt, bottom=1.5pt,
  fontupper=\sffamily\bfseries\scriptsize,
  colupper=#1 
}

\newtcbox{\pill}[1][gray]{
  on line, 
  arc=3pt, 
  colback=#1, 
  colframe=#1, 
  baseline=-0.4ex,
  before upper={\rule[-0.2ex]{0pt}{2ex}}, 
  boxrule=0.8pt, 
  boxsep=0pt, 
  left=2pt, right=2pt, top=1.5pt, bottom=1.5pt,
  fontupper=\sffamily\bfseries\scriptsize,
  colupper=white 
}

\newcommand{\rpill}[2][gray]{\raisebox{-0.6ex}{\pill[#1]{#2}}}
\newcommand{\rborderpill}[2][gray]{\raisebox{-0.6ex}{\borderpill[#1]{#2}}}

\newtcolorbox{academicbox}[1]{%
    colback=gray!5!white,
    colframe=gray!60!black,
    arc=1mm,
    boxrule=0.5pt, 
}

\newcommand{\pipeline}[5]{%
\begin{tikzpicture}[baseline={([yshift=-0.6ex]current bounding box.center)}, scale=0.8, every node/.style={transform shape}]
    
    \tikzset{
        lbl/.style={text=white, font=\sffamily\bfseries\small, anchor=center},
        lblOff/.style={text=okabe_3, font=\sffamily\bfseries\small, anchor=center} 
    }

    \def\wUI{0.7cm}
    \def\wR{0.7cm}
    \def\wDI{0.7cm}
    \def\wDA{0.7cm}
    \def\wO{0.7cm}
    \def\h{0.6cm}      
    \def\g{0.05cm}      
    \def\totalW{3.5cm} 

    \pgfmathsetmacro{\totalWWithGaps}{\totalW + 4*\g}

    \begin{scope}
        \clip[rounded corners=4pt] (0,0) rectangle (\totalWWithGaps pt,\h);

        \ifnum#1=1 \fill[okabe_3] (0,0) rectangle (\wUI,\h); \node[lbl] at (0.5*\wUI, 0.5*\h) {UI};
        \else \fill[white] (0,0) rectangle (\wUI,\h); \fi

        \coordinate (X1) at ($(\wUI,0) + (\g,0)$);
        \ifnum#2=1 \fill[okabe_3] (X1) rectangle ++(\wR,\h); \node[lbl] at ($(X1)+(0.5*\wR, 0.5*\h)$) {R};
        \else \fill[white] (X1) rectangle ++(\wR,\h); \fi

        \coordinate (X2) at ($(X1)+(\wR,0)+(\g,0)$);
        \ifnum#3=1 \fill[okabe_3] (X2) rectangle ++(\wDI,\h); \node[lbl] at ($(X2)+(0.5*\wDI, 0.5*\h)$) {DI};
        \else \fill[white] (X2) rectangle ++(\wDI,\h); \fi

        \coordinate (X3) at ($(X2)+(\wDI,0)+(\g,0)$);
        \ifnum#4=1 \fill[okabe_3] (X3) rectangle ++(\wDA,\h); \node[lbl] at ($(X3)+(0.5*\wDA, 0.5*\h)$) {DA};
        \else \fill[white] (X3) rectangle ++(\wDA,\h); \fi

        \coordinate (X4) at ($(X3)+(\wDA,0)+(\g,0)$);
        \ifnum#5=1 \fill[okabe_3] (X4) rectangle ++(\wO,\h); \node[lbl] at ($(X4)+(0.5*\wO, 0.5*\h)$) {O};
        \else \fill[white] (X4) rectangle ++(\wO,\h); \fi
        
    \end{scope}

    \draw[okabe_3, line width=1.5pt, rounded corners=4pt] (0,0) rectangle (\totalWWithGaps pt,\h);

\end{tikzpicture}%
}

\newcolumntype{C}[1]{>{\centering\arraybackslash}m{#1}}

\newlength{\figwidth}
\newcommand{\openSettingIcon}{%
  \raisebox{-0.5ex}{\includegraphics[height=2.5ex]{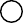}}%
}

\newcommand{\setSettingIcon}{%
  \raisebox{-0.5ex}{\includegraphics[height=2.5ex]{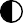}}%
}

\newcommand{\multiSettingIcon}{%
  \raisebox{-0.5ex}{\includegraphics[height=2.5ex]{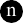}}%
}

\newcommand{\singleSettingIcon}{%
  \raisebox{-0.5ex}{\includegraphics[height=2.5ex]{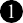}}%
}

\newcommand{\SQLIcon}{%
  \raisebox{-0.5ex}{\includegraphics[height=2.5ex]{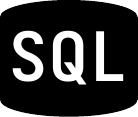}}%
}

\newif\ifpdflscape
\pdflscapetrue

\newenvironment{fptablerot}{\sidewaystable[p]}{\endsidewaystable}
\newcommand{\fptablelscbody}[1]{%
  \begin{landscape}%
    \vspace*{\fill}%
    \begingroup
      \centering
      #1%
      \par
    \endgroup
    \vspace*{\fill}%
  \end{landscape}%
}
\ifpdflscape
  \NewEnviron{fullpagetable}{%
    \expandafter\afterpage\expandafter{\expandafter\fptablelscbody\expandafter{\BODY}}%
  }
  \newcommand{\fullpagecaption}[1]{\captionof{table}{#1}}
\else
   
  \newcommand{\fullpagecaption}[1]{\caption{#1}}
\fi

\makeglossaries

\newacronym{task}{OpenTI}{Open Tabular Insight Extraction}
\newacronym{llm}{LLM}{Large Language Model}
\newacronym{ir}{IR}{Information Retrieval}
\newacronym{dag}{DAG}{Directed Acyclic Graph}
\newacronym{dsl}{DSL}{Domain-Specific Language}
\newacronym{gpl}{GPL}{General-Purpose Language}
\newacronym{rag}{RAG}{Retrieval Augmented Generation}
\glsdisablehyper

\title{Open Tabular Insight Extraction: Where Do We Stand, and Where Should We Go?}

\author{\name Daniel Gomm \email daniel.gomm@cwi.nl \\
      \addr Centrum Wiskunde \& Informatica\\
      University of Amsterdam
      \AND
      \name Maarten de Rijke \email m.derijke@uva.nl \\
      \addr University of Amsterdam
      \AND
      \name Madelon Hulsebos \email madelon.hulsebos@cwi.nl\\
      \addr Centrum Wiskunde \& Informatica
      }

\def\mymonth{MM}  
\def\myyear{YYYY} 
\def\openreview{\url{https://openreview.net/forum?id=XXXX}} 

\begin{document}

\maketitle

\begin{abstract}
    Democratizing access to the knowledge held in large corpora of tables such as data lakes is emerging as a central research challenge. Research in this space is advancing and broadening in scope, increasingly supplying the components to satisfy a person's insight need end-to-end. Yet these efforts remain fragmented across communities that frame the problem under their own conventions, such as table question answering, text-to-SQL, and data analysis agents, with works six times as likely to cite within the same task label as across labels. To bring these communities onto common ground, we establish a holistic framework for this pursuit, which we refer to as \textit{Open Tabular Insight Extraction (OpenTI)}. We formalize OpenTI from first principles around the analytical knowledge a person needs, the procedure for deriving it from a corpus of tables, and how well a result serves the person who sought it. In doing so we consolidate frameworks and terminology across information retrieval, natural language processing, machine learning, databases, and human-computer interaction, and apply this grounding in a systematic review and analysis of systems and benchmarks that work towards OpenTI.
    We find that current systems do not cover the end-to-end scope of OpenTI, mainly focusing on the analysis itself, and that benchmarks are largely unfit for evaluations in an open setting as inputs presuppose knowledge of tables, and validation mechanisms do not match the setup.
    Finally, we distill a research agenda towards OpenTI systems, evaluation, and interaction paradigms that surface the insights users need. An interactive companion to our paper is available at \href{https://open-tabular-insight-extraction.github.io}{https://open-tabular-insight-extraction.github.io}.

\end{abstract}

\newpage

\setcounter{tocdepth}{2}
\begingroup
\hypersetup{linkcolor=black}
\tableofcontents
\endgroup

\section{Introduction}
\label{sec:introduction}
Tabular data, found in relational databases, spreadsheets, data lakes, on the web, and as standalone files, is among the most pervasive forms of data in our digital society~\citep{borisov_deep_2024, shwartz-ziv_tabular_2022, cafarella_webtables_2008}. Organizations across the public and private sectors accumulate it at scale and depend on it to make decisions~\citep{jagadish_big_2014}. Tables owe this prevalence to their compact and semantically rich structure that encodes entities, their attributes, and the relationships among them in a form that is both machine-readable and amenable to analysis~\citep{codd_relational_1970}. Yet the knowledge that makes this data valuable does not just reside in the cells themselves. A table of measurements does not state how a quantity develops over time, where it concentrates, or how two variables relate. Instead, such knowledge is implicit, surfacing only while relevant data is located, integrated, and methodically subjected to computations~\citep{jagadish_big_2014, herzig_open_2021, ji_target_2024}. What people require to act is therefore not the tabular data itself but the knowledge derived from it.

Deriving this knowledge is demanding~\citep{debie_automating_2022}, and not only because of the analysis it requires. Organizations often hold vast collections of tables without a precise account of what they contain or where any particular information resides, which makes locating the relevant data itself a substantial task~\citep{bogatu_dataset_2020, nargesian_data_2019}. Both locating the data and analyzing it require skills, effort, and familiarity with the data~\citep{jagadish_big_2014} that those who have a need for that knowledge may lack. This is characteristic of the open setting in organizations, where individuals do not know the location, contents, structure, and coverage of the required data in advance and where time, resource, and skill constraints are practical hurdles to extracting knowledge.

For instance, consider a public-health officer who wants to know how nitrogen dioxide (\ce{NO2}) pollution varies across a city's boroughs to inform a future traffic policy. Measurements are collected by the city's sensor network, yet the officer does not know where in the municipality's data lake they are stored, how they are organized, and how complete they are. The measurements are tagged with coordinates, so further data on administrative boundaries needs to be located and combined to analyze the pollution per borough.

Analytical knowledge remains effectively locked behind the discovery and analysis needed to surface it. Democratizing access to it could do for the knowledge held in tables what web search does for information held in documents, letting anyone with a need obtain insights without a background in data analysis or programming. Realizing this potential calls for treating the problem end-to-end, spanning the full path from the need a person has to presenting the knowledge in a form that delivers insights.

We refer to the pursuit of making the knowledge held in tabular data accessible to those who require it as \acrfull{task}. We understand end-to-end \acrshort{task} as a destination that research is advancing toward and whose foundations this work develops.

\begin{academicbox}{Open Tabular Insight Extraction}
    \textbf{\acrfull{task}} is the task of satisfying a sufficiently directed insight need by deriving and presenting the analytical knowledge that most fully serves it, over a corpus of tabular data in an open setting where corpus contents are potentially diverse and not known when the need is expressed.
\end{academicbox}

The term \textit{\acrlong{task}} captures the characteristics of the task. It is \textit{open} in that the need is expressed over a corpus whose contents, schema, and coverage are unknown in advance, requiring exploration of the available data. It concerns \textit{tabular} data, which enables analysis and shapes how data is discovered and combined. It targets \textit{insights} because the underlying need is satisfied through analytical knowledge that, interpreted within a person's contextual understanding, yields an insight. And it is \textit{extraction} because this knowledge is derived from the discovered data through analytical operations. 

\begin{figure}[ht]
    \centering
    \begin{subfigure}[t]{6.5cm}
        \includegraphics[width=6.4cm]{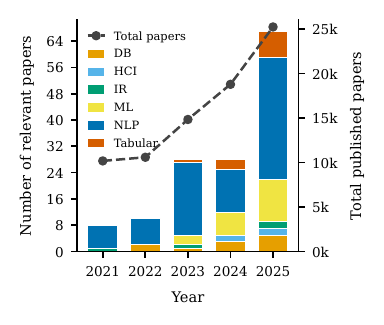}
        \vspace{-0.2cm}
        \caption{Total number of papers, and papers we select as relevant, across research communities, illustrating increased attention in the topic of OpenTI.}
        \label{fig:stats:papers_by_year}
    \end{subfigure}
    \hspace{0.4cm}
    \begin{subfigure}[t]{9.2cm}
        \includegraphics[width=9.05cm]{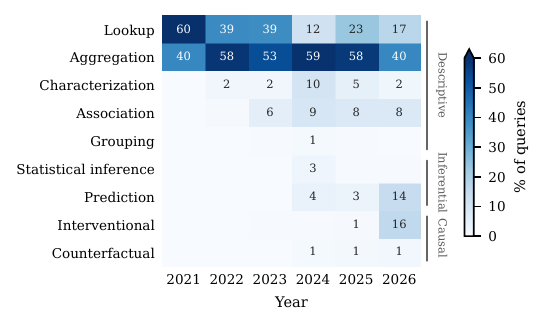}
        \vspace{-0.2cm}
        \caption{Prevalence of different types of analytical tasks in benchmarks per year, as the share of inputs averaged with equal weight per benchmark, showing a trend towards more diverse and complex tasks.}
        \label{fig:stats:insight_types_trend}
    \end{subfigure}
    \caption{Growth and diversification of research relevant to \acrshort{task}. Benchmark statistics cover 2026 only partially.}
    \label{fig:stats}
\end{figure}

Several research communities work toward easing access to the knowledge held in tabular data, each from a different starting point. Database (DB) research has long pursued dataset search to surface relevant tables within large collections~\citep{chapman_dataset_2020, nargesian_data_2019}, while semantic parsing of natural language into logical forms such as SQL~\citep{liu_survey_2025} sits at the intersection of DB and natural language processing (NLP). NLP has further focused on answering questions over tables and reasoning about their structure~\citep{zhou_table_2026, lu_large_2025}. Information retrieval (IR) has advanced the retrieval of tables relevant to a query~\citep{wang_retrieving_2021, chen_table_2024, chen_can_2025}, building on the open-domain tradition of retrieving evidence before answering~\citep{chen_open_2020, herzig_open_2021}. Recent research on agents assembles foundation models into systems that plan and execute analytical work over data~\citep{chen_large_2025, tang_llm_2025}. Human-computer interaction (HCI) has studied visual presentations of analytical results and the interfaces for specifying and refining the analysis~\citep{srinivasan_natural_2017, setlur_eviza_2016}. Each direction advances a complementary facet of \acrshort{task}, yet none spans the whole.

This activity is broad and accelerating faster than surrounding fields. We identify 160 relevant works from 18 venues spanning ML, IR, HCI, DB, and NLP in a systematic literature review, as detailed in Appendix~\ref{s:appendix:review_method}.
Between 2021 and 2025 the number of works relevant to \acrshort{task} rose roughly eightfold, far outpacing the overall growth at the same venues (Figure~\ref{fig:stats:papers_by_year}). Simultaneously, relevant works have increasingly appeared at cross-cutting machine learning venues and specialized workshops, indicating that interest is not limited to any single originating community. \acrshort{task} is thus emerging as an area in its own right.

That this is happening now reflects that the capabilities needed to holistically address the problem are maturing. Advances in retrieval and learned representations of tables enable uncovering relevant data from large corpora~\citep{wang_retrieving_2021, chen_table_2024, herzig_tapas_2020}, while \acrlong{llm}s and the agentic methods built upon them provide flexible means to address the full path from the expression of a need to presenting an insight~\citep{chen_large_2025, tang_llm_2025}. Capabilities traditionally pursued in isolation can thus be assembled into systems that attempt insight extraction end-to-end. Simultaneously, the range of data analysis workloads that systems are expected to cover is broadening. Figure~\ref{fig:stats:insight_types_trend} exemplifies this, showing a clear trend of the scope of benchmarks broadening over time, and only recently expanding beyond lookup and aggregation workloads.
The problem is thus outgrowing the narrow framings under which its facets have been studied. The components and the demand exist, yet the work remains scattered across communities without a shared account of the problem they collectively address.

While the systems proposed across these communities have grown more capable, they remain largely optimized for a narrow target, the information directly queried for, with success measured by how accurately an exact answer is returned. This only partly captures the requirements on surfacing knowledge from tabular data. Information extracted from tables is seldom wanted for its own sake but for the \textit{insight} it enables for the one who sought it.
What is requested and the insight actually sought need not align. The public-health officer from above may ask for a simple average of \ce{NO2} per borough, yet that figure reflects where sensors sit rather than where people are exposed, misleading the very policy decision the officer needs to inform.

Existing efforts to systematize the problem approach parts of it but fall short of providing a holistic account that spans the full end-to-end scope. Systematic efforts have organized works on open-domain question answering~\citep{zhang_survey_2023}, retrieval augmented generation~\citep{zhao_retrieval_2024}, and \acrshort{llm}-agents~\citep{kapoor_ai_2024, sumers_cognitive_2024}, establishing a structured perspective on systems that perform complex tasks over unseen information, yet none address tabular data. Those that do remain largely confined to surveying works in a narrow descriptive manner, organizing works within the conventions of specific research communities under labels such as text-to-SQL, table question answering, or data science agents~\citep{fang_large_2024, lu_large_2025, liu_survey_2025, zhou_table_2026, tang_llm_2025, tian_realworld_2026, chen_large_2025}, which we compare against in detail in Appendix~\ref{s:appendix:related_work}. We argue that these labels reflect facets of one problem, that treating them separately obscures what has been established and what remains open, and that a unified account of that problem is achievable from first principles. This separation is reflected in how works engage with one another, where works cite others that use the same label at six times the rate they cite across labels (Section~\ref{sec:problem:spectrum}). We therefore develop \acrshort{task} as a field of its own, providing the shared vocabulary and problem definition needed to transfer knowledge between communities and to compare systems and benchmarks beyond the conventions of the community each originates from. Overall, we make the following contributions:

\begin{enumerate}
    \item \textbf{Conceptual foundation.} We establish principled foundations for \acrshort{task} by formalizing it as a unified problem setting that addresses \textit{insight needs} with \textit{analytical knowledge} derived through \textit{realizations} over corpora of tabular data (Section~\ref{sec:problem}). This formulation allows deriving the functional capabilities systems must exercise (Section~\ref{sec:orchestration}), and it separates establishing that a realization is valid from measuring how well it serves the need, revealing that current benchmarks approximate the former while leaving the latter largely unmeasured (Section~\ref{sec:eval}). We use this framework to position systems and benchmarks from different research communities within a single scheme (Tables~\ref{tab:functional_capabilities_overview} and~\ref{tab:datasets}).
    \item \textbf{Review and analysis.} We systematically review how works from different research areas work towards \acrshort{task}. We apply the conceptual framework to position systems by the functional capabilities they exercise and the way that they orchestrate the interplay between them (Section~\ref{sec:orchestration}) and analyze benchmarks against desired criteria for robust evaluations (Section~\ref{sec:eval}), showing that current systems and benchmarks only partially cover the scope and requirements of OpenTI.
    \item \textbf{Synthesis.} We consolidate fragmented work on the human-side dimensions of \acrshort{task} under a framing of cooperative interaction. We systematize interaction modes and means and examine interpretability, personalization, and contextualization as means of aligning system realizations with users' latent insight needs (Section~\ref{sec:user_facing}).
    \item \textbf{Open challenges and research agenda.} We derive a research agenda for \acrshort{task} in which directions are anchored in the gaps and limitations surfaced in the review, structured along key questions towards reliable and accessible insight extraction (Section~\ref{sec:research_agenda}).
\end{enumerate}

We find a consistent pattern across existing research. Effort concentrates on performing limited types of analysis over data that is already at hand, while complementary aspects within the end-to-end scope remain thin. 41 of the 58 systems we review lack any capability to retrieve tables, while outputs are predominantly text and rarely contextualized for specific insights. Similarly, we find current benchmarks to be largely unfit for evaluations under an open setting. We analyze inputs from benchmarks and find that they routinely reference columns, files, and values that a user in an open setting could not know, and that they can be interpreted in multiple ways in the majority of instances of all but two benchmarks while largely assessing the validity of outputs or the analysis against a single reference.

\section{Open Tabular Insight Extraction}
\label{sec:problem}
Since research towards \acrshort{task} originates in different communities that address overlapping problems with divergent terminology and assumptions, this section develops a formal framework for \acrfull{task}. We lay out this shared understanding by establishing the core concepts and formalizing \acrshort{task} (Section~\ref{sec:problem:conceptual_foundations}). We then examine the conditions under which insight needs can be satisfied (Section~\ref{sec:problem:sufficiency}). Finally, we cover the spectrum of insight needs that \acrshort{task} addresses (Section~\ref{sec:problem:spectrum}).

\begin{figure}[ht]
    \centering
    \includegraphics[width=0.95\linewidth]{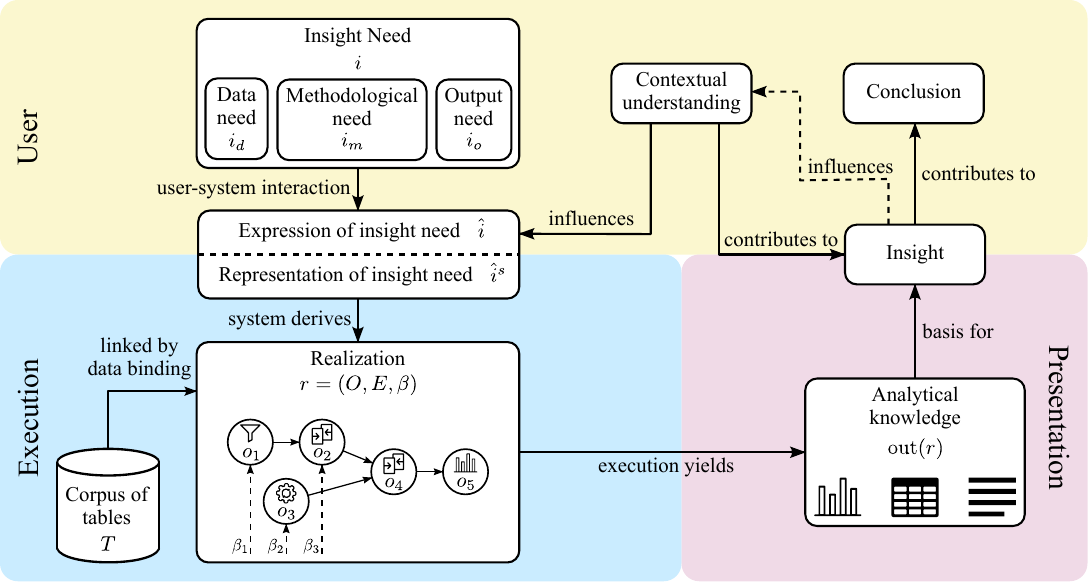}
        \caption{Relationships and dependencies among core concepts of \acrshort{task}. A user holds an insight need $i$, consisting of a data need $i_d$, methodological need $i_m$, and output need $i_o$, and expresses it as $\hat{i}$. A system maps $\hat{i}$ to a realization $r=(O,E,\beta)$, a graph of operations $O$ connected by dataflow edges $E$ with a binding $\beta$ that links operations to tables from the corpus $T$. Executing $r$ yields analytical knowledge as output $\out(r)$, which is the basis for an insight.}
    \label{fig:insight_need}
\end{figure}

\subsection{Conceptual Foundation}
\label{sec:problem:conceptual_foundations}

We develop the conceptual foundation of \acrshort{task} from first principles. We first ground the general notion of an \textit{insight} as the form of knowledge to be delivered (Section~\ref{sec:problem:conceptual_foundations:insights}). We then operationalize this into \textit{insight needs}, characterizing what users seek in tabular data and along which dimensions it varies (Section~\ref{sec:problem:conceptual_foundations:insight_needs}). Building on this, we formulate the problem of \acrshort{task} (Section~\ref{sec:problem:conceptual_foundations:open_tix}), show how the analytical knowledge that serves an insight need derives from data through realizations, executable compositions of operations bound to data (Section~\ref{sec:problem:conceptual_foundations:realizations}), and finally formulate \acrshort{task} as an optimization problem (Section~\ref{sec:problem:conceptual_foundations:problem_formulation}). Figure~\ref{fig:insight_need} provides an overview of the relationships between the concepts introduced throughout this section.

\subsubsection{Insights}
\label{sec:problem:conceptual_foundations:insights}

\textit{Insights} have been defined and discussed in various ways across visualization research~\citep{north_measuring_2006, chang_defining_2009, saraiya_insightbased_2005}, cognitive science~\citep{sternberg_nature_1994, mai_aha_2004}, and data analytics~\citep{mikalef_big_2018, tang_extracting_2017, ding_quickinsights_2019}, with ongoing debate about their defining characteristics, including whether insights arise unexpectedly by definition or not~\citep{north_measuring_2006, chang_defining_2009, battle_what_2024}. Adapting the recent synthesis of \citet{battle_what_2024}, which consolidates these perspectives, we understand an insight as a \textit{collection of knowledge} that links \textit{analytical knowledge} derived from data with a person's pre-existing \textit{contextual understanding}. Analytical knowledge is the direct output of computational operations on data, whereas contextual understanding denotes a person's pre-existing knowledge, including domain expertise, organizational background, prior experiences, and previously acquired information~\citep{pirolli_sensemaking_2005}. As such, insights exhibit the following characteristics:

\begin{itemize}
    \item \textit{Grounded in data.} An insight is grounded in analytical knowledge extracted from data through analytical operations, such as transformations, statistical modeling, or other computations~\citep{battle_what_2024}.
    \item \textit{Contextual.}
    An insight is not the analytical knowledge itself but its interpretation within a person's contextual understanding~\citep{battle_what_2024, karer_insight_2021}.
    The role of contextual understanding ranges from minimal, when the analytical knowledge is largely self-explanatory (e.g., a retrieved fact), to substantial, where domain expertise fundamentally shapes interpretation (e.g., in understanding ROC curves or residual plots).
    \item \textit{Independent of presentation.} Since insights emerge from interpretation, they are not bound to a specific form of presenting analytical knowledge~\citep{battle_what_2024}. Instead, different forms of presenting analytical knowledge may invoke the same insights.
    \item \textit{Compositional.} Insights contribute to a person's contextual understanding, and subsequent insights can build on this enriched context~\citep{battle_what_2024}. This hierarchical structure means that complex insights may build on the context of prior insights~\citep{north_measuring_2006}, that is, there exist insights that compose a set of sub-insights, which may themselves be composed of sub-insights.
\end{itemize}

\subsubsection{Insight Needs}
\label{sec:problem:conceptual_foundations:insight_needs}

An \textit{insight need} arises when a person\footnote{We refer to a \textit{person} when characterizing insight needs conceptually and to a \textit{user} when referring to the entity interacting with a system. We note that this entity need not be human and may instead be an automated agent acting on an insight need.} requires analytical knowledge that is derived from data. We ground this concept in the established notion of \textit{information needs} from information retrieval, where an information need represents the gap in a person's knowledge that motivates interaction with an information system~\citep{taylor_process_1962, belkin_anomalous_1980}. An insight need specializes this to the analytical domain by targeting a knowledge gap concerning analytical conclusions residing in data, where satisfying this need requires computationally deriving knowledge rather than retrieving pre-existing information alone.

\begin{definition}
    \label{def:insight_need}
    An \textbf{insight need} $i$ is a person's requirement for analytical knowledge which, when interpreted within their contextual understanding, constitutes an insight.
\end{definition}

Insight needs are \textit{analytical} and \textit{insight-centered}. They seek analytically derived knowledge (i.e., facts, patterns, relationships, predictions, or other analytical conclusions), rather than to produce an artifact. For instance, a need for a trained machine learning model is not an insight need, though artifacts may be instrumental in delivering insights, such as when a model is applied to make a desired prediction.
In addition to these characteristics, insight needs vary along the following \textit{dimensions}:

\begin{itemize}
    \item \textit{Directedness.} Insight needs range from highly directed, targeting specific phenomena, entities, or relationships (e.g., ``What is the correlation between drug dosage and patient recovery time?''), to loosely directed, where the analytical perspective remains open (e.g., ``What patterns exist in hospital readmission data?'')~\citep{marchionini_exploratory_2006}. The degree of directedness determines whether it is possible to evaluate whether candidate analytical knowledge satisfies the insight need.
    \item \textit{Latency.} The actual insight need $i$ exists in a person's mind and may not be entirely specified in their expression of it~\citep{zhu_automated_2025}. A person can express an approximation $\hat{i}$ through utterances and contextual signals. This gap between latent need and expressed need mirrors the classical model of information need formation by \citet{taylor_process_1962}, which traces how needs progress from an unexpressed visceral sense through conscious and formalized stages to a compromised expression adapted to system capabilities. For insight needs, the complexity of the need is compounded by the analytical requirements of selecting which operations should be performed on which data.
    Latency is rooted in the person's contextual understanding, including tacit knowledge the person cannot fully articulate~\citep{polanyi_tacit_1967}, assumptions the person considers obvious, and preferences the person has not consciously examined~\citep{belkin_anomalous_1980, taylor_process_1962}.
    \item \textit{Compositionality.} Following the compositional nature of insights, insight needs can also be composed of atomic needs. They may decompose into sub-needs that jointly describe the insight need, which may themselves decompose recursively, allowing arbitrarily complex compositions of insight needs.
\end{itemize}

\subsubsection{Open Tabular Insight Extraction}
\label{sec:problem:conceptual_foundations:open_tix}

Insight needs may target any form of data and any analytical domain. \acrlong{task} scopes this to satisfying insight needs by deriving analytical knowledge from a corpus of tables, where users may have an incomplete knowledge of its contents, schemas, and coverage.

\begin{definition}
    \label{def:opentix}
    \textbf{\acrfull{task}} is the task of satisfying a sufficiently directed insight need by deriving and presenting the analytical knowledge that most fully serves it, over a corpus of tabular data in an open setting where corpus contents are potentially diverse and not known when the need is expressed.

\end{definition}

Recalling the directedness dimension introduced above, \acrshort{task} addresses insight needs that are sufficiently directed so that it is determinable whether candidate analytical knowledge satisfies the need. Since the need itself is latent, no party can check this condition directly. We thus operationalize this sufficient specification on the observable expression of the insight need $\hat{i}$ in Section~\ref{sec:problem:sufficiency:specification}.

\paragraph{Tabular Data.} 
The task operates on a collection of tabular data. A table $t$ consists of a header, an ordered set of column names that defines the table's schema, a data matrix holding the core data values, where a row typically corresponds to an entity and a cell holds the value of the attribute named by its column, and metadata that may provide contextual information such as titles, captions, column descriptions, or surrounding text~\citep{liu_rethinking_2024}. A corpus $T = \{t_1, t_2,...,t_n\} \subset \mathcal{T}$ comprises $n$ such tables and can correspond to any large-scale collection of tables like a data lake, warehouse, or repository. In practice, tabular data largely exists in relational databases~\citep{codd_relational_1970}, which add interconnections between tables through explicit schema relationships such as primary and foreign keys that define how entities relate across tables and how distributed information can be combined. In contrast, data lakes~\citep{nargesian_data_2019} and collections of web tables~\citep{cafarella_webtables_2008, hulsebos_gittables_2023} aggregate tables from heterogeneous sources without such explicit relationships, exhibiting variation in schema quality, naming conventions, and completeness. The existence or absence of these relationships places different requirements on systems~(Section~\ref{sec:problem:conceptual_foundations:realizations}).

\paragraph{Open Setting.}
\acrshort{task} operates under an \textit{open} setting in which users express insight needs over a corpus whose contents, schemas, and coverage are unknown to them. This contrasts with closed settings~\citep{yu_spider_2018, li_can_2024}, where users formulate requests with implicit knowledge of the available data. Users therefore cannot be expected to express insight needs in terms that align with the data's vocabulary, structure, or scope, requiring systems to identify relevant data from $T$. As under the open-world assumption~\citep{reiter_closed_1981}, users do not know whether relevant data is captured in the corpus, meaning the system has to establish whether $T$ covers a given insight need. As in open-domain (tabular) question answering and retrieval~\citep{voorhees_trec_2001, kong_opentab_2023, chen_open_2020, herzig_open_2021}, the relevant data must be retrieved from $T$, though without being contingent on topical breadth, meaning that a large single-domain data lake is also considered open under \acrshort{task}.

\paragraph{Decomposition of Insight Needs.}
We decompose insight needs into data, methodological, and output factors following task typologies in visualization, which distinguish the dimensions of ``what'', ``how'', and ``why''~\citep{brehmer_multilevel_2013}, and mirroring the implicit structure of database queries, which separately specify data selection, analytical operations, and output formatting:

\begin{enumerate}
    \item \textbf{Data need} ($i_d$): The specification of what data entities, temporal scope, and domain boundaries are relevant for the analysis. This may be explicitly stated (e.g., ``revenue of Fortune 500 companies in 2023'') or implicitly indicated through contextual cues and conventions (e.g., ``GDP of Liechtenstein last year'', if the current year is known by the system).
    \item \textbf{Methodological need} ($i_m$): The specification of what analytical operations and procedures should be applied to extract the desired analytical knowledge. This ranges from explicit methodological directives (e.g., ``calculate the Pearson correlation'') to high-level analytical goals that delegate methodological choices to the system (e.g., ``analyze the relationship between\ldots'').
    \item \textbf{Output need} ($i_o$): The specification of how results should be presented. This may remain completely unspecified by a user that expects the system to select an appropriate presentation.
\end{enumerate}

\subsubsection{Realizations}
\label{sec:problem:conceptual_foundations:realizations}
We formalize the relationship between data and the derived analytical knowledge as \textit{realizations}:

\begin{definition}
    \label{def:realization}
    A \textbf{realization} $r=(O,E,\beta)$ is a complete, executable specification of computational logic with a binding of specific tables to that logic, which, when executed, yields analytical knowledge as output $\out(r)$.
\end{definition}

We adopt the term from systems theory, where a realization denotes a concrete implementation of an input-output behavior~\citep{kailath_linear_1980, kalman_mathematical_1963}. In \acrshort{task}, a realization captures a derivation path from data in the corpus to analytical knowledge end-to-end. Whether the analytical knowledge a realization derives satisfies a given need is a property of the realization relative to that need.

We represent realizations as a \acrfull{dag} of operations connected by edges denoting data flow, similar to representations of computations established in scientific workflow specifications~\citep{ludascher_scientific_2006, deelman_workflows_2009} and database query execution plans~\citep{graefe_volcano_1994}. We adapt this representation to insight extraction, where computational operations are anchored to data discovered from a corpus rather than pre-specified inputs. Unlike a database query, a realization is often a multistep workflow involving data discovery, integration, transformation, analysis, and presentation.
Formally, a realization $r \in \mathcal{R}$ is a tuple $r = (O, E, \beta)$ describing a converging directed acyclic graph of operations:

\begin{itemize}
    \item $O = \{o_1, o_2, \ldots, o_m\}$ is the set of fully parameterized operations applied in the derivation.
    \item $E \subset O \times O$ is a set of directed edges connecting the output of operations to the inputs of subsequent operations, defining the data flow.
    \item $\beta \subset \mathcal{T} \times O$ is the data binding that connects specific tables $T_r \subseteq \mathcal{T}$ to the source nodes of the graph.
\end{itemize}

\begin{figure}[ht]
    \centering
    \includegraphics[width=0.9\linewidth]{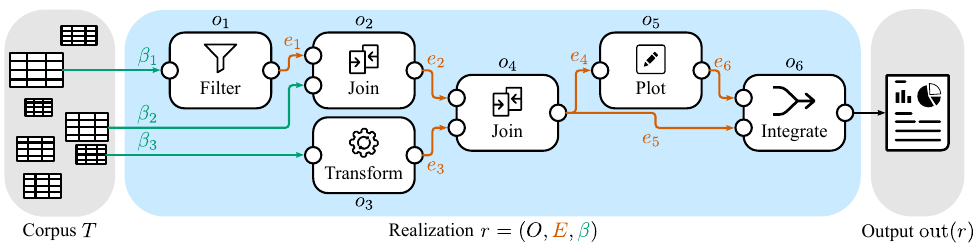}
    \caption{Example of an abstracted realization $r=(O,E,\beta)$. Nodes are fully parameterized operations $O$, edges $E$ denote dataflow, and the binding $\beta$ anchors source operations to corpus tables. The graph converges to a single sink operation producing the output.}
    \label{fig:formalization_realization}
\end{figure}

An \textit{operation} $o$ represents a discrete functional unit, which can range from granular primitives (e.g., filter, join, \ldots) to high-level analytical functions (e.g., fit a regression model, solve an optimization problem, \ldots). Each operation in $O$ is fully parameterized, meaning all relevant choices, such as aggregation functions, model specification, or visualization parameters, are determined. The edges $E$ capture the compositional structure of the derivation, specifying how intermediate results flow between operations. The data binding $\beta$ anchors the computation to concrete tables, specifying which tables serve as inputs to which operations. The \acrshort{dag} converges into a single sink operation, which produces the final output resulting from the realization. Figure~\ref{fig:formalization_realization} exemplifies an abstracted realization that derives analytical knowledge to address an insight need.

Importantly, a realization is a declarative account of \textit{how} analytical knowledge is derived from data. It does not capture the process by which a system arrived at this specification, which may include exploratory search, retrieval failures, user interactions, and iterative refinement. This parallels the established differentiation between \textit{prospective provenance} (the specification of a computational task) and \textit{retrospective provenance} (the record of what was actually executed) in scientific workflow research~\citep{davidsonProvenanceScientificWorkflows2008, freire_provenance_2008}. Section~\ref{sec:orchestration} examines the process by which systems derive realizations.

\paragraph{Multiple Realizations and Utility.}
A single insight need may admit many realizations. When users express a methodological goal like ``analyze the relationship between X and Y,'' different operational compositions (e.g., Pearson correlation, Spearman correlation, regression analysis) may each constitute a valid way to address this intent. Similarly, different combinations of tables from $T$ could support distinct realizations. This one-to-many relationship reflects a division of labor in cooperative interaction between the user and the system~\citep{gomm_are_2025, horvitz_principles_1999}, where users grant systems agency over the aspects of their needs they leave open, requiring the system to select among the resulting alternatives. 
These alternatives, however, are rarely equal in value. A realization drawing on more recent or more complete data, employing more robust analytical methods, or producing clearer output may substantially better serve the user's goals. We model this qualitative dimension through a utility function $u_i$.

\begin{definition}
    \label{def:utility} 
    For a given insight need $i$, the \textbf{realization utility function} $u_i: \mathcal{R} \to \mathbb{R}$ maps possible realizations to a real-valued score, where a higher value indicates a realization that more effectively, accurately, or suitably satisfies the user's insight need.
\end{definition}

The utility function aggregates the quality with which a realization addresses data, methodological, and output needs, capturing not just the value of the final outputs but the full realization performed~\citep{ghallab_automated_2004}. For the data component, utility reflects the relevance of the data~\citep{saracevic_relevance_2007, cooper_definition_1971}, reflecting factors such as recency, completeness, and reliability~\citep{wang_accuracy_1996}. For the methodological structure, utility may capture the appropriateness and rigor of analytical procedures given the analytical intent. For the output presentation, utility may reflect clarity, informativeness, as well as the effectiveness and appropriateness of visualizations~\citep{lei_dacomp_2025}. While users do not explicitly specify what contributes to their utility functions~\citep{borlund_concept_2003}, these functions embody the implicit quality criteria that distinguish superior realizations from adequate ones. Utility only captures how well a realization satisfies an insight need, regardless of how it was derived. Computational effort, latency, and the interaction burden placed on the user are properties of the process by which a system constructs a realization, which influence the usefulness of a system but do not influence the utility of a realization.

\subsubsection{Problem Formulation}
\label{sec:problem:conceptual_foundations:problem_formulation}

A realization is \textit{valid} for an insight need $i$ if the tables bound through $\beta$ provide the information the data need $i_d$ mandates, its operations and their composition correctly address the methodological need $i_m$, and its output delivers the kind of result the output need $i_o$ calls for. Validity is a property of the realization relative to the insight need, judged on the content of the bound tables and the analytical logic applied to them, independent of whether the bound tables actually exist in the corpus. We denote the set of valid realizations for an insight need $i$ as $R_{\text{valid}}(i) \subset \mathcal{R}$. Validity and utility are complementary. Validity is a binary criterion that separates realizations addressing the insight need from those that do not, whereas the utility function $u_i$ grades how well realizations serve it. A realization can thus be valid yet exhibit low utility, for instance when its output delivers the required analytical knowledge but presents it poorly.

A realization is \textit{feasible} given a corpus $T$ if the tables it requires are present in the corpus, that is, all tables bound via $\beta$ exist in $T$, $T_r \subseteq T$. Feasibility is a property of the data binding relative to the corpus, independent of whether the realization addresses any particular insight need or whether a given system possesses the capabilities to combine the bound tables. We denote the set of feasible realizations as $R_{\text{feasible}}(T) \subset \mathcal{R}$.

The set of \textit{candidate realizations} $R_{\text{cand}}(i, T) = R_{\text{valid}}(i) \cap R_{\text{feasible}}(T)$ captures realizations that both address the insight need and are grounded in available data. The realization that optimally satisfies the insight need is then the candidate realization $r^*$ that maximizes utility:

\begin{equation}
    \label{eq:objective}
    r^* = \begin{cases}
    \arg\max_{r \in R_{\text{cand}}(i, T)} u_i(r), & \text{if}\ R_{\text{cand}}(i, T) \neq \varnothing \\
    \bot, & \text{if}\ R_{\text{cand}}(i, T) = \varnothing,
    \end{cases}
\end{equation}

where $\bot$ represents a distinct failure state indicating that the insight need cannot be resolved given the corpus.
This objective is defined relative to the insight need $i$ and thus characterizes an idealized target. A system never has direct access to $i$, and consequently not to $R_{\text{valid}}(i)$ or $u_i$, observing only the expressed need $\hat{i}$. To construct an approximation of $r^*$, a system must construct its own representation of the need, denoted as $\hat{i}^s$, from the expression $\hat{i}$ and available contextual information, bringing $\hat{i}^{s}$ into closer correspondence with $i$ (see Section~\ref{sec:user_facing}). The realization space $\mathcal{R}$ is combinatorially vast, as the number of possible operation compositions, parameterizations, and data bindings grows with the scale of the corpus and the expressiveness of the available operations. Neither an analytical solution nor enumeration is practical. Section~\ref{sec:orchestration} thus explores how systems can navigate this space effectively to construct utility-maximizing realizations.

\subsection{Sufficiency Criteria}
\label{sec:problem:sufficiency}

Not all insight needs can be satisfied in \acrshort{task}. As implied in the formalization, a system can only produce an output that satisfies the insight need if the candidate set $R_{\text{cand}}(i, T) = R_{\text{valid}}(i) \cap R_{\text{feasible}}(T)$ is non-empty. This requires that the expression of the insight need $\hat{i}$ is sufficiently specified for a system to delineate $R_{\text{valid}}(i)$ (Section~\ref{sec:problem:sufficiency:specification}), and that the corpus $T$ contains sufficient data so that at least one valid realization is feasible, i.e., $R_{\text{valid}}(i) \cap R_{\text{feasible}}(T)$ is non-empty (Section~\ref{sec:problem:sufficiency:data}).

\subsubsection{Sufficient Specification}
\label{sec:problem:sufficiency:specification}

Sufficient specification describes the expressed need $\hat{i}$ relative to the underlying need $i$. $\hat{i}$ needs to allow delineating $R_{\text{valid}}(i)$, separating realizations that satisfy the insight need from those that do not. This requires that the data, methodological, and output components $i_d$, $i_m$, $i_o$ are specified or can be inferred accordingly. For instance, ``What is the average temperature in Amsterdam in January?'' is sufficiently specified since the implicit methodological details can be resolved through conventions and reasonable defaults. In contrast, ``How does construction waste compare?'' is insufficient since no convention fixes what to compare against, by what metric, or in what region, leaving valid and invalid realizations indistinguishable. We discuss how systems close such gaps autonomously and in collaboration with the user in Section~\ref{sec:user_facing}.

\subsubsection{Data Sufficiency}
\label{sec:problem:sufficiency:data}

Even a fully specified insight need is impossible to satisfy if the corpus lacks the required data, which is particularly relevant when users are unaware of the contents and structure of $T$. Satisfiability requires that the corpus $T$ supports at least one valid realization. We formalize this as sufficient sets of tables:

\begin{definition}
    \label{def:sufficient_data}
    \textit{A set of tables $T_s \subseteq T$ is \textbf{sufficient} for an insight need $i$ if there exists at least one valid realization $r=(O, E, \beta) \in R_{\text{valid}}(i)$ where the set of tables $T_r$ bound via $\beta$ satisfies $T_r \subseteq T_s$.}
\end{definition}

When no sufficient set exists within the corpus, that is $\nexists T_s \subseteq T$ sufficient for $i$, the candidate set is empty and the insight need cannot be resolved, an instance of \textit{insufficient data coverage} for $i$ in $T$. Within a sufficient set, tables align both with the insight need and with each other. Following \citet{kothyari_crush4sql_2023} and \citet{chen_table_2024}, we distinguish two forms of alignment:

\begin{enumerate}
    \item \textbf{Semantic alignment (need-table):} The tables in $T_s$ collectively cover all semantic concepts specified in the data need $i_d$.
    \item \textbf{Structural alignment (table-table):} Tables holding complementary information are combinable through the operations available to the system (e.g., joinable via shared keys, unionable,\ldots).
\end{enumerate}

The two forms are independent. A set may cover all relevant semantic concepts yet be structurally incompatible, and tables may be perfectly joinable yet lack the required semantic coverage, both rendering the set insufficient. The insight need defines how semantic components must relate, while a system's operational capabilities determine how information can be combined across tables.

\begin{figure}[ht]
    \centering
    \begin{tikzpicture}[font=\small]
    \def\dx{5.5cm}

    \node[draw, thick, rounded corners, fill=gray!10, inner sep=1mm, text width=15.5cm, align=left] at (1*\dx, 4.2) {
    \textbf{Example Query:} \\
    \texttt{Plot the relationship between the \colorbox{okabe_4}{population} and \colorbox{okabe_3}{GDP per capita} of each \colorbox{okabe_1}{EU country} in \colorbox{okabe_2}{2021}.}
    };
        
    \node[align=center, text width=5cm] at (0,2.5) {\textbf{Not semantically \& structurally aligned}};
    \node[align=center, text width=5cm] at (\dx,2.5) {\textbf{Not structurally aligned}};
    \node[align=center, text width=5cm] at (2*\dx,2.5) {\textbf{Semantically \& structurally aligned}};
    
    \draw[dashed] (\dx/2,-4.25) -- (\dx/2,3);
    \draw[dashed] (1.5*\dx,-4.25) -- (1.5*\dx,3);
    
    \begin{scope}[xshift=0cm, yshift=0.5cm]
      \node[draw, thick, inner sep=1mm] (t11) {
        \begin{tabular}{ll}
        \textbf{\colorbox{okabe_1}{Country}} & \textbf{\colorbox{okabe_4}{Population}} \\
        \hline
        Germany & 83 200 000 \\
        France  & 67 400 000 \\
        $\cdots$ & $\cdots$
        \end{tabular}
      };
      \node[above=0cm of t11] (t11_t) {\texttt{population\_by\_country\_\colorbox{okabe_2}{2021}}};
      \node[draw, thick, inner sep=1mm, below=1cm of t11, , minimum height=1.5cm] (t12) {
        \begin{tabular}{ll}
        \textbf{Region} & \textbf{EmploymentRate} \\
        \hline
        DE11       & 75.5\% \\
        DE12       & 66.2\% \\
        $\cdots$ & $\cdots$
        \end{tabular}
      };
      \node[above=0cm of t12] (t12_t) {\texttt{employment\_\colorbox{okabe_2}{2021}}};
      \node[align=center, text width=5cm, below=0.25cm of t12] (t1_desc) {GDP information is missing and tables cannot be combined.};
    \end{scope}

    \begin{scope}[xshift=\dx, yshift=0.5cm]
      \node[draw, thick, inner sep=1mm] (t21) {
        \begin{tabular}{ll}
        \textbf{\colorbox{okabe_1}{Country}} & \textbf{\colorbox{okabe_4}{Population}} \\
        \hline
        Germany & 83 200 000 \\
        France  & 67 400 000 \\
        $\cdots$ & $\cdots$
        \end{tabular}
      };
      \node[above=0cm of t21] (t21_t) {\texttt{population\_by\_country\_\colorbox{okabe_2}{2021}}};
      \node[draw, thick, inner sep=1mm, below=1cm of t21] (t22) {
        \begin{tabular}{ll}
        \textbf{\colorbox{okabe_1}{Pays}}       & \textbf{\colorbox{okabe_3}{PIB}} \\
        \hline
        Allemagne  & €46 000 \\
        France     & €39 000 \\
        $\cdots$ & $\cdots$
        \end{tabular}
      };
      \node[above=0cm of t22] (t22_t) {\texttt{\colorbox{okabe_3}{pib\_par\_habitant}\_\colorbox{okabe_2}{2021}}};
      \node[align=center, text width=5cm, below=0.25cm of t22] (t2_desc) {All attributes present but cannot be combined because of missmatch in country keys.};
    \end{scope}

    \begin{scope}[xshift=2*\dx, yshift=0.5cm]
      \node[draw, thick, inner sep=1mm] (t31) {
        \begin{tabular}{ll}
        \textbf{\colorbox{okabe_1}{CountryCode}} & \textbf{\colorbox{okabe_4}{Population}} \\
        \hline
        DE          & 83 200 000 \\
        FR          & 67 400 000 \\
        $\cdots$ & $\cdots$
        \end{tabular}
      };
      \node[above=0cm of t31] (t31_t) {\texttt{country\_population\_\colorbox{okabe_2}{2021}}};
      \node[draw, thick, inner sep=1mm, below=1cm of t31] (t32) {
        \begin{tabular}{ll}
        \textbf{\colorbox{okabe_1}{CountryCode}} & \textbf{\colorbox{okabe_3}{GDP}} \\
        \hline
        DE          & €46 000 \\
        FR          & €39 000 \\
        $\cdots$ & $\cdots$
        \end{tabular}
      };
      \node[above=0cm of t32] (t32_t) {\texttt{\colorbox{okabe_3}{gdp\_per\_capita}\_\colorbox{okabe_2}{2021}}};
      \node[align=center, text width=5cm, below=0.25cm of t32] (t3_desc) {All attributes present and joinable on CountryCode.};
    \end{scope}
\end{tikzpicture}
    \vspace{-0.75cm}
    \caption{Example of semantic and structural alignment. A candidate set is semantically aligned when its tables jointly cover the required concepts and structurally aligned when they are combinable, here via a join on CountryCode. Both must hold for sufficiency.}
    \label{f:alignment_example}
\end{figure}

Figure~\ref{f:alignment_example} exemplifies these conditions. Where the required information does not reside in a single table, it must be integrated across multiple tables. As shown, a sufficient set must cover population counts, GDP values, country identifiers, and the year 2021 (semantic alignment), and these must be integrable through the means a system possesses, here through shared keys like CountryCode (structural alignment). The conditions for structural alignment depend on the capabilities of the system processing the data. For a system capable of aligning French and English contents, for example, the middle column would be structurally aligned. How systems surface such sets and navigate the search over candidate sets that both conditions induce is examined under the data retrieval capability in Section~\ref{sec:orchestration:capabilities:retrieval}.

\subsection{Taxonomy of Insight Types}
\label{sec:problem:spectrum}

Deriving the analytical knowledge that satisfies an insight need requires substantially different assumptions and capabilities depending on which kind of knowledge is sought. We organize this space by \textit{insight type}, the kind of analytical knowledge a realization derives, rather than by insight need. An insight need $i$ is latent and only partially captured by its expression $\hat{i}$ (Section~\ref{sec:problem:conceptual_foundations:insight_needs}), whereas the insight type is observable on the realization and most directly determines the capabilities a system requires. The operations $O$ and their composition $E$ differ categorically across types, such as between reading a stored value and estimating a causal effect, while data and output complexity vary within a type without imposing qualitatively different requirements. Figure~\ref{fig:insight_types_taxonomy} organizes insight types into three tiers ordered by what must be assumed, beyond locating the relevant data, for the derived knowledge to be justified, following the established distinction between descriptive, inferential, and causal analysis~\citep{tukey_exploratory_1977, pearl_causality_2009} applied to \acrshort{task}.

\begin{figure}[ht!]
    \centering
    \footnotesize
    \begin{forest}
    where={level()<2}{}{draw, very thick,minimum width=15.5cm,text width=15.3cm,inner sep=2pt, rounded corners=4pt},
    for tree={grow'=0,
    folder,
    align=left,
    font=\footnotesize,
    s sep=1mm,
    inner sep=2pt,
    fork sep=3mm}
    [\textbf{Insight Types}
        [{\textbf{Descriptive}\\\scriptsize Knowledge about the observed data, requiring no assumptions beyond it.}
            [{\textbf{Lookup}\\
            \scriptsize Reading specific values or records present in the corpus, covering both attribute values for identified entities and sets of\\\scriptsize entities matching stated criteria.\\
            \scriptsize\textit{``What was the population of Germany in 2022?''}\\
            \scriptsize\textit{``Which countries joined the European Union after 2000?''}},draw=okabe_1
            ]
            [{\textbf{Aggregation}\\
            \scriptsize Computing summary values over a set of records, including counts, sums, extrema, and rankings.\\
            \scriptsize\textit{``Which Mercosur member country has the highest public debt as a share of GDP?''}},draw=okabe_1]
            [{\textbf{Characterization}\\
            \scriptsize Describing the statistical distribution of a variable, including its central tendency, spread, and shape.\\
            \scriptsize\textit{``How are household incomes distributed in New Zealand?''}},draw=okabe_1]
            [{\textbf{Association}\\
            \scriptsize Measuring how two or more variables co-vary in the observed data, ranging from correlation to general dependence.\\
            \scriptsize\textit{``What is the correlation between public education spending and literacy rates across countries?''}},draw=okabe_1]
            [{\textbf{Grouping}\\
            \scriptsize Discovering segments of entities or records that share latent structure across attributes, without a predefined target\\ \scriptsize variable.\\
            \scriptsize\textit{``What clusters of countries exist based on their energy-mix profiles?''}},draw=okabe_1] 
        ]
        [{\textbf{Inferential}\\\scriptsize Knowledge about the data-generating process behind the observed data, requiring assumptions that link the two.}
            [{\textbf{Statistical Inference}\\
            \scriptsize Determining whether an observed pattern reflects a stable feature of the underlying process rather than noise, with \\\scriptsize parameter estimates and quantified uncertainty.\\
            \scriptsize\textit{``Is there evidence of a positive association between public education spending and literacy rates across countries?''}},draw=okabe_3
            ]
            [{\textbf{Prediction}\\
            \scriptsize Estimating unknown or future values for unobserved instances by generalizing observed patterns, spanning regression, \\\scriptsize classification, and forecasting.\\
            \scriptsize\textit{``What is the expected population of Algeria in 2040?''}},draw=okabe_3]
        ]
        [{\textbf{Causal}\\\scriptsize Knowledge about the effects of interventions or counterfactual states, requiring assumptions on the underlying causal structure.}
            [{\textbf{Interventional}\\
            \scriptsize Determining the effect of an action or policy at the population level.\\
            \scriptsize\textit{``How would commuting mode choices change if fuel prices rose by 25\%?''}},draw=okabe_5
            ]
            [{\textbf{Counterfactual}\\
            \scriptsize Determining the outcome for a specific observed instance under a condition contrary to the known facts.\\
            \scriptsize\textit{``Had the price not increased, would this customer have churned?''}},draw=okabe_5]
        ]
    ]
\end{forest}
    \caption{Taxonomy of insight types in \acrshort{task}, organized by the assumptions required for the derived analytical knowledge. Each tier encompasses representative insight types with examples.}
    \label{fig:insight_types_taxonomy}
\end{figure}

\textit{Descriptive} insight types concern the observed data itself and require no assumptions about anything beyond it. They form the most varied and, in current research, the most common tier (see Figure~\ref{fig:stats:insight_types_trend}), ranging from exact computed facts to summary characterizations of the data. Retrieving a cell value or selecting entities that meet a criterion (\textit{Lookup}), summarizing values over records such as counts, sums, extrema, and rankings (\textit{Aggregation}), describing the distribution of a variable (\textit{Characterization}), measuring how variables co-vary (\textit{Association}), and surfacing latent groups among records (\textit{Grouping}) all describe what is present in the data and commit to nothing further. Many such needs, like computing a key performance indicator or ranking entities by an aggregate, have a single correct answer fixed by the data, yet deriving them can demand considerable retrieval, integration, and computation across a corpus. Others, like characterizing a distribution or quantifying an association, produce summary statistics, but these still describe the actual records rather than generalizing beyond them.
\textit{Inferential} insight types concern the data-generating process behind the observed data, making claims that are meant to hold for instances the corpus does not contain and therefore relying on assumptions that link the data to that process. Determining whether an observed pattern reflects a stable feature of the process rather than noise (\textit{Statistical Inference}) and estimating values for instances the corpus does not contain, including future states (\textit{Prediction}), both reach beyond the data at hand, differing in whether the goal is to characterize the process or to forecast values~\citep{shmueli_explain_2010, breiman_statistical_2001}.
\textit{Causal} insight types make the strongest claims, concerning not what the process looks like but how it would respond to intervention. Estimating the effect of an action or policy at the population level (\textit{Interventional}) and determining the outcome of a specific case under a condition contrary to fact (\textit{Counterfactual}) cannot be settled from observed patterns alone and require assumptions about the causes behind the data, such as a structural causal model~\citep{pearl_causality_2009}.

These tiers stand in an asymmetric relationship to each other. Knowledge from a higher tier entails regularities on a lower tier, but not vice versa. A model of the process entails the patterns in any data it generates, and a causal account of the process entails both those patterns and the effects of intervening on it. On the other hand, no observed association, however strong, implies how the process would respond to intervention, and no body of observations uniquely determines the process that produced them. For example, a structural causal model fixes the distribution of the data it generates and the effect of intervening on any variable, whereas that same distribution is consistent with many distinct causal models that disagree about those effects. The familiar maxim that correlation does not imply causation is an example of this relationship.

The tiers in the taxonomy shown in Figure~\ref{fig:insight_types_taxonomy} span the full spectrum of possible insight types based on analytical knowledge extracted from tabular data. An analytical claim either concerns the observed data or generalizes beyond it to the process that produced it. A claim about that process either concerns the process as observed or under intervention~\citep{pearl_causality_2009}. Any analytical outcome thus falls into exactly one tier, and no further level of assumptions exists between them. The insight types listed within each tier are representative but not necessarily exhaustive, capturing the categories most relevant in \acrshort{task}. We ground the descriptive types in established accounts of low-level analytical primitives~\citep{amar_lowlevel_2005}, the inferential split in the distinction between explanatory and predictive modeling~\citep{shmueli_explain_2010, breiman_statistical_2001}, and the causal tier in the causal hierarchy~\citep{pearl_causality_2009}.

The insight types in the taxonomy are atomic. Since insights are inherently compositional (Section~\ref{sec:problem:conceptual_foundations:insights}), an insight may compose an arbitrary number of atomic insights of the same or different types, and many common analytical tasks represent such compositions. For instance, comparing two means composes two aggregations when it asks which is larger, and becomes a statistical inference when it asks whether the difference reflects the underlying process rather than noise. Similarly, analyzing a trend decomposes into a descriptive slope over observed values or an inferential forecast. At the far end of this spectrum, insight needs may demand elaborate compositions that layer substantial methodology onto atomic types, as in scientific workflows that apply optimization, computational modeling, or simulation, all of which inherit the tier of the knowledge they operate over rather than forming a tier of their own. The latent nature of insight needs means that the type implied by an expression need not match the type that satisfies the underlying need. The same comparison, expressed as ``Does X or Y have the larger mean Z?'', resolves on its surface to a composition of aggregations, yet the user may actually seek to know whether the two means differ meaningfully, which is a question of statistical inference. We develop this methodological latency in Sections~\ref{sec:user_facing} and~\ref{sec:research_agenda}.

\begin{figure}[ht]
    \centering
    \begin{subfigure}[t]{7.62cm}
        \centering
        \includegraphics[width=7.62cm]{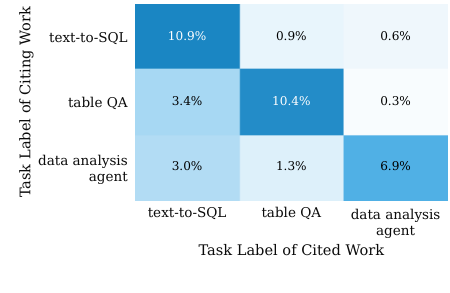}
        \caption{Share of chronologically feasible citations between works by their task label.}
        \label{fig:fragmentation:matrix}
    \end{subfigure}%
    \hspace{0.35cm}%
    \begin{subfigure}[t]{8.00cm}
        \centering
        \includegraphics[width=8.00cm]{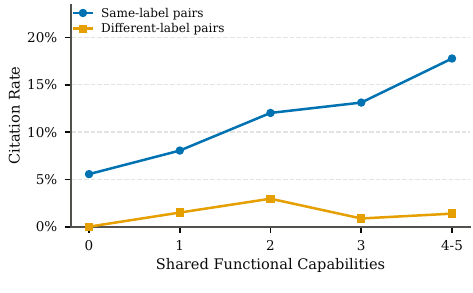}
        \caption{Citation rate among reviewed systems by the number of capabilities two systems have in common.}
        \label{fig:fragmentation:capabilities}
    \end{subfigure}
    \caption{Citation rates between reviewed works against how they label the task they target. Appendix~\ref{s:appendix:citation_analysis} details the analysis.}
    \label{fig:fragmentation}
\end{figure}

This taxonomy situates \acrshort{task} within a landscape of related but independently studied tasks. A substantial body of work covers deriving knowledge from tables, yet this work originates from different research communities that frame the challenge in diverging terminology and without an overarching direction. \textit{Tabular question answering}~\citep{zhang_reactable_2024, zhao_tapera_2024, zhu_autotqa_2024}, \textit{table-based reasoning}~\citep{kong_opentab_2023, wang_chainoftable_2024, lu_tart_2025}, \textit{text-to-SQL}~\citep{zhang_murre_2025, kothyari_crush4sql_2023, wang_dbcopilot_2025}, \textit{data science and analysis agents}~\citep{huang_dacode_2024, hu_infiagentdabench_2024, guo_dsagent_2024}, and various further characterizations~\citep{wang_aop_2025, chen_tablerag_2024, li_are_2025} all address the problem space of \acrshort{task}, facing similar challenges and often arriving at substantially similar system designs. Looking at how works across these task terminologies cite each other shows this fragmentation. Figure~\ref{fig:fragmentation:matrix} shows that across the reviewed works, works cite $10.6\%$ of works that are chronologically possible within the same label against just $1.8\%$ of the works of a different label.

These labels primarily reflect conventions, such as the assumed input format or data setting, rather than the covered analytical scope. A tabular question answering system may address only \textit{Lookup}, or it may extend across the descriptive tier and into inferential or causal types, which demand categorically different capabilities, and the label alone does not reveal which. Conversely, systems operating under different labels may address the same insight types with similar capabilities. As shown in Figure~\ref{fig:fragmentation:capabilities}, the more capabilities two systems share, the more likely they are to cite one another when they share the same label, but they are barely more likely to do so if they do not. The taxonomy of insight types makes the analytical scope explicit. Together with the preceding formalization, which centers the insight need as the object a system seeks to satisfy, it provides a unified vocabulary for describing what a given work actually concerns. The subsequent sections employ this framework to examine how systems construct realizations (Section~\ref{sec:orchestration}), the possibilities of user-system interaction (Section~\ref{sec:user_facing}), how evaluation methodology can assess system capabilities (Section~\ref{sec:eval}), and what research is missing on the path toward capable \acrshort{task} systems (Section~\ref{sec:research_agenda}).

\section{Anatomy of OpenTI Systems}
\label{sec:orchestration}
Section~\ref{sec:problem} formalized the aim of \acrshort{task} systems to produce realizations that capture how analytical knowledge derives from data as an optimization problem over a combinatorially vast space of operations. In this section, we examine \textit{how} \acrshort{task} systems derive this analytical knowledge. We frame \acrshort{task} systems as agents that navigate the realization space under partial observability and uncertainty (Section~\ref{sec:orchestration:problem}). We then examine how agents orchestrate the analytical process (Section~\ref{sec:orchestration:strategies}), and then decompose the functional capabilities that agents must exercise to construct realizations (Section~\ref{sec:orchestration:capabilities}). For details on the literature review protocol and the inclusion criteria for systems in Table~\ref{tab:functional_capabilities_overview}, consult Appendix~\ref{s:appendix:review_method}.

\subsection{OpenTI Systems}
\label{sec:orchestration:problem}

Constructing a realization requires a sequence of mutually dependent decisions, like interpreting the insight need, acquiring relevant data from the corpus, selecting and composing analytical operations, and assessing whether the emerging result satisfies the user's need. These decisions are mutually dependent since data availability constrains what methods can be applied, the choice of method determines what data is needed, and both depend on the interpretation of the insight need, which itself may need to adapt as data and analytical constraints surface. This makes constructing a realization an instance of a coupled decision problem as studied in automated planning~\citep{ghallab_automated_2004}, where consequences of early actions are uncertain and affect which later actions are appropriate. A system must moreover make these decisions with incomplete knowledge of the user's latent insight need $i$ and the corpus's contents and structure. \acrshort{task} is therefore a problem of \textit{sequential decision-making under mutual dependencies and partial observability}~\citep{ghallab_automated_2004, russell_artificial_2022, sutton_reinforcement_2018, kaelbling_planning_1998}. Accordingly, we understand \acrshort{task} systems as agents that interact with their environment to construct a realization that maximizes the utility.

\begin{definition}
    An \textbf{\acrshort{task} system} is an agent that constructs a realization $r$ from a corpus of tables $T$, which maximizes the utility for an expressed insight need $\hat{i}$ by \textit{observing} the state of its \textit{environment}, taking \textit{actions} that progressively construct the realization, and directing this process according to a \textit{policy}.
\end{definition}

We borrow this terminology from the study of sequential decision-making under partial observability~\citep{kaelbling_planning_1998, sutton_reinforcement_2018} to characterize the functional role of \acrshort{task} systems without prescribing whether these functions are realized through language model reasoning, learned neural models, classical algorithms, or other mechanisms. While current implementations largely employ \acrshort{llm}s, the design space extends beyond. In \acrshort{task}, the \textit{environment}, \textit{observations}, \textit{actions}, and \textit{policy} have specific instantiations that shape the setting and the space of possible systems.

\textbf{Environment.} The agent's environment consists of everything outside the agent that it interacts with~\citep{sutton_reinforcement_2018, russell_artificial_2022}. Following \citet{sutton_reinforcement_2018}, we understand everything the agent cannot change arbitrarily as part of the environment. Thus, the corpus $T$ and insight need $i$ are necessary components of the environment in \acrshort{task}. Agents can explore and use $T$, but cannot alter it. Similarly, they can observe the expressed insight need $\hat{i}$ but not influence the underlying need $i$. The environment may extend beyond these necessary components to data relationships, contextual resources like documentation or ontologies that reduce uncertainty, interaction histories, or user profiles (see Sections~\ref{sec:user_facing} and~\ref{sec:research_agenda}).

\textbf{Observations.} While deriving a realization, the agent continually makes observations that inform its subsequent actions~\citep{sutton_reinforcement_2018}. In \acrshort{task}, these observations comprise the expressed insight need and the intermediate analytical state. In interactive settings users may specify the expression of their insight need throughout the analytical process, for instance through clarification, feedback, or analytical guidance (see Section~\ref{sec:user_facing:interaction_modes}). The observed expression $\hat{i}$ thus accumulates during the process as the user contributes to it, whereas in non-interactive settings it remains fixed to what the user provided initially.
The intermediate analytical state captures the partial realization the agent has constructed so far, the outcomes of executed operations, and the intermediate results it has produced. Since realizations are converging \acrshort{dag}s, it can span multiple independent analytical strands that have not yet converged. The environment is only partially observable since the agent cannot access the latent insight need $i$, the corpus $T$ has to be explored, and the agent can only estimate the utility against the expression of the insight need it has observed so far.

\textbf{Actions and functional capabilities.} At any time, an agent takes an action within its action space~\citep{sutton_reinforcement_2018}. Actions are the agent's means of influencing and exploring the environment~\citep{russell_artificial_2022}. In \acrshort{task}, actions span the \textit{functional capabilities} required to derive a realization. The concrete actions available to an agent are determined by its specific implementation and may, for instance, include retrieving tables, and composing and executing analytical operations on the tables. The outcomes of actions in \acrshort{task} are uncertain. For example, retrieval may surface irrelevant data and analytical operations may produce errors. Section~\ref{sec:orchestration:capabilities} examines these functional capabilities.

The agent's policy governs how it selects actions based on its observations~\citep{sutton_reinforcement_2018, russell_artificial_2022}. In \acrshort{task}, the policy determines how the agent navigates the realization space, how it allocates effort across functional capabilities, and how it selects specific actions within each capability. The policy affects all choices the agent makes, from high-level decisions such as which capability to invoke next to fine-grained ones such as what analytical operations to apply to a table. The policy itself is often bound up in model parameters, prompt design, or hardcoded logic, making it only indirectly observable. \acrshort{task} agents can instead be characterized by higher-level \textit{orchestration strategies}, which Section~\ref{sec:orchestration:strategies} examines and structures.

\subsection{Orchestration Strategies}
\label{sec:orchestration:strategies}

The orchestration strategy describes the structural pattern of how an \acrshort{task} agent composes its functional capabilities. Thereby, it provides the high-level scaffolding for the policy applied by the agent. This subsection develops the space of orchestration strategies and positions existing systems within it.

\setlength{\tabcolsep}{2mm}
\newlength{\rcol}\setlength{\rcol}{\dimexpr\linewidth-\figwidth-2\tabcolsep\relax}

\tcbset{refbox/.style={colback=white,colframe=gray!55,boxrule=0.4pt,arc=1pt,
   boxsep=0pt,left=3pt,right=3pt,top=3pt,bottom=3pt,width=\rcol}}

\newcommand{\refcell}[1]{\mbox{\begin{tcolorbox}[refbox]\scriptsize #1\end{tcolorbox}}}

\begin{figure}[ht]
\centering
\setlength{\tabcolsep}{1mm}
\renewcommand{\arraystretch}{1.6}
 
\begin{tabular}{@{}
    m{\figwidth}
    m{\dimexpr\linewidth-\figwidth-2\tabcolsep\relax}
  @{}}
 
\includegraphics[width=\figwidth]{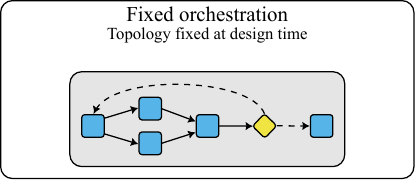} &
\refcell{
\textbf{CQR-SQL}~\citeyearpar{xiao_cqrsql_2022}, \textbf{\citet{cao_apiassisted_2023}}, \textbf{OpenTab}~\citeyearpar{kong_opentab_2023}, \textbf{ToolWriter}~\citeyearpar{gemmell_toolwriter_2023}, \textbf{MITQA}~\citeyearpar{kumar_multirow_2023}, \textbf{CRUSH4SQL}~\citeyearpar{kothyari_crush4sql_2023}, \textbf{ITR}~(\citeyear{lin_inner_2023}), \textbf{DataQue}~\citeyearpar{kochedykov_conversing_2023}, \textbf{S3HQA}~\citeyearpar{lei_s3hqa_2023}, \textbf{DIN-SQL}~\citeyearpar{pourreza_dinsql_2023}, \textbf{MultiTabQA}~\citeyearpar{pal_multitabqa_2023}, \textbf{Dater}~\citeyearpar{ye_large_2023}, \textbf{ReFSQL}~\citeyearpar{zhang_refsql_2023}, \textbf{TabSQLify}~\citeyearpar{nahid_tabsqlify_2024}, \textbf{ROUTE}~\citeyearpar{qin_route_2024}, \textbf{PoTable}~\citeyearpar{mao_potable_2024}, \textbf{SynTQA}~\citeyearpar{zhang_syntqa_2024}, \textbf{TabDSR}~\citeyearpar{jiang_tabdsr_2025}, \textbf{CHASE-SQL}~\citeyearpar{pourreza_chasesql_2025}, \textbf{ACR}~\citeyearpar{li_are_2025}, \textbf{ALTER}~\citeyearpar{zhang_alter_2025}, \textbf{TART}~\citeyearpar{lu_tart_2025}, \textbf{SGAM}~\citeyearpar{wang_plugging_2025}, \textbf{TAMO}~\citeyearpar{li_table_2025}, \textbf{FISQL}~\citeyearpar{menon_fisql_2025}, \textbf{STaR-SQL}~\citeyearpar{he_starsql_2025}, \textbf{Pi-SQL}~\citeyearpar{chi_pisql_2025}, \textbf{H-STAR}~\citeyearpar{abhyankar_hstar_2025}, \textbf{SAFE-SQL}~\citeyearpar{lee_safesql_2025}, \textbf{TiInsight}~\citeyearpar{zhu_automated_2025}, \textbf{TableLLM}~\citeyearpar{wu_tablebench_2025}, \textbf{QCMA-SQL}~\citeyearpar{shao_enhancing_2025}, \textbf{DAgent}~\citeyearpar{xu_dagent_2025}, \textbf{Table-Critic}~\citeyearpar{yu_tablecritic_2025}, \textbf{GTR}~\citeyearpar{zou_rag_2025}, \textbf{UCS-SQL}~\citeyearpar{wu_ucssql_2025}, \textbf{RTS++}~\citeyearpar{chen_reliable_2026}, \textbf{LitE-SQL}~\citeyearpar{piao_litesql_2026}, \textbf{DS-GURU}~\citeyearpar{lai_kramabench_2026}
}
\\
 
\includegraphics[width=\figwidth]{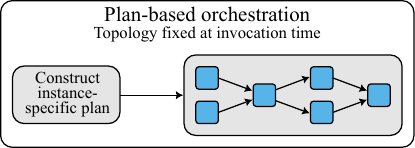} &
\refcell{
\textbf{Weaver}~\citeyearpar{khoja_weaver_2025}, \textbf{AixelAsk}~\citeyearpar{zhang_aixelask_2025}
}
\\
 
\includegraphics[width=\figwidth]{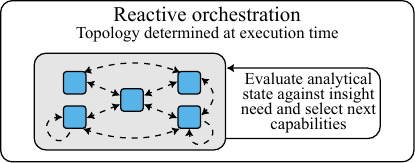} &
\refcell{
\textbf{DAAgent}~\citeyearpar{hu_infiagentdabench_2024}, \textbf{DA-Agent}~\citeyearpar{huang_dacode_2024}, \textbf{Chain-of-Table}~\citeyearpar{wang_chainoftable_2024}, \textbf{ReAcTable}~\citeyearpar{zhang_reactable_2024}, \textbf{TALON}~\citeyearpar{jin_talon_2025}, \textbf{DataMind}~\citeyearpar{qiao_scaling_2025}, \textbf{MACT}~\citeyearpar{zhou_efficient_2025}, \textbf{DACO}~\citeyearpar{wu_daco_2024a}
} \\
 
\includegraphics[width=\figwidth]{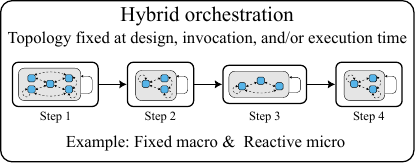} &
\refcell{
\textbf{TableRAG}~\citeyearpar{chen_tablerag_2024}, \textbf{TableGPT2}~\citeyearpar{su_tablegpt2_2024}, \textbf{TaPERA}~\citeyearpar{zhao_tapera_2024}, \textbf{AutoTQA}~\citeyearpar{zhu_autotqa_2024}, \textbf{AOP}~\citeyearpar{wang_aop_2025}
} \\
 
\end{tabular}

\includegraphics[width=\linewidth]{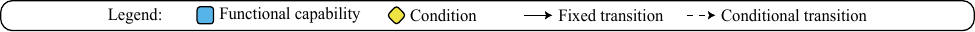}

\caption{Overview of orchestration strategies alongside works that employ the respective strategies. The schematic in each row depicts the corresponding control flow.}
\label{fig:orchestration_strategies}
\end{figure}

The underlying question of when and how the structure of a process is determined has been studied across several traditions. In planning under uncertainty, approaches range from offline derivation of complete policies before actions are taken, through online methods that interleave deliberation with execution and replan as observations are received, to reactive strategies that select actions based solely on the current belief state~\citep{ghallab_automated_2004, kaelbling_planning_1998}. Research into intelligent agent architectures~\citep{wooldridge_intelligent_1995} distinguishes \textit{deliberate} agents, which maintain explicit models of the world and plan before acting, from \textit{reactive} agents, which dynamically decide on actions based on immediate observations, and hybrid architectures that layer both. Database systems differentiate between static optimization, where the execution plan is fixed before any data is processed, and adaptive query processing, which tunes an initial execution plan based on data-dependent feedback~\citep{deshpande_adaptive_2007}. Across these traditions, the fundamental design choice concerns the degree to which the structure of the computational process is fixed before and during the execution of operations. Put differently, it concerns the influence execution-time information, like the actual instance and observations made during processing, has on not-yet-executed operations.
Following these traditions, we categorize orchestration strategies by when the topology of the composition of functional capabilities is determined. In \textit{fixed} orchestration, the topology is fixed at design time based on \textit{assumptions} about the analytical process. \textit{Plan-based} orchestration fixes the topology at invocation time, generating an insight-need-specific topology based on \textit{anticipation} of required actions for the instance. In \textit{reactive} orchestration the topology is not predetermined or planned but dynamically composed at execution time based on \textit{observations} of the analytical environment. These orchestration strategies can also be combined in a \textit{hybrid} orchestration strategy, for instance by employing different strategies at different levels of granularity, like fixing a macro-level structure while allowing incremental decisions within each phase.\footnote{The main source of ambiguity in classifying the orchestration of systems originates in chain-of-thought prompting. For replicability, we understand chain-of-thought prompting for code generation as a single functional capability of analytical composition and execution, while it could also be interpreted as inherently reactive, where the chain-of-thought interleaves analytical composition with the interpretation of self-induced observations.}

Figure~\ref{fig:orchestration_strategies} organizes the systems surfaced in our literature review along this categorization of orchestration strategies, showing 54 of the 58 systems that orchestrate multiple functional capabilities, with the remaining four only commanding a single capability, requiring no orchestration. We classify most studied systems under the fixed orchestration category, a smaller number as reactively orchestrated or following a hybrid orchestration strategy, and only two as purely plan-based.

\subsubsection{Fixed Orchestration}
\label{sec:orchestration:strategies:fixed}
Fixed orchestration follows a topology of capability invocations that is determined at design time. The parameterization, for instance which retrieval query to employ or what code to generate, is determined at execution time. The simplest form of fixed orchestration is following a single sequence of capabilities that is applied identically to every instance. Beyond this, fixed orchestration may also employ conditionals and branching with multiple predetermined paths. This strategy resolves mutual dependencies between functional capabilities by \textit{assumption} as the designer commits to a fixed ordering based on expectations about how insight needs are typically structured and processed, assuming that this ordering is adequate for the range of instances the system will encounter.

A fixed topology provides a well-defined context for every invocation of a functional capability, since the inputs to each capability are determined by its predecessors and its outputs feed its successors. This bounded structure enables independent optimization of individual capabilities and makes a system's behavior predictable and easier to trace, which supports interpretability and simplifies fine-grained evaluation of individual capabilities. It also lets the designer of a method encode domain knowledge about effective analytical workflows directly into the topology, drawing on established practices such as the data analysis pipelines common in data science~\citep{martinez-plumed_crispdm_2021}, for instance by hard-wiring an analyst-inspired sequence of stages~\citep{mao_potable_2024, lai_kramabench_2026}.

These benefits come at the cost of adaptability. A fixed topology cannot adjust to insight needs that deviate from the assumed processing pattern, to unexpected data characteristics discovered during execution, or to errors that fall outside predefined recovery paths. Since the agent operates in an uncertain and only partially observable environment (Section~\ref{sec:orchestration:problem}), the assumptions underlying the fixed structure may be violated for any given instance. A strategy that performs all data retrieval before analytical composition, for example, cannot retrieve tables that become relevant only after initial analysis. The more diverse the insight needs and the more heterogeneous the corpus, the more likely these assumptions are to break.

Many works mirror established retriever-reader~\citep{chen_reading_2017} and \acrfull{rag}~\citep{lewis_retrievalaugmented_2020} architectures, separating an upfront retrieval stage from a downstream data processing step~\citep{lin_lirage_2023, zhang_murre_2025, zou_rag_2025, wang_dbcopilot_2025}, often extended with interpretation of the inputs as a first step~\citep{kothyari_crush4sql_2023, chen_table_2024, kong_opentab_2023} and output synthesis as a last step~\citep{kong_opentab_2023, nahid_tabsqlify_2024, menon_fisql_2025, abhyankar_hstar_2025}. While most methods traverse these fixed stages one-by-one, other systems parallelize the analytical process by decomposing it and following a fixed structure for each sub-analysis~\citep{zhang_alter_2025, xu_dagent_2025}, resulting in a more complex, fixed topology of functional capabilities.

\subsubsection{Plan-Based Orchestration}
\label{sec:orchestration:strategies:plan_based}
In plan-based orchestration, the topology of capabilities is determined at the invocation of the agent, which generates an instance-specific topology before starting the execution. In contrast to fixed orchestration, the agent constructs a plan that specifies which capabilities to invoke, in what order, and with what dependencies based on the expressed insight need $\hat{i}$. Execution follows this plan, resembling a fixed topology, which may include specified branches and error recovery mechanisms. This strategy resolves mutual dependencies between functional capabilities by \textit{anticipation}, as the agent seeks to factor in the likely requirements of the insight need and the expected behavior of its capabilities to produce a topology that addresses the anticipated dependencies upfront.

Since each plan is constructed for a specific insight need, it provides an end-to-end perspective that can account for instance-specific requirements for the analytical process before any operation is executed. This perspective lets the agent assess a plan's coherence, identify potential bottlenecks, and optimize the structure, for instance by expressing independent steps as a dependency graph that can be executed in parallel~\citep{zhang_aixelask_2025} or by generating and evaluating alternative candidate plans against estimated quality criteria~\citep{wang_aop_2025}. As an explicit artifact, the plan can also be verified before execution~\citep{khoja_weaver_2025}, surfaced to users for intervention, and used to support interpretability.

The effectiveness of this strategy relies on the quality of the plan, which is generated without execution feedback. The plan rests on anticipations that may not hold, as the required data may have an unexpected structure, an analytical method may prove inappropriate once applied to the actual data, or intermediate results may reveal that the insight need calls for a different approach than planned. Recovery from such deviations and from errors is possible only to the extent that it was anticipated and encoded in the plan.

The two systems with a plan-based orchestration differ in how expressive their plans are. Weaver~\citep{khoja_weaver_2025} generates a linear plan of interleaved SQL and language model steps that a second model verifies and refines before the plan is executed step by step. AixelAsk~\citep{zhang_aixelask_2025} instead constructs a directed acyclic graph of retrieval and reasoning nodes with explicit dependencies, which exposes independent branches for parallel execution. Beyond these, AOP~\citep{wang_aop_2025} illustrates the optimization potential of an explicit plan, compiling operators into a graph that enables parallelization and prefetching.

\subsubsection{Reactive Orchestration}
\label{sec:orchestration:strategies:reactive}
Agents with a reactive orchestration strategy select each subsequent action based on their current observations, which can include the partial realization constructed so far, the outcomes of previous actions, and the expressed insight need at that time. The process structure emerges from the accumulation of these step-by-step decisions. This strategy resolves mutual dependencies by \textit{reaction} as they manifest. When the analysis reveals a data gap, the agent can invoke data retrieval, and when retrieval returns unexpected data, it can adjust its analytical approach without being locked into a plan that assumed otherwise.

Reactive orchestration leverages observations of the uncertain environment that \acrshort{task} agents operate in. By conditioning each decision on the observation history, the agent responds to the actual state of the analysis rather than to assumptions or anticipations that may no longer hold, which gives it inherent capacity for recovery and adaptation, as seen in agents that condition continued execution on observed errors to debug and retry their own operations~\citep{hu_infiagentdabench_2024, qiao_scaling_2025, zhang_reactable_2024}.
This adaptivity foregoes an inherent end-to-end perspective on the process. Purely step-by-step decision-making risks pursuing locally reasonable actions that are globally suboptimal, for example selecting data that is adequate in isolation but ill-suited to operations performed later in the process, mirroring the principle from planning under uncertainty that optimal behavior requires accounting for long-range consequences~\citep{kaelbling_planning_1998}. How well a reactive agent avoids such trajectories depends on how much anticipation its underlying mechanism brings to each decision, which some systems reinforce by adding an explicit planning or critic component to the reactive loop~\citep{zhou_efficient_2025, jin_talon_2025}.

In the reviewed works, reactive orchestration rests exclusively on a language model backbone, with most systems following the paradigm of interleaved reasoning and acting introduced by ReAct~\citep{yao_react_2023, zhang_reactable_2024, huang_dacode_2024, hu_infiagentdabench_2024, jin_talon_2025, qiao_scaling_2025}. A variation is to distribute the process across multiple language model agents, where a planning agent reactively invokes further agents that embody functional capabilities~\citep{zhou_efficient_2025}.

\subsubsection{Hybrid Orchestration}
\label{sec:orchestration:strategies:hybrid}
Hybrid orchestration combines other orchestration strategies, typically by applying different strategies at different levels of granularity or different phases of the analytical process. The general aim is to use the strengths of the combined strategies so that they counteract their respective shortcomings. Many such combinations are conceivable, and the reviewed systems instantiate three forms that differ in which strategy governs the macro structure and which governs the steps nested within it.

One hybridization is to embed a reactive loop within an otherwise fixed multi-step pipeline, combining an informed macro structure with local adaptivity inside each step. In TableRAG~\citep{chen_tablerag_2024}, a fixed sequence of query expansion and retrieval provides the inputs to a reactive program-aided solver that iterates over generated code and its observed outputs. TableGPT2~\citep{su_tablegpt2_2024} likewise wraps a reactive code and tool-calling agent inside a fixed pipeline of input preparation, agent execution, and answer generation. The intuition behind this combination is to bound the overall process with a reliable skeleton while letting the least predictable stage, the analytical composition, adapt to intermediate results.

Similarly, embedding reactive behavior in an instance-specific plan allows treating the plan as the primary strategy and reactive orchestration as a fallback, so the system retains the global coherence and efficiency of an optimized plan while avoiding the rigidity that would otherwise make a plan fail once observations deviate from what was anticipated in the plan. AOP~\citep{wang_aop_2025} compiles operators into an optimized graph that supports parallelization and prefetching, then executes it as planned and halts to re-plan the remaining pipeline when an operator fails or returns unsatisfactory results.

Nesting a plan inside a reactive loop that governs the macro structure inverts this combination. Instead of committing to a plan, such systems treat plans as provisional and reactively decide when to extend or revise them. TaPERA~\citep{zhao_tapera_2024} iteratively builds and refines a plan based on sub-questions, deriving sub-answers and deciding whether to finalize or refine the plan from them, while AutoTQA~\citep{zhu_autotqa_2024} uses a critic that judges intermediate results and directs the planner to revise it whenever it detects gaps.

\subsection{Functional Capabilities}
\label{sec:orchestration:capabilities}

The functional capabilities of \acrshort{task} systems describe the means with which they need to be equipped to derive a realization. We derive the necessary functional capabilities from the problem formulation (Section~\ref{sec:problem:conceptual_foundations:problem_formulation}), which requires constructing a realization $r = (O, E, \beta) \in R_{\text{valid}}(i) \cap R_{\text{feasible}}(T)$ for an insight need $i$ over a corpus of tables $T$.
Determining which realizations are valid, that is, defining $R_{\text{valid}}(i)$, requires mapping the expressed insight need $\hat{i}$ to a system-side representation $\hat{i}^{s}$ that constrains which realizations satisfy $i$, which in turn requires specifying and disambiguating its data, methodological, and output aspects $i_d, i_m, i_o$. We summarize this capability as \textit{interpretation} (Section~\ref{sec:orchestration:capabilities:interpretation}). Identifying feasible realizations $R_{\text{feasible}}(T)$ and constructing data bindings $\beta$ requires exploring $T$ to discover and bind relevant tables, a capability we refer to as \textit{data retrieval} (Section~\ref{sec:orchestration:capabilities:retrieval}). Beyond this, \acrshort{task} agents compose the structure of operations $(O, E)$ by selecting, parameterizing, and combining the analytical operations that derive knowledge from the data, which we capture as \textit{analytical composition and execution} (Section~\ref{sec:orchestration:capabilities:analytical_composition}). Finally, the agent produces outputs that satisfy the insight need. While formally part of $(O, E)$, producing this output addresses qualitatively distinct concerns like selecting output modalities, contextualization, and the form of presenting the analytical knowledge, which correspond to the output need $i_o$ rather than the analytical derivation itself. We thus distinguish \textit{output synthesis} (Section~\ref{sec:orchestration:capabilities:output}) as a separate functional capability.

In combination, these functional capabilities cover all necessary aspects of constructing a realization and are thus jointly sufficient for constructing some $r \in R_{\text{cand}}$. However, \acrshort{task} seeks a realization that maximizes the utility $r^* = \arg\max_{r \in R_{\text{cand}}}u_i(r)$ in a combinatorially vast space. The necessary functional capabilities introduced above all constitute \textit{object-level} computation~\citep{davis_metalevel_1977, davis_metarules_1980}, directly producing components that constitute $r$. They do not explicitly cover navigating the realization space towards high-utility realizations, which requires \textit{meta-level} capabilities~\citep{davis_metalevel_1977, davis_metarules_1980, russell_principles_1991}. Meta-level capabilities concern reasoning about the process of constructing $r$ itself, aiming to work towards the optimization objective but leaving no direct trace in $r$.

The process of constructing a realization can diverge substantially from the clean \acrshort{dag} of the resulting realization. The \textit{execution trace} of an agent may include cycles (e.g., retrying after errors), branches (e.g., exploring alternative analytical strategies), dead ends (e.g., retrieval returning insufficient data), and backtracking (e.g., revising a strategy based on intermediate results). Meta-level capabilities let the agent navigate this process by assessing whether intermediate results are on course towards the insight need, validating the quality of partial realizations, detecting and recovering from errors, and deciding when to backtrack or pursue alternatives. 
These capabilities shape which realization gets constructed without themselves appearing as nodes in $r$, which parallels the distinction between reasoning and action in \acrshort{llm} agent frameworks~\citep{yao_react_2023}. While the orchestration strategy (Section~\ref{sec:orchestration:strategies}) provides the structural pattern within which the agent operates, meta-level capabilities are the functional means the agent exercises within that structure to navigate toward high-utility realizations. We examine these under \textit{process governance} (Section~\ref{sec:orchestration:capabilities:governance}).

Table~\ref{tab:functional_capabilities_overview} provides an overview of systems with a broad coverage of functional capabilities (selection criteria in Appendix~\ref{s:appendix:review_method:overview}). Our discussion of functional capabilities is based on the full set of 58 systems, extended with capability-specific literature. Few systems cover all functional capabilities. Most works lack retrieval capabilities entirely (41 of 58), operating on pre-specified data rather than discovering relevant tables from a corpus. Interpretation often remains implicit (18 of 58), relying on inherent interpretation capabilities of language models rather than explicit mechanisms. Output modalities are largely limited to text or code execution results (46 of 58), with few systems capable of producing tables (15 of 58) or visualizations (6 of 58), despite the diversity of analytical results that may benefit from such presentation.

\begin{fullpagetable}
    \footnotesize
    \centering
    \fullpagecaption{Functional capabilities of the reviewed systems. Columns correspond to the five functional capabilities of Section~\ref{sec:orchestration:capabilities}, with \textbf{Analysis} abbreviating Analytical Composition and Execution and \textbf{Output} abbreviating Output Synthesis. Under \textbf{Retrieval}, \singleSettingIcon~(single table) indicate that no retrieval is performed, \setSettingIcon~indicates that a limited set of relevant and irrelevant tables is provided (e.g., a small database), requiring selection of necessary tables, and \openSettingIcon~indicates an open setting, requiring retrieval. Under \textbf{Analysis}, \faPython{} denotes general-purpose code, \SQLIcon{} domain-specific code (e.g., SQL), \faWrench{} tool calling, and \faRobot{} direct model inference. Under \textbf{Output}, \faAlignLeft{} denotes text, \faTable{} tables, \faChartBar{} visualizations, \faCommentDots{} contextualization, and \faRandom{} modality selection. Under \textbf{Process Governance}, Error Handling abbreviates Error Detection and Recovery and Validation abbreviates Result Validation (Section~\ref{sec:orchestration:capabilities:governance}).}

    \renewcommand{\arraystretch}{1.25}
\begin{tabularx}{\linewidth}{@{\extracolsep{\fill}} l l l l l l @{}}
\toprule
\textbf{System/Name} & \textbf{Interpretation} & \textbf{Retrieval} & \textbf{Analysis} & \textbf{Output} & \textbf{Process Governance} \\
\midrule \arrayrulecolor{lightline}
Dater~\citeyear{ye_large_2023} & Decompose, Augment, Scope & \singleSettingIcon & \SQLIcon~\faRobot & \faAlignLeft & Ensembling \\ \hline
ITR~\citeyear{lin_inner_2023} & Scope & \singleSettingIcon & \SQLIcon~\faRobot & \faAlignLeft & Validation, Ensembling \\ \hline
ROUTE~\citeyear{qin_route_2024} & Augment & \setSettingIcon~& \SQLIcon & \faAlignLeft & Error Handling, Validation \\ \hline
ReAcTable~\citeyear{zhang_reactable_2024} & Decompose & \singleSettingIcon & \faPython~\SQLIcon & \faAlignLeft~\faTable & Error Handling, Ensembling \\ \hline
TableRAG~\citeyear{chen_tablerag_2024} & Augment, Scope & \singleSettingIcon & \faPython & \faAlignLeft & Error Handling, Ensembling \\ \hline
FISQL~\citeyear{menon_fisql_2025} & Augment, Clarify & \setSettingIcon~& \SQLIcon & \faAlignLeft~\faTable~\faCommentDots &  \\ \hline
H-STAR~\citeyear{abhyankar_hstar_2025} & Decompose, Scope & \singleSettingIcon & \SQLIcon~\faRobot & \faAlignLeft & Validation, Ensembling \\ \hline
SAFE-SQL~\citeyear{lee_safesql_2025} & Augment, Scope, Context & \setSettingIcon~& \SQLIcon & \faAlignLeft & Validation \\ \hline
TabDSR~\citeyear{jiang_tabdsr_2025} & Decompose & \singleSettingIcon & \faPython & \faAlignLeft & Error Handling, Validation \\ \hline
TALON~\citeyear{jin_talon_2025} & Scope & \singleSettingIcon & \faPython~\faWrench & \faAlignLeft & Error Handling, Adaptive Steering \\ \hline
ALTER~\citeyear{zhang_alter_2025} & Decompose, Augment, Scope & \singleSettingIcon & \SQLIcon~\faRobot & \faAlignLeft & Ensembling \\ \hline
TART~\citeyear{lu_tart_2025} & Augment, Scope & \singleSettingIcon & \faPython & \faAlignLeft~\faCommentDots & Error Handling \\ \hline
TabSQLify~\citeyear{nahid_tabsqlify_2024} & Scope & \singleSettingIcon & \SQLIcon~\faRobot & \faAlignLeft & Error Handling, Validation \\ \hline
DAgent~\citeyear{xu_dagent_2025} & Decompose, Context & \setSettingIcon~Embedding & \SQLIcon~\faRobot & \faAlignLeft~\faCommentDots & Validation \\ \hline
DS-GURU~\citeyear{lai_kramabench_2026} & Augment & \setSettingIcon~LLM-select & \faPython & \faAlignLeft & Error Handling, Validation \\ \hline
MACT~\citeyear{zhou_efficient_2025} & Decompose, Augment & \singleSettingIcon & \faPython~\faWrench & \faAlignLeft & Error Handling, Ensembling \\ \hline
Pi-SQL~\citeyear{chi_pisql_2025} & Decompose, Scope & \setSettingIcon~& \faPython~\SQLIcon & \faAlignLeft & Error Handling, Validation, Ensembling \\ \hline
QCMA-SQL~\citeyear{shao_enhancing_2025} & Decompose, Augment, Context & \setSettingIcon~& \SQLIcon & \faAlignLeft & Error Handling \\ \hline
Weaver~\citeyear{khoja_weaver_2025} & Decompose, Augment, Scope & \singleSettingIcon & \SQLIcon~\faRobot & \faAlignLeft & Error Handling, Adaptive Steering \\ \hline
RTS++~\citeyear{chen_reliable_2026} & Scope, Clarify & \setSettingIcon~& \SQLIcon & \faAlignLeft & Validation, Ensembling \\ \hline
AutoTQA~\citeyear{zhu_autotqa_2024} & Decompose, Augment, Context & \setSettingIcon~LLM-select & \faPython~\SQLIcon & \faAlignLeft~\faTable~\faCommentDots & Error Handling, Validation, Adaptive Steering \\ \hline
TaPERA~\citeyear{zhao_tapera_2024} & Augment & \singleSettingIcon & \faPython & \faAlignLeft~\faCommentDots & Error Handling, Adaptive Steering \\ \hline
ACR~\citeyear{li_are_2025} & Augment, Clarify, Context & \singleSettingIcon & \faPython & \faAlignLeft~\faTable~\faChartBar~\faCommentDots & Error Handling \\ \hline
SGAM~\citeyear{wang_plugging_2025} & Decompose, Augment & \setSettingIcon~Embedding & \SQLIcon~\faRobot & \faAlignLeft &  \\ \hline
DataQue~\citeyear{kochedykov_conversing_2023} & Augment, Clarify, Context & \setSettingIcon~Classifier & \SQLIcon~\faWrench & \faTable~\faChartBar~\faCommentDots~\faRandom & Error Handling, Validation \\ \hline
OpenTab~\citeyear{kong_opentab_2023} & Scope & \openSettingIcon~Lexical & \SQLIcon~\faRobot & \faAlignLeft & Error Handling, Validation \\ \hline
TableGPT2~\citeyear{su_tablegpt2_2024} & Scope, Clarify, Context & \openSettingIcon~Embedding & \faPython~\SQLIcon~\faWrench & \faAlignLeft~\faTable~\faChartBar~\faCommentDots~\faRandom & Error Handling, Adaptive Steering \\ \hline
AOP~\citeyear{wang_aop_2025} & Decompose, Augment & \openSettingIcon~Embedding & \SQLIcon~\faWrench~\faRobot & \faAlignLeft~\faCommentDots & Error Handling, Validation, Adaptive Steering, Ensembling \\ \hline
TiInsight~\citeyear{zhu_automated_2025} & Decompose, Augment, Scope, Context & \openSettingIcon~Embedding & \SQLIcon & \faTable~\faChartBar~\faRandom & Error Handling, Validation \\
\arrayrulecolor{black}\bottomrule
\end{tabularx}
    \label{tab:functional_capabilities_overview}
\end{fullpagetable}

\subsubsection{Interpretation}
\label{sec:orchestration:capabilities:interpretation}

Interpretation allows agents to determine which realizations satisfy the insight need, that is, which realizations are part of $R_{\text{valid}}(i)$ by mapping the expressed insight need $\hat{i}$ to the system's representation $\hat{i}^{s}$. The latency of insight needs introduces a gap between $\hat{i}$ and $i$. Users leave aspects implicit because they employ conventions, delegate choices to the system, or lack awareness of what must be specified (Section~\ref{sec:user_facing}). Interpretation bridges this gap by specifying implicit aspects and resolving ambiguities. All systems perform some interpretation before acting, yet many of the reviewed systems lack explicit mechanisms and instead do so implicitly in a neural model, mostly through the language understanding of an underlying language model. Interpretation may also recur during insight extraction, triggered when data retrieval, analytical execution, or further user inputs reveal misalignment.

\paragraph{Input Decomposition and Augmentation.}
Decomposing and augmenting inputs seeks to transform and specify $\hat{i}^s$ to better serve downstream capabilities. Decomposition breaks the expressed need into parts. Augmentation enriches or rewrites it to increase its specification or alignment with the data environment.

The data need $i_d$ is often augmented by mapping inputs to data-level representations, for instance by generating hypothetical schemas of the required data~\citep{kothyari_crush4sql_2023, zhang_murre_2025} or extracting concept-attribute pairs~\citep{chen_table_2024, chen_tablerag_2024, wu_ucssql_2025} that serve retrieval and analytical execution, and, after retrieval, by schema linking that aligns concepts mentioned in the expressed need with concrete schema elements~\citep{pourreza_dinsql_2023, qin_route_2024, shao_enhancing_2025}, a technique originating in text-to-SQL systems~\citep{liu_survey_2025}. In contrast, the methodological need $i_m$ is augmented by transforming inputs into an actionable plan for the analytical process~\citep{zhao_tapera_2024, lai_kramabench_2026, lu_tart_2025, zhu_autotqa_2024, zhou_efficient_2025, zhang_aixelask_2025}, bridging the inputs with retrieval and analytical execution. \citet{zhang_alter_2025} propose an alternative by applying step-back augmentation~\citep{zheng_take_2024}, which abstracts the expressed need to a higher level of generality before deriving sub-needs, aiming to surface implicit aspects that a direct transformation of the input would miss.

Reflecting the compositional nature of insight needs, decomposition is commonly employed to extract logically dependent~\citep{zhao_tapera_2024, lai_kramabench_2026, zhu_autotqa_2024, pourreza_dinsql_2023, abhyankar_hstar_2025, jiang_tabdsr_2025} and independent~\citep{xu_dagent_2025, zhang_alter_2025, perez_llmbased_2025} sub-needs from the expressed insight need, making the downstream analysis less complex to compose and execute. Decomposition and augmentation are the most widespread explicit interpretation mechanisms, listed as ``Decompose'' and ``Augment'' in the \textit{Interpretation} column of Table~\ref{tab:functional_capabilities_overview}.

\paragraph{Contextualization.}
Contextualization aims to inform the interpretation of the insight need with knowledge that is not directly expressed in the input. This contextual knowledge can originate from different sources. \textit{Interaction-histories} can serve as shared context, resolving references to previous results and interpreting follow-up needs in the light of prior results~\citep{li_are_2025, kochedykov_conversing_2023, xiao_cqrsql_2022}. Similarly, \textit{demonstration-retrieval} can contextualize the analytical process in prior instances of processed insight needs~\citep{lee_safesql_2025, zhang_refsql_2023, pourreza_chasesql_2025, chen_pandora_2025, gu_structured_2025} or historical planning paths~\citep{xu_dagent_2025}. Additionally, \textit{domain-knowledge} is used to resolve terminology~\citep{zhu_automated_2025} and conventions specific to the analytical domain~\citep{su_tablegpt2_2024}, such as standard operating procedures from business operations~\citep{zhu_autotqa_2024}, directly addressing cases where users omit detail because a default interpretation exists in the domain. Systems integrate such external expert knowledge through different mechanisms, retrieving it from an index via retrieval augmented generation~\citep{zhao_chat2data_2024a}, integrating a dedicated data-expert model that supplies knowledge left implicit in the schema and the expressed need~\citep{hong_knowledgetosql_2024}, or materializing it as supplementary data that extends the retrieved tables with the missing information~\citep{liu_augment_2025}.

\paragraph{Scoping.}
Part of interpreting the data need $i_d$ is interpreting it against specific tables. Scoping aims to interpret a table and scope its contents to those that are actually relevant to $i_d$ by selecting only the relevant rows, columns, or cells.
Approaches vary in granularity, from selecting relevant rows and columns based on semantic similarity to the expressed insight need~\citep{zhang_alter_2025, kong_opentab_2023, patnaik_cabinet_2023}, to retrieving individual schema elements and cell values~\citep{chen_tablerag_2024}. Following a different approach, \citet{gemmell_toolwriter_2023} generate code on demand to filter rows programmatically before a downstream model computes the answer. The efficacy of an explicit scoping step is disputed, with \citet{mao_potable_2024} finding a reduction in accuracy.

\paragraph{Engaging Users.}
Beyond automatically resolving ambiguities, systems may involve users by soliciting clarification or presenting alternative interpretations. This is particularly relevant when the expressed need has multiple sensible interpretations that cannot be resolved through conventions or context alone~\citep{saparina_disambiguate_2025, gomm_are_2025}. To address this, systems may engage users by asking targeted follow-up questions~\citep{su_tablegpt2_2024, li_are_2025, menon_fisql_2025, chen_reliable_2026, kochedykov_conversing_2023}, or otherwise soliciting their input to specify the expressed insight need. Choosing when to autonomously handle specification and when to solicit clarification from the user introduces a trade-off between burdening the user with interactions and reducing the risk of misalignments. We assess this trade-off and the design of interactions in Section~\ref{sec:user_facing} in more detail.

\paragraph{Interpretation Throughout Processing.}
The interpretation mechanisms in the reviewed works are predominantly applied before the analytical process begins, operating on the expressed insight need in isolation from data and analytical outcomes that emerge during execution. Yet the environment is only partially observable, so information relevant to interpretation may surface only during processing, for example when a seemingly unambiguous entity matches multiple retrieved tables with different semantics. Nonetheless, explicit re-interpretation during processing, that is, revising $\hat{i}^{s}$ in light of new observations, remains limited, except for iterative retrieval approaches~\citep{zhang_murre_2025, boutaleb_exploring_2025} that progressively refine the data need $i_d$ based on intermediate retrieval results, and iterative revisions of analytical plans that re-interpret the insight need under intermediate results~\citep{zhao_tapera_2024}.

The degree to which re-interpretation is possible depends on the orchestration strategy. Fixed orchestration confines interpretation to designated positions in the topology, preventing revisions based on downstream observations. Plan-based orchestration fixes interpretation at planning time, making re-interpretation possible only if explicitly anticipated. Reactive orchestration lets the agent condition each decision on prior observations, though whether this constitutes re-interpretation depends on whether the agent's representation $\hat{i}^{s}$ is itself revised rather than merely adjusting execution within an unchanged interpretation.

\subsubsection{Data Retrieval}
\label{sec:orchestration:capabilities:retrieval}
Data retrieval capabilities allow agents to explore the corpus and surface sufficient data to address an insight need, enabling the agent to determine which realizations the corpus supports, $R_{\text{feasible}}(T)$, and to retrieve tables for data bindings $\beta$ of candidate realizations. This requires mapping the data need $i_d$ to a set of tables for the data bindings $\beta$, and may additionally be constrained by the methodological need $i_m$, for instance when an analytical method requires a minimum sample size to yield statistically valid results.

When an insight need can be satisfied from a single table, retrieval concerns identifying the most relevant table. However, insight needs may require integrating information from multiple tables, requiring retrieval to surface a sufficient set (Definition~\ref{def:sufficient_data}) that satisfies both semantic alignment between the insight need and the retrieved tables and structural alignment among the retrieved tables themselves. The \textit{Retrieval} column in Table~\ref{tab:functional_capabilities_overview} shows the data setting each system operates in, separating systems that operate over a pre-selected set of tables from those that select among a bounded set of tables and those that retrieve from an open corpus. Out of the systems in Table~\ref{tab:functional_capabilities_overview}, only four target an open setting. We examine how existing work represents tables and inputs, the granularities at which retrieval is performed, how sets of multiple tables are composed, and how retrieval integrates within the broader analytical process.

\paragraph{Table and Input Representations.}
\label{sec:orchestration:capabilities:retrieval:representations}

Index-based retrieval maintains an explicit collection of table representations and scores table relevance against an input at retrieval time. Bridging the semantic gap between input and table representations requires appropriate representations of both that make them comparable.

Tables have predominantly been represented through linearization into text, which enables applying retrieval methods developed for text documents. The design space for linearization is large~\citep{zhang_same_2025}, ranging from schema-only representations using column headers~\citep{kong_opentab_2023} to full-content serializations including row-level data~\citep{pan_endtoend_2022}, representations enriched with surrounding contextual text~\citep{chen_open_2020, huang_mixedmodality_2022}, and \acrshort{llm}-generated natural language descriptions of table contents~\citep{balaka_pneuma_2025, zhang_same_2025}. These linearization choices have been shown to significantly influence retrieval performance~\citep{gomm_metadata_2025}. However, linearizing tables into text has been criticized for ignoring semantic dependencies and structural information~\citep{trabelsi_strubert_2022, wang_retrieving_2021}, limiting the amount of tabular content that can be represented~\citep{gomm_metadata_2025}, and being unable to handle different cell datatypes individually~\citep{wang_retrieving_2021}. This has motivated table-specific encoding approaches that modify transformer attention mechanisms to account for tabular structure~\citep{herzig_tapas_2020, trabelsi_strubert_2022}, apply graph-based encoders to capture structural properties~\citep{wang_retrieving_2021}, or operate on images of tables using multimodal language models~\citep{xuEfficientTableRetrieval2026}. Despite these efforts, linearization to text with general-purpose encoders remains the dominant approach in research.

Orthogonally, the semantic gap can be narrowed from the input side by transforming the expressed insight need into a form that better matches the table representations. One approach that has been explored is to transform inputs into text that is structurally similar to the linearizations of the indexed tables, for instance by generating a hypothetical table schema from the input and retrieving against the actual schemas~\citep{kothyari_crush4sql_2023, zhang_murre_2025}, as well as query decomposition and expansion techniques discussed further in the context of semantic and structural alignment. More broadly, retrieval is often deeply intertwined with interpretation capabilities (Section~\ref{sec:orchestration:capabilities:interpretation}) that expand, decompose, or rewrite inputs to improve alignment.

Given these representations, the reviewed works apply established retrieval paradigms including lexical methods~\citep{kong_opentab_2023, ji_target_2024}, bi-encoder dense retrieval~\citep{karpukhin_dense_2020, pan_endtoend_2022, zhang_murre_2025, kothyari_crush4sql_2023, chen_table_2024, xu_dagent_2025, zhu_automated_2025, wang_plugging_2025}, late interaction mechanisms~\citep{khattab_colbert_2020, lin_lirage_2023}, cross-encoder re-ranking~\citep{luoDecompositionDrivenMultiTableRetrieval2026}, and hybrid combinations of lexical and dense signals~\citep{balaka_pneuma_2025, li_tailoring_2025}. For systems that perform retrieval, the \textit{Retrieval} column in Table~\ref{tab:functional_capabilities_overview} annotates their retrieval mechanism. Retrieval mechanisms are applied largely independently of the representations of tables and inputs. Rather than which retrieval paradigm to apply, research has mostly focused on what representations effectively bridge the semantic gap and how to incorporate sufficiency criteria into retrieval.

\paragraph{Retrieval Granularity.}
\label{sec:orchestration:capabilities:retrieval:granularity}

Besides representation and retrieval mechanisms, approaches vary in the granularity at which tables are indexed. Most index at the table level, producing one representation per table~\citep{kong_opentab_2023, herzig_open_2021, balaka_pneuma_2025}, which limits the detail that can be captured but preserves cross-column semantics that convey what a table is about as a whole. In contrast, column-level indexing~\citep{kothyari_crush4sql_2023, chen_table_2024} enables finer-grained matching between individual attributes and components of the data need, at the cost of a substantially larger index and the potential loss of inter-column context. This finer granularity is particularly relevant for composing sufficient sets of multiple tables, where alignment must be assessed per individual attribute rather than per table. Row-level indexing~\citep{kumar_multirow_2023, ji_target_2024} has also been explored, yielding fine-grained representations that capture specific records in the tables, facilitating retrieval in settings where relevance is concentrated in specific records. Since these granularities capture complementary signals, approaches may also combine them, for instance, leveraging coarse-grained representations for initial filtering and fine-grained representations for precise matching~\citep{chen_table_2024, wang_aop_2025, kong_opentab_2023, zhu_automated_2025}.

\paragraph{Semantic and Structural Alignment.}
\label{sec:orchestration:capabilities:retrieval:alignment}
When the data required to satisfy an insight need spans multiple tables, retrieving individually relevant tables does not guarantee a sufficient set (Definition~\ref{def:sufficient_data}). The retrieved set must be \textit{semantically aligned} with the data need, collectively covering all required concepts, and \textit{structurally aligned} among its tables, enabling their integration through operations available to the system (Section~\ref{sec:problem:sufficiency:data}). Composing such sets can be understood as a search over candidate table combinations~\citep{boutaleb_exploring_2025}, where candidate sets are assessed along both alignment dimensions. Approaches differ in how they track semantic coverage, assess structural compatibility, and navigate the search space.

Semantic alignment is primarily addressed by decomposing the data need into sub-components and matching which of them are covered by retrieved tables. \citet{kothyari_crush4sql_2023} generate a hypothetical minimal set of table schemas from the insight need, while \citet{chen_table_2024} decompose it into concept-attribute pairs, each targeting a specific data requirement. Both decompose in a single step without considering which tables are actually available, then track coverage during a subsequent selection phase. These contrast with iterative retrieval, where retrieval steps are interleaved with an assessment of the alignment of the tables retrieved so far~\citep{boutaleb_exploring_2025}. \citet{zhang_murre_2025} retrieve one table per iteration and rewrite the input to remove aspects already covered, continuing until all concepts in the data need are covered. The iterative structure allows adapting to the actual structure of the data and implicitly determining the size of the sufficient set, at the cost of computational overhead that scales with the number of iterations. \citet{chen_can_2025} instead align the input with the organization of the corpus by exploring relationships among its tables, retrieving a sufficient set in a single pass rather than decomposing the need without regard to available data or querying iteratively.

Structural alignment requires assessing whether retrieved tables can be integrated, for instance through join or union operations. Approaches estimate structural compatibility from signals intrinsic to table pairs, derived from an offline schema graph~\citep{kothyari_crush4sql_2023}, from online value overlap such as Jaccard similarity over columns~\citep{chen_table_2024}, or from a search for join paths that connect the matched attributes~\citep{wang_plugging_2025}. These signals can then be combined with semantic coverage into a joint selection objective~\citep{chen_table_2024, kothyari_crush4sql_2023}, or folded directly into the per-step selection of iterative retrieval, avoiding a separate optimization phase~\citep{boutaleb_exploring_2025}. These column relationships are also applied to expand retrieved sets of tables with joinable ones to include tables that have a low semantic alignment~\citep{agarwal_rear_2025}. Across these approaches, compatibility is judged on properties intrinsic to table pairs, such as key relationships or value overlap. Yet structural compatibility is ultimately determined by the agent's capabilities for analytical composition and execution, since capable agents may construct integration paths through transformations such as schema matching or feature engineering. The assessment of structural compatibility should thus depend on the capabilities of the agent.

\citet{zou_rag_2025} take a different approach. Instead of a regular index, they construct a hypergraph of tables based on semantic, structural, and heuristic similarity features. They then apply clustering and PageRank to surface groups of related tables. Unlike the approaches above, which compose sets that align with the specific data need, this approach captures general table relatedness within the corpus. It can surface structurally similar tables but does not explicitly optimize for semantic coverage and structural compatibility.

Notably, none of the discussed approaches actively handles cases in which no sufficient set of tables exists within the corpus for the given data need. When $R_{\text{cand}} = \varnothing$, most retrieval methods would still return a set of tables, leaving the insufficiency undetected.

\paragraph{Generative Retrieval.}
\label{sec:orchestration:capabilities:retrieval:generative}

Generative retrieval presents an alternative to maintaining an explicit index by encoding the corpus into the parameters of a sequence-to-sequence model that directly generates identifiers of relevant tables given an input~\citep{tayTransformerMemoryDifferentiable2022}, unifying indexing and search into a single differentiable process. \citet{guo_birdie_2025} apply this to individual tables, assigning each a semantic identifier derived from hierarchical clustering of metadata and cell values, while \citet{wang_dbcopilot_2025} extend it to multi-table settings by serializing a schema graph into the generation target and constraining decoding to valid connected table sets. Both train on synthetic queries generated over the target schemas.

Since generative retrieval learns the index directly it can capture semantic and structural properties jointly, which, depending on the training, enables deeper input-table interaction than bi-encoders that encode queries and tables independently~\citep{guo_birdie_2025}. Its effectiveness is limited by the quality of the synthetic training data, which may propagate hallucinations or distributional biases into retrieval errors~\citep{wang_dbcopilot_2025}. Additionally, since the index is learned, corpus changes require new or additional training, with parameter-isolation strategies for new tables explored but not validated at scale~\citep{guo_birdie_2025}.

\paragraph{Dynamic Retrieval.}
\label{sec:orchestration:capabilities:retrieval:dynamic}

Most \acrshort{task} systems treat retrieval as a separate phase preceding the analytical process, constructing the complete data bindings $\beta$ upfront. This assumes the data need $i_d$ can be fully determined from the initial input. Yet intermediate results may reveal requirements for additional data, such as linking tables for data integration, reference data for normalization, or variables whose relevance becomes apparent only after initial analysis. Further, decoupling retrieval makes recovery from retrieval misses impossible~\citep{risis_complex_2025}. Works on text documents have thus explored conditioning retrieval on the processing state~\citep{jiang_active_2023, asai_selfrag_2024}. In the tabular context, iterative retrieval methods~\citep{zhang_murre_2025, boutaleb_exploring_2025} interleave retrieval and interpretation capabilities, but still operate within a dedicated retrieval phase. \citet{wang_aop_2025} take an initial step towards integrating retrieval with other capabilities by treating it as an operator that is invocable during execution, though without systematic methods for leveraging the intermediate analytical state to inform retrieval decisions.

\subsubsection{Analytical Composition and Execution}
\label{sec:orchestration:capabilities:analytical_composition}
Analytical composition and execution capabilities allow \acrshort{task} agents to construct, parameterize, and execute the operational structure $(O, E)$ of a realization. They thus enable \acrshort{task} agents to derive the analytical knowledge to satisfy the user's insight need. We describe them based on the \textit{mechanisms} that are employed to specify and execute operations, the \textit{analytical composition}, that is, the manner in which operations are composed, and the role of \textit{data preparation} in this process.

\paragraph{Mechanisms.}
The mechanisms through which an agent executes operations determine what operations it can express and, therefore, which insight needs the agent can address. \acrshort{task} systems may employ code generation in domain-specific and general-purpose languages, tool calling over defined functionalities, and direct inference techniques as mechanisms for executing operations. The \textit{Analysis} column of Table~\ref{tab:functional_capabilities_overview} shows which of these mechanisms each system employs, showing that most systems rely on domain-specific and general-purpose code, whereas fewer systems make use of direct inference and tool calling.

\textit{\acrfull{dsl} generation}, mostly studied in the form of SQL generation, is the most extensively researched mechanism for deriving knowledge from tabular data, building on a substantial body of research into text-to-SQL generation~\citep{liu_survey_2025}. Several systems apply SQL generation as their sole analytical mechanism~\citep{pourreza_dinsql_2023, lee_safesql_2025, menon_fisql_2025, xu_dagent_2025, chen_reliable_2026}, while others combine it with other mechanisms~\citep{zhang_alter_2025, xu_dagent_2025, kong_opentab_2023, wang_aop_2025, zhang_reactable_2024, huang_dacode_2024, zhu_autotqa_2024, su_tablegpt2_2024}. \acrshort{dsl}s like SQL are designed for a particular application domain. In SQL, this specialization provides formal semantics grounded in relational algebra~\citep{codd_relational_1970}, enabling a well-defined execution behavior and a constrained output space. However, the specialization of \acrshort{dsl}s bounds what operations can be expressed. SQL has limited capacity to express statistical modeling, machine learning, optimization, or visualization operations~\citep{zhao_tapera_2024, kong_opentab_2023}, confining systems that rely on it as their sole execution mechanism largely to descriptive insight needs, with inferential and causal needs falling outside its expressive scope.
Systems that employ SQL generation alongside other mechanisms typically delegate operations outside the scope of SQL to complementary mechanisms, such as direct inference for result synthesis~\citep{zhang_alter_2025, xu_dagent_2025, kong_opentab_2023} or general-purpose code for data transformations~\citep{zhang_reactable_2024, zhu_autotqa_2024, huang_dacode_2024, su_tablegpt2_2024}.

\textit{\acrfull{gpl} generation}, mostly studied in the form of Python generation, generates code in a general-purpose language to interact with data~\citep{gao_pal_2023}. While \acrshort{gpl}s can express arbitrary computations, making them theoretically capable of addressing the full spectrum of insight needs, they provide less domain-specific scaffolding than \acrshort{dsl}s, resulting in a less constrained output space and a greater range of potential errors. In \acrshort{task}, the extensive ecosystem of Python libraries for data manipulation, statistical analysis, machine learning, and visualization~\citep{zhang_reactable_2024, chen_pandora_2025} supplies this scaffolding within a \acrshort{gpl}, letting systems leverage these abstractions while retaining the expressiveness to address arbitrary operations. Many of the reviewed works therefore employ Python generation as a primary~\citep{zhao_tapera_2024, mao_potable_2024, hu_infiagentdabench_2024, chen_tablerag_2024, chen_pandora_2025, lai_kramabench_2026, li_are_2025} or complementary mechanism~\citep{zhang_reactable_2024, zhu_autotqa_2024, huang_dacode_2024}. Despite the theoretical breadth, these systems mostly demonstrate their Python code generation on descriptive needs that largely overlap with tasks also addressable through SQL, with only a few extending to inferential needs such as statistical inference and prediction~\citep{hu_infiagentdabench_2024, huang_dacode_2024, qiao_scaling_2025, li_are_2025}, leaving the practical effectiveness for more complex analytical intents largely undemonstrated. Moreover, the unconstrained output space has motivated the application of safety measures like containerization for executing operations~\citep{hu_infiagentdabench_2024, huang_dacode_2024, su_tablegpt2_2024}.

\textit{Tool calling} employs functions with specified semantics that the agent selects and parameterizes at execution time. Unlike code generation, where the operation is specified in full by the generated code, tool calling constrains the agent to invoking specific functions from a limited set. Abstracting operations into predefined callable functions allows tighter design-time control over the operations applied to the data, at the cost of limiting what an agent can express, making the expressiveness of tool-calling systems mostly dependent on the scope of the tool set. Methods vary in the number and complexity of tools they have access to, from small sets of simplistic table operations like select rows and group by~\citep{wang_chainoftable_2024, jin_talon_2025}, to larger sets of more expressive operations~\citep{wang_aop_2025, kochedykov_conversing_2023}.
Automatically generating missing tools for specific inputs~\citep{gemmell_toolwriter_2023, lu_tart_2025} has been proposed to rectify the inflexibility of predefined tools.

\textit{Direct inference} employs models directly to perform operations over tabular data and intermediate results. Unlike code generation and tool calling, where operations are explicitly specified and executed by an external interpreter, direct inference performs computations within a model itself, making the operation implicit in its learned behavior. The expressiveness of this mechanism depends on the capabilities of the employed models. In principle, a sufficiently capable model could map directly from raw data to analytical knowledge in a single step, subsuming all intermediate operations. In practice, systems apply direct inference at varying levels of granularity, ranging from atomic operations such as cell selection or fact verification~\citep{wang_chainoftable_2024, zou_rag_2025, lin_lirage_2023} to synthesizing free-form responses from intermediate results produced by other mechanisms~\citep{kong_opentab_2023, zhang_alter_2025, xu_dagent_2025}.
In direct inference, operations are not externalizable, meaning no executable specification can be inspected, verified, or reused independently of the model, limiting the interpretability of how analytical knowledge is derived. Moreover, the reliability of computational operations such as arithmetic, aggregation, or statistical computation relies on the model's precision rather than formal execution semantics. Studies into the application of language models for numerical computations show degradations~\citep{wu_tablebench_2025, wolff_how_2025, zhou_frebtqa_2024}, particularly as input size increases~\citep{zou_rag_2025}. Most reviewed works thus combine direct inference with other mechanisms, relegating direct inference to higher-level functions such as aggregating results across sub-needs~\citep{zhang_alter_2025, wang_aop_2025}, synthesizing free-form responses from execution outputs~\citep{kong_opentab_2023, xu_dagent_2025}, or performing simple extraction tasks that do not require precise computation~\citep{zhang_syntqa_2024, zou_rag_2025, lin_lirage_2023}.

\paragraph{Analytical Composition.}
Analytical composition concerns how the analytical logic $(O, E)$ is assembled, that is, the manner in which operations $O$ are selected, connected and parameterized. Analytical composition is coupled with the orchestration strategy, which determines when and how often the agent invokes analytical composition, constraining the possible composition patterns. In plan-based orchestration, the operational structure may be largely determined during plan generation, integrating composition directly into orchestration~\citep{wang_aop_2025}. In reactive orchestration, composition decisions are distributed across successive actions during execution, making the capability more clearly distinguishable.

Approaches mostly differ in how much of the analysis they specify at a time. In \textit{one-shot composition}, agents specify the analytical logic in a single step, for instance a full SQL query~\citep{kothyari_crush4sql_2023, zhang_murre_2025, wang_dbcopilot_2025, chen_table_2024} or a complete Python script~\citep{chen_pandora_2025, lai_kramabench_2026}, committing to the operational structure without observing intermediate states. In \textit{incremental composition}, the operational structure is built over successive actions, each producing an intermediate result the agent can observe before determining the next operation~\citep{zhang_reactable_2024, mao_potable_2024, wang_chainoftable_2024, zhou_efficient_2025, huang_dacode_2024, chen_tablerag_2024, zhu_autotqa_2024, hu_infiagentdabench_2024}, allowing it to detect unexpected results, adjust the analytical approach, or refine the scope of operations based on the data. Works report improved accuracy over one-shot generation~\citep{zhang_reactable_2024, wang_chainoftable_2024}, contingent on the agent's ability to effectively use intermediate state information.
When insight needs are decomposed into sub-needs, each sub-need can be treated individually and integrated with the others downstream~\citep{zhao_tapera_2024, zhang_alter_2025, xu_dagent_2025, wang_aop_2025}, which also lets independent sub-needs be processed in parallel~\citep{wang_aop_2025, zhang_alter_2025}.

Execution mechanisms and composition patterns are largely coupled, though not rigidly. The declarative nature of \acrshort{dsl}s favors one-shot composition, as a single query typically encodes the full analytical logic, yet \acrshort{dsl} generation is also used incrementally~\citep{zhang_reactable_2024, mao_potable_2024, chen_tablerag_2024}. Tool calling is inherently incremental, since each call is a separate action with observable intermediate state~\citep{wang_chainoftable_2024, zhou_efficient_2025}, whereas general-purpose code generation spans both one-shot~\citep{chen_pandora_2025, lai_kramabench_2026} and incremental settings and allows arbitrarily sized actions.

\paragraph{Data Preparation.}
Data preparation aims to select, clean, and transform retrieved data into a form that enables the analytical operations that generate the analytical knowledge. It forms a substantial component of manual analytical workflows~\citep{martinez-plumed_crispdm_2021}, addressing the selection of relevant data, data quality issues such as inconsistent formatting, missing values, and structural irregularities~\citep{rahm_data_2000}, as well as structural mismatches between data and the requirements of intended operations~\citep{wickham_tidy_2014}. Insufficient data preparation has been identified as a significant contributor to failures of systems operating in open environments with messy, real-world data~\citep{lai_kramabench_2026}. Nonetheless, it has seen limited coverage in the reviewed works. Several systems integrate specific data cleaning stages~\citep{mao_potable_2024, zhang_alter_2025, su_tablegpt2_2024}, sanitizing data before applying analytical operations. Beyond these, other approaches may apply ad-hoc cleaning, such as type coercions and string normalization~\citep{chen_tablerag_2024, chen_pandora_2025, zhang_reactable_2024}. Departing from these programmatic approaches, \citet{lu_tart_2025} explore fine-tuning a language model to perform data cleaning as a sequence-to-sequence task.

\subsubsection{Output Synthesis}
\label{sec:orchestration:capabilities:output}
Output synthesis capabilities enable \acrshort{task} agents to produce the final user-facing output from the analytical knowledge derived in the realization. Formally, this corresponds to executing the sink node of the realization \acrshort{dag} to address the output need $i_o$. While this is part of the operation structure $(O, E)$, output synthesis addresses how analytical knowledge is communicated to the user rather than how it is computed. This distinction matters because the utility of a realization depends not only on the soundness of its analytical derivation but also on how effectively its results are presented~\citep{munznerVisualizationAnalysisDesign2014, moritz_formalizing_2019}.

\acrshort{task} agents perform output synthesis to different degrees. At the minimal end, a system returns solely the raw computational result (e.g., a retrieved or calculated value, a result table). At the other end, systems select appropriate modalities, integrate results from multiple analytical branches into a coherent response, and add interpretative framing that contextualizes the analytical knowledge. In practice, the output need $i_o$ is often underspecified or entirely absent from the input, as reflected in current benchmarks where the expected modality is typically specified in the evaluation protocol rather than in the input (Section~\ref{sec:eval:input}). This means that systems need to compose a fitting presentation~\citep{gomm_are_2025, mackinlay_automating_1986}. The degree to which a system performs output synthesis thus determines to what extent it can make the analytical knowledge accessible to the user.

\paragraph{Output Modalities.} The output modality determines the form the analytical knowledge is presented in. Outputs may consist of natural language text, structured data such as tables, visualizations, and multimodal combinations thereof. As shown in the \textit{Output} column of Table~\ref{tab:functional_capabilities_overview}, the majority of the reviewed systems only output text. Within these, however, the outputs range considerably, from single discrete values concerning descriptive insight needs~\citep{zhou_efficient_2025, mao_potable_2024, kong_opentab_2023}, through short natural language responses that verbalize computational results~\citep{zou_rag_2025, wang_chainoftable_2024}, to paragraph-length answers that narrate analytical findings~\citep{zhao_tapera_2024} and long-form analytical reports that synthesize complex results~\citep{xu_dagent_2025}. Systems that return raw execution results, whether that is SQL result tables~\citep{zhang_murre_2025, kothyari_crush4sql_2023, chen_table_2024, wang_dbcopilot_2025} or code output values~\citep{chen_pandora_2025, lai_kramabench_2026}, surface the computational result without verbalization or interpretation, requiring that users understand the analytical context to interpret the raw results directly.

Beyond text and tables, systems have explored visual outputs, either as a separate modality, selected based on the input and standing on its own~\citep{huang_dacode_2024, li_are_2025, zhu_automated_2025, kochedykov_conversing_2023}, or contextualized in a multimodal response~\citep{su_tablegpt2_2024}. Dedicated capabilities for generating visualizations from data and natural language descriptions exist~\citep{yang_matplotagent_2024} but have not been widely integrated into the reviewed \acrshort{task} systems. Generally, different types of analytical knowledge benefit from different modalities. Retrieved facts and summary statistics, for example, are well served in text form, whereas trends and distributions may benefit from visualizations.

\paragraph{Result Integration.} If insight needs decompose into sub-needs, particularly independent sub-needs that are processed in parallel (Section~\ref{sec:orchestration:capabilities:interpretation}), the output must integrate the results of these sub-needs into a coherent output that addresses the original need. Systems that handle result integration specifically employ \acrshort{llm}s to compose a unified output from the results of sub-needs~\citep{zhang_alter_2025, xu_dagent_2025, wang_aop_2025, zhao_tapera_2024}.

\paragraph{Modality Selection.} To best address an output need $i_o$ and maximize the utility, a \acrshort{task} system must select an appropriate modality, or combination of modalities, for outputting the analytical knowledge. The appropriate modality depends on the nature of the analytical result, the insight need, and the user's ability to interpret different forms of presentation~\citep{munznerVisualizationAnalysisDesign2014}. However, in most reviewed systems, the selection is fixed to an evaluation-task-specific modality. In systems that support multiple modalities~\citep{huang_dacode_2024, li_are_2025}, the modality for a given instance is typically specified in the task instruction rather than determined by the system. 
A few systems do select the modality themselves. TableGPT2~\citep{su_tablegpt2_2024} dynamically selects it based on learned behavior in training, while TiInsight~\citep{zhu_automated_2025} and DataQue~\citep{kochedykov_conversing_2023} apply explicit rules and heuristics that map the result's data characteristics to a chart type or fall back to a table. Adjacent works on visualizations have addressed modality selection through methods that map data characteristics and analytical tasks to effective visualizations~\citep{mackinlay_automating_1986, moritz_formalizing_2019, mackinlay_show_2007}. However, among the reviewed systems, similar efforts have thus far not been undertaken.

\paragraph{Contextualization.} Recalling the definition of an insight as a collection of knowledge that links analytical knowledge with a person's contextual understanding (Section~\ref{sec:problem:conceptual_foundations:insights}), the output of an \acrshort{task} system may integrate the presented analytical knowledge with contextual information to improve the insight the user gets. This may involve relating findings back to the insight need, noting limitations of the underlying data, qualifying the scope of the results, or explaining what the analytical knowledge means in context. Yet contextualization is largely absent from output synthesis in current systems. A few systems target it explicitly. For instance, DAgent~\citep{xu_dagent_2025} generates reports that include interpretations alongside analytical results, DACO~\citep{wu_daco_2024a} is trained to translate quantitative findings into strategic, domain-specific recommendations, and AOP~\citep{wang_aop_2025} attaches generated reasoning to its results through a dedicated explanation operator. In most other systems, contextualization is incidental rather than a designed capability, as in TableGPT2~\citep{su_tablegpt2_2024}, which occasionally notes data availability limitations. The absence of contextualization of outputs is also reflected in evaluation practice, which mainly assesses the correctness of final values and treats additional verbalization as noise rather than value (Section~\ref{sec:eval:validation}).

\subsubsection{Process Governance}
\label{sec:orchestration:capabilities:governance}

Process governance capabilities operate at the meta-level~\citep{davis_metalevel_1977, davis_metarules_1980}, shaping which realization $r$ gets constructed without directly contributing executable operations to it. 
This includes error detection and recovery, validating intermediate and final results, reflecting on and revising the analytical strategy, and determining when the process is complete. As shown in Table~\ref{tab:functional_capabilities_overview}, many systems lack explicit process governance or only handle errors. Other systems employ a range of process governance techniques extending to result validation, adaptive steering, and ensembling.

\paragraph{Error Detection and Recovery.} \textit{Error detection} and \textit{error recovery} deal with failure states in the analytical process. This requires detecting failure states and exerting appropriate corrections, which may range from rigid fallbacks that follow predetermined alternatives to adaptive corrections informed by the observations of the error.
In the reviewed works, error detection is universally implemented as programmatic checks on code execution, using runtime exceptions from Python and SQL execution as a signal for failure states. In contrast, error recovery approaches follow different patterns. The most rigid form consists of applying predetermined fallbacks that are fixed at design time, such as cascading to executing simpler code~\citep{kong_opentab_2023}, switching to an alternative tool~\citep{zhou_efficient_2025}, or falling back to a different mode of generating the output~\citep{lu_tart_2025}. More adaptive recovery feeds the error message back to a generating \acrshort{llm} to re-generate the failing operation. This pattern is rooted in self-debugging~\citep{chenTeachingLargeLanguage2023} and widely adopted in the reviewed works~\citep{chen_pandora_2025, lai_kramabench_2026, mao_potable_2024, huang_dacode_2024, hu_infiagentdabench_2024, li_are_2025, su_tablegpt2_2024, zhao_tapera_2024, chen_tablerag_2024, jin_talon_2025, jiang_tabdsr_2025, qin_route_2024, pourreza_chasesql_2025}, typically bound by a fixed limit for retries~\citep{chen_pandora_2025, mao_potable_2024}, or a global step budget in ReAct-based systems~\citep{huang_dacode_2024, hu_infiagentdabench_2024}. In contrast, \citet{zhang_reactable_2024} explore error recovery by retrying failed SQL queries arising in a ReAct step against intermediate tables from earlier steps.

\paragraph{Result Validation.} Result validation assesses whether intermediate or final results meet quality criteria beyond successful code execution, since code can run error-free yet yield results that are empty, irrelevant, or misaligned with the insight need. Validation differs in depth. At a surface level, systems verify that execution produces non-empty or structurally compliant outputs and trigger error recovery otherwise~\citep{chen_pandora_2025, lai_kramabench_2026}. At a deeper level, systems assess whether a result is relevant to the insight need, either before execution by scoring or rewriting generated queries against the analytical requirements~\citep{xu_dagent_2025, chen_reliable_2026, chi_pisql_2025} or after execution by filtering out results obtained from irrelevant tables~\citep{kong_opentab_2023}. At the process level, validation can assess whether accumulated results collectively address the insight need, as in AOP's~\citep{wang_aop_2025} operators that search external knowledge for supporting or contradicting evidence and AutoTQA's~\citep{zhu_autotqa_2024} critic agent that compares results against the original need and triggers replanning where sub-needs remain unaddressed.

\paragraph{Adaptive Steering.} Adaptive steering covers capabilities to diagnose whether the analytical process is on track to yield the desired result and to adjust course where it is not. This mirrors the monitoring and replanning cycle in automated planning under uncertainty~\citep{ghallab_automated_2004}, where agents detect when observations invalidate the assumptions underlying their current strategy and adapt accordingly. Adaptations can range from assessments and alterations within individual steps, to backtracking, the exploration of different analytical paths, and, most comprehensively, to strategic revisions of the full analytical plan. Language model-based systems can perform the diagnostic component through \textit{self-reflection}, reasoning about the analytical process as a whole as part of the generative process. In ReAct-based systems~\citep{huang_dacode_2024, hu_infiagentdabench_2024, chen_tablerag_2024, su_tablegpt2_2024, wang_chainoftable_2024}, the interleaved reasoning steps provide an implicit form of reflection, where the agent reasons about its observations before selecting the next action. AOP~\citep{wang_aop_2025} handles self-reflection in a dedicated operator that generates justifications for intermediate results, making the agent's reasoning explicit and available for subsequent integration across analytical paths.
For correcting the analytical course, some systems implement \textit{re-planning}, adjusting an overarching analytical plan to the results of self-reflection. TaPERA~\citep{zhao_tapera_2024} and AutoTQA~\citep{zhu_autotqa_2024} perform a post-hoc analysis of the analytical process, informing re-planning of a further iteration of analysis. In contrast, AOP~\citep{wang_aop_2025} adapts its generated plan forward by evaluating intermediate results after each execution layer and pruning branches where operations have failed, re-invoking pipeline generation conditioned on the results gathered so far. Table-Critic~\citep{yu_tablecritic_2025} structures correction as a multi-agent loop in which a judge identifies errors in the reasoning chain, a critic suggests fixes, and a refiner revises the affected steps.

\paragraph{Ensembling.} Ensembling mitigates the variance inherent in stochastic processes, such as language model generation, by sampling multiple analytical trajectories and aggregating their results. A common approach is majority voting over independent runs of the full analytical process~\citep{chen_tablerag_2024, zhang_reactable_2024, sun_sqlprompt_2023}, selecting the most frequent answer. Alternatively, a trained selector can be applied to select among candidates~\citep{zhang_syntqa_2024, pourreza_chasesql_2025}. Ensembling is also applied to the analytical operations themselves, as CHASE-SQL~\citep{pourreza_chasesql_2025} selects between candidate SQL queries and AOP~\citep{wang_aop_2025} generates multiple candidate pipelines with justifications and has a language model combine them by assessing the soundness of each analytical process. Instead of comparing outputs or complete realizations, \citet{zhang_reactable_2024} explore branching at each analytical step and grouping intermediate steps that produce computationally equivalent results, finally selecting by aggregated log-probability of the respective output generation, grounding the selection in computational outcome rather than surface-level agreement. On a finer granularity, MACT~\citep{zhou_efficient_2025} samples multiple candidate actions per step and selects by majority vote.

\paragraph{Governance Mechanisms.} The mechanisms discussed above vary in how far meta-level governance capabilities are separated from object-level analytical capabilities. Separating the component that assesses results from the one that produces them is a well-established principle, originating in actor-critic architectures in reinforcement learning~\citep{konda_actorcitic_1999}. For \acrshort{llm}-based systems, which dominate current \acrshort{task} implementations, this separation is particularly relevant as self-correction by language models has been shown to be unreliable without external verification~\citep{huang_large_2023}, though structured verification procedures can partially mitigate this~\citep{wu_large_2024}. The majority of the reviewed works use the same language model that performs the analytical work also for process governance~\citep{chen_pandora_2025, lai_kramabench_2026, mao_potable_2024, zhao_tapera_2024, huang_dacode_2024, hu_infiagentdabench_2024}. Some systems introduce a separation by instructing the same underlying language model for different roles, as in AutoTQA's~\citep{zhu_autotqa_2024} critic agent and AOP's~\citep{wang_aop_2025} validation operators. In contrast, other systems use distinct models for a stronger separation, such as a pretrained cross-encoder for relevance assessment~\citep{kong_opentab_2023}, a fine-tuned evaluator for query quality~\citep{xu_dagent_2025}, or multiple agents in different roles that mutually validate one another~\citep{zhou_efficient_2025}. Complementing these, programmatic checks for code execution errors and results provide deterministic governance signals~\citep{chenTeachingLargeLanguage2023}.

\section{Interaction and Interpretability in OpenTI}
\label{sec:user_facing}
The preceding section examined how systems orchestrate functional capabilities to derive insight realizations from tabular corpora. Yet the quality of any realization ultimately also depends on how accurately the system interprets what the user, be that a person or an agent, seeks in the first place. This section shifts the focus from the mechanics of insight extraction to the interface between users and systems, examining how insight needs are communicated, interpreted, and refined through interaction. Compared to other sections in this paper, the discussion here is more conceptual and draws extensively on adjacent work in human-computer interaction and information retrieval beyond the systems and benchmarks reviewed in Sections~\ref{sec:orchestration} and~\ref{sec:eval}.

Requiring users to fully specify all choices involved in selecting data ($i_d$), defining the analytical methodology ($i_m$), and presenting results ($i_o$) is both impractical and often impossible. Complex insight needs entail numerous interdependent choices that users cannot reasonably enumerate upfront, and doing so would place a prohibitive burden on interactions. More fundamentally, insight needs are inherently latent, as established in Section~\ref{sec:problem:conceptual_foundations:insight_needs}.
The expressed need $\hat{i}$ is a compromised formulation of the latent need $i$, shaped by what users can articulate and believe the system can accommodate~\citep{taylor_process_1962, belkin_anomalous_1980}. The system consequently never has direct access to $i$, and can only construct an approximation $\hat{i}^{s}$ from the expressed input and available contextual information.
The fundamental challenge of designing user-system interactions is thus aligning the system's representation of the insight need $\hat{i}^{s}$ with the user's latent insight need $i$. How well this gap can be closed depends on the mechanisms for interpreting and specifying inputs available to systems and their users. Instead of demanding exhaustive specification of the insight need $i = (i_d, i_m, i_o)$, \acrshort{task} systems should treat users as cooperative communicators and interpret their inputs accordingly.

\citet{grice_logic_1975} posits that participants in cooperative communication provide sufficient but not excessive information (maxim of quantity) while remaining truthful (maxim of quality). In the context of insight extraction, this implies that users are intentional about which aspects of their insight needs they specify explicitly and which they leave implicit. When a user asks ``What is the development of global renewable energy installations since 2010?'', they specify the core semantic concepts but leave the exact data to retrieve, the methodological implementation, and the output format unspecified. This underspecification is not necessarily a failure of articulation but can be interpreted as a delegation of authority~\citep{clarkGroundingCommunication1991, gomm_are_2025}, where the user relies on the system to infer reasonable interpretations for unspecified components, and the system accepts this delegation as its contribution to the grounding process.

Extending the argument of \citet{gomm_are_2025}, whether and how a system can resolve a given instance of underspecification depends on the character of that instance. We propose to distinguish three ways of resolving underspecification, illustrated along the running example of a user asking ``What is the average income in Paris?'':

\begin{itemize}
    \item \textit{Conventional resolution.} Where strong conventions or domain defaults exist, the system can resolve unspecified aspects autonomously by assuming the most reasonable interpretation. The entity ``Paris'' is conventionally interpreted as the capital of France rather than any of the towns bearing the same name in the US or elsewhere, and the system can proceed on this basis, potentially disclosing the interpretation to the user.
    \item \textit{Selective resolution.} Where the expressed need admits multiple valid realizations but the user has expressed no preference among them, the system may exercise agency in selecting one. In the same example, the boundaries of Paris are not uniquely determined by the input, as the city of Paris proper, the greater Paris metropolitan area, and the Île-de-France region are all defensible interpretations. The system may select one based on what data is available or what is most commonly intended, but such choices must be disclosed, since users cannot otherwise judge whether the resulting realization reflects their intent and may mistake a contingent analytical decision for an objective finding.
    \item \textit{User engagement.} Where neither convention nor reasonable selection is applicable, the system must engage the user directly. This can take one of two forms. First, straightforward underspecification, where the user has not provided enough information to differentiate valid from invalid realizations. A user asking ``What is the average income?'' without any geographic or demographic boundary illustrates this, as there exists no convention to select a default scope, no principled set of alternatives can be derived from the input alone, and the system must elicit the missing specification before proceeding. Second, unresolvable underspecification, where the user cannot specify because they do not know the methodological possibilities of the system or the data it has access to. In this case, no clarifying question can be meaningfully answered until the system first makes something about its data and capabilities visible to the user.
\end{itemize}

This distinction parallels the obligations of user revealment and system revealment formulated by \citet{radlinski_theoretical_2017}, applied to \acrshort{task} systems. \textit{User revealment} requires the system to actively support users in articulating and refining their insight need over the course of interaction, through mechanisms that elicit, disambiguate, and extend the expressed need. \textit{System revealment} requires the system to continuously make its data coverage, analytical capabilities, and the choices it has made sufficiently transparent that users can form accurate expectations, update their inputs, and assess whether the system's contributions meet the grounding criterion. System revealment is not a one-time disclosure at the start of an interaction but an ongoing obligation that runs throughout the analytical process, as the system retrieves data, selects methods, and constructs realizations whose basis the user cannot otherwise inspect.

The extent to which conventional and selective resolution can align $\hat{i}^{s}$ with $i$ depends on the contextual knowledge the system can draw on when interpreting $\hat{i}$. Systems that do not adapt to specific users are limited to population-level conventions. Yet users bring distinct expertise, roles, and analytical perspectives~\citep{zhang_how_2020}, and they create a shared context with the system through continued interactions, where prior interactions serve as context for subsequent ones~\citep{yin_natural_2023}, so the same expressed need may warrant different interpretations and realizations depending on who expresses it and in what context. \textit{Personalization} aims to address this by conditioning interpretation on contextual information such as user profiles and interaction histories, a mechanism with a long tradition in information retrieval~\citep{teevan_personalizing_2005} that has more recently been applied to conditioning language model outputs on retrieved user context~\citep{salemi_lamp_2024, salemi_optimization_2024}. The mechanisms for contextualization reviewed in Section~\ref{sec:orchestration:capabilities:interpretation} partially address this conditioning, yet they ground interpretation exclusively in the domain, the corpus, and prior interactions within a session, whereas persistent adaptation to the individual user remains unexplored in the reviewed systems. Successful contextualization enables autonomous resolution of inputs that would otherwise require user engagement, reducing the interaction burden. At the same time, adaptation that is neither visible nor controllable risks resolutions that are misaligned in ways users cannot detect~\citep{jameson_pros_2002}, extending the disclosure obligations of system revealment to personalized interpretations.

This section explores the mechanisms through which user-system interactions support this dynamic. We characterize user-system interactions in \acrshort{task} along the \textit{interaction mode} (Section~\ref{sec:user_facing:interaction_modes}), which governs when and by whom input is provided during the insight extraction process, and the \textit{interaction means} (Section~\ref{sec:user_facing:interaction_means}), which describe the mechanisms through which input is conveyed. Together, interaction modes and means shape how insight needs are communicated, refined, and aligned between users and systems to enable appropriate realizations. Beyond these mechanisms, Section~\ref{sec:user_facing:interpretability} examines interpretability, required for users to verify the system's analytical derivations.

\subsection{Interaction Modes}
\label{sec:user_facing:interaction_modes}

\begin{figure}[ht]
    \centering
    \includegraphics[width=\linewidth]{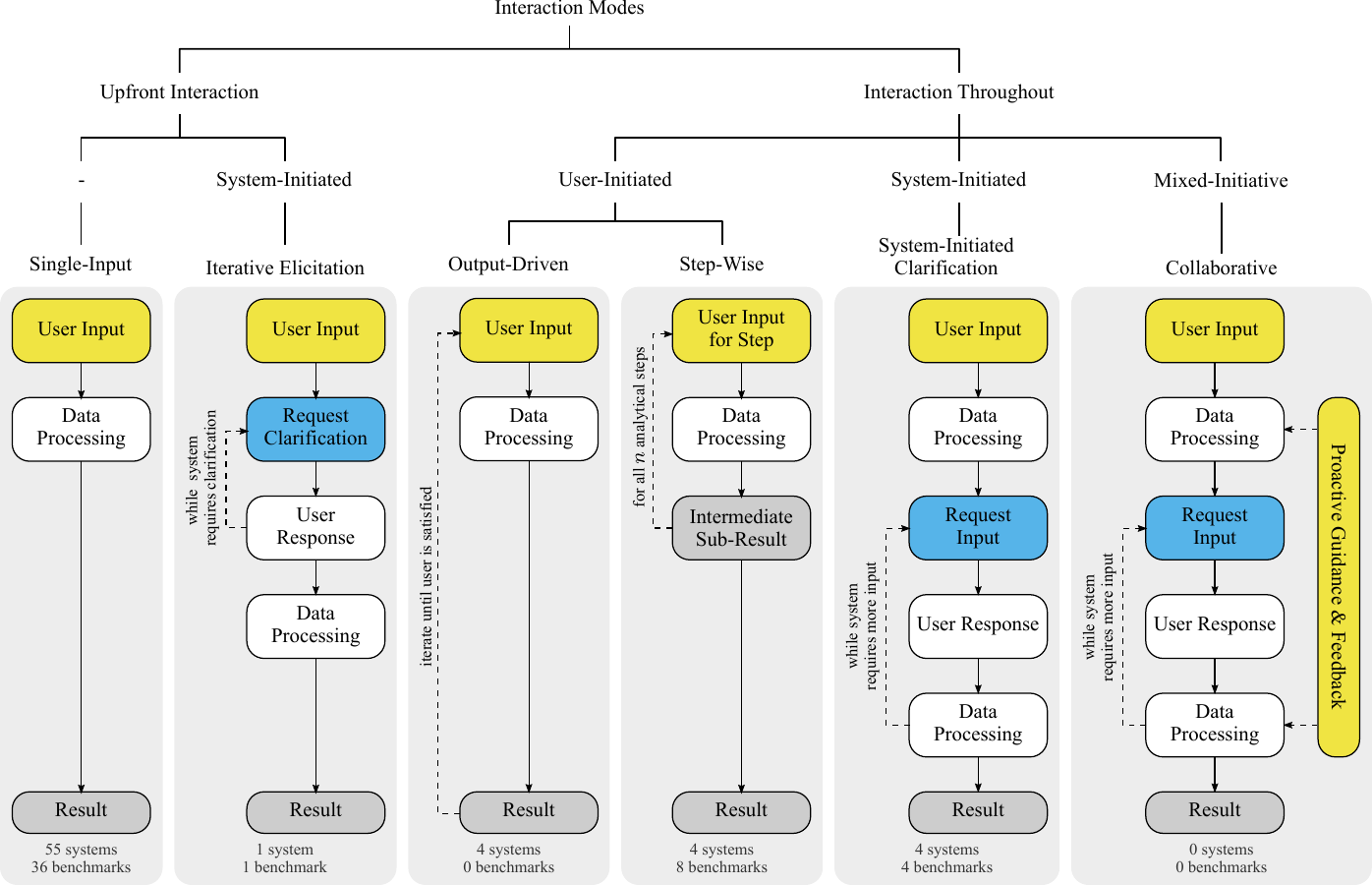}
    \caption{Interaction modes organized by interaction timing, whether input occurs only upfront or continues during execution, and interaction initiative. Counts state how many of the 58 reviewed systems and 42 benchmarks support the respective mode. Counts are not mutually exclusive; 7 systems and 7 benchmarks support more than one mode.}
    \label{fig:interaction_modes}
\end{figure}

Interaction modes define the control flow between user and system throughout the insight extraction process. As illustrated in Figure~\ref{fig:interaction_modes}, we propose distinguishing interaction modes by \textit{interaction timing}, differentiating whether user input occurs solely before data processing begins or continues throughout, and by the \textit{interaction initiative}, differentiating whether the system or the user takes the lead in steering the insight extraction and alignment with the user's intent. The initiative dimension draws on the spectrum of mixed-initiative interaction established by \citet{allen_mixedinitiative_1999}, ranging from purely user-led interactions through increasing degrees of system involvement to fully negotiated interactions in which neither party has a fixed role and both take active roles in steering the interaction. The timing dimension distinguishes modes in which user-system interaction is confined to an upfront stage before the system performs any data processing, from those in which interactions unfold throughout the execution, which is particularly relevant in \acrshort{task} settings where users cannot anticipate the data and analytical choices the system will confront during processing~\citep{horvitz_principles_1999, radlinski_theoretical_2017}.

Most systems (Section~\ref{sec:orchestration}) and benchmarks (Section~\ref{sec:eval}) assume input occurs solely before processing. The simplest case is \textbf{single-input} interaction, shown as the left-most mode in Figure~\ref{fig:interaction_modes}, where a single user input triggers the full extraction process without intermediate exchange, corresponding to purely user-initiative interaction. This mode minimizes user involvement and relies entirely on conventional and selective resolution to derive $\hat{i}^{s}$, but provides no mechanism to resolve remaining underspecifications or adapt to data particularities that users are not aware of, resulting in the propagation of misalignments into the final output. This places a high burden on the initial input to provide sufficient specification~\citep{yu_cosql_2019}, which may be particularly challenging the more complex the required analysis~\citep{grammel_how_2010}. \textbf{Iterative elicitation} (second from the left in Figure~\ref{fig:interaction_modes}) extends single-input by engaging users in pre-processing dialogue to refine $\hat{i}^{s}$ before data binding or analysis begins, mirroring commercial\footnote{E.g., \href{https://gemini.google/overview/deep-research}{Google Gemini Deep Research}, \href{https://mistral.ai/news/le-chat-dives-deep}{Mistral Le Chat}.} and academic~\citep{huang_deep_2025} ``deep research'' systems that ask clarifying questions to stabilize intent. This can surface and correct misconceptions before committing computational resources, but adds interaction overhead before users see any results and cannot address ambiguities arising from data selection or corpus-specific particularities encountered during execution.

While only taking input upfront minimizes user interactions, it does not enable users to react to and steer the insight extraction process, requiring them to be highly aware of and capable of expressing their insight need with sufficient specification. In contrast, integrating user input throughout processing enables reactive steering based on intermediate results, at the cost of increasing the interaction complexity. This provides users with the opportunity to disambiguate, verify intermediate results~\citep{yu_cosql_2019}, contribute domain knowledge, specify latent preferences, or intervene on critical decisions~\citep{shao_collaborative_2026}. The simplest such extension is \textbf{output-driven} interaction (third from the left in Figure~\ref{fig:interaction_modes}), where users observe system outputs and provide refinements, triggering iterative cycles until the realization $r$ satisfies their need~\citep{xie_waitgpt_2024, menon_fisql_2025}. This allows users to recognize retroactively when their initial specification led to misaligned realizations, correcting poor selective specification by the system as it becomes apparent upon seeing concrete results. Yet, achieving a satisfactory realization may require many iterations if the initial $\hat{i}$ diverges significantly from $i$, and requires full processing to complete before feedback can be incorporated. \textbf{Step-wise} interaction~\citep{li_are_2025, yin_natural_2023, dong_practiq_2025, yu_cosql_2019, xiao_cqrsql_2022}, shown as fourth from the left in Figure~\ref{fig:interaction_modes}, grants finer control by having users provide instructions step-by-step, directly specifying individual analytical steps rather than the complete, potentially high-level, insight need upfront. This is still a user-initiative interaction, but applied at the granularity of individual operations rather than the full extraction task. It suits more exploratory needs where $i$ emerges for users through data engagement~\citep{marchionini_exploratory_2006}, but shifts the orchestration burden to the user, requiring greater analytical expertise and end-to-end planning.

In \textbf{system-initiated clarification} (second from the right in Figure~\ref{fig:interaction_modes}), initiative shifts to the system, which solicits user input when encountering low-confidence decisions or missing information~\citep{wu_need_2024} required to resolve $\hat{i}$, that is, when selective resolution has low confidence or when information is genuinely missing. Rather than proceeding with uncertain interpretations, the system explicitly engages the user to resolve ambiguities, operationalizing the cooperative principle by exercising agency where it can while deferring when alignment is uncertain~\citep{horvitz_principles_1999}. This, however, requires a system that reliably estimates its uncertainty and may interrupt user workflows at unpredictable points. \textbf{Collaborative} interaction (right-most in Figure~\ref{fig:interaction_modes}) generalizes this to a bidirectional exchange throughout processing~\citep{chen_dango_2025, freund_flowco_2025}. The system requests clarification while users can also intervene proactively, observing operations and providing guidance without explicit solicitation~\citep{allen_mixedinitiative_1999}. This flexibility lets collaborative interaction address all three resolution challenges bidirectionally, but introduces implementation complexity and may overwhelm users unfamiliar with the analytical process. System-initiated and collaborative modes are only emerging for \acrshort{task}, building on established mixed-initiative principles from HCI research~\citep{horvitz_principles_1999, allen_mixedinitiative_1999}.

These modes are not discrete alternatives but form a continuum, and aspects of different modes may be combined, for instance by applying step-wise control with iterative refinement at each step. The appropriate mode depends on user expertise and the complexity and requirements of the analytical task~\citep{jameson_pros_2002}. Empirical evidence from information retrieval suggests that the degree to which an insight need is directed elicits measurably different interaction behaviors~\citep{athukorala_exploratory_2016}. Needs that are less directed at the start, which iteratively become more concrete through engagement with data~\citep{marchionini_exploratory_2006, white_exploratory_2009}, require sustained interaction throughout processing and tend to be poorly served by upfront specification modes, whereas needs that are more concrete and directed from the start place lower demands on continued interaction. Overall, interaction modes should accurately approximate the insight need while minimizing the burden placed on the user~\citep{wu_need_2024}.

The interaction modes described above assume the system commits to a single realization $r$ at each step, with iteration correcting misalignments between $r$ and the user's latent $i$. An alternative, compatible with several of the modes above, is for the system to surface multiple candidate realizations $\{r_1, \ldots, r_k\}$ simultaneously~\citep{bhaskar_benchmarking_2023}, making selective resolution explicit by presenting users with the space of valid alternatives rather than having the system choose implicitly. \citet{saparina_disambiguate_2025} generate multiple natural language interpretations of ambiguous queries before parsing, allowing users to select among semantically distinct readings. Similarly, multiverse analysis interfaces~\citep{sarma_milliways_2024} present multiple valid analytical paths arising from defensible methodological choices, enabling users to evaluate how results vary across reasonable alternatives. These approaches are particularly relevant in upfront specification modes where post-processing iteration is limited, making selective resolution explicit and preserving user agency over choices that would otherwise be delegated to the system. However, presenting too many alternatives risks overwhelming users, while too few may miss valid interpretations~\citep{saparina_disambiguate_2025}. Balancing comprehensiveness with cognitive manageability remains an open design challenge.

\subsection{Interaction Means}
\label{sec:user_facing:interaction_means}
Orthogonal to the interaction mode, the \textit{means} of interaction describe the mechanisms by which users provide input. Human-computer interaction research has established a progression of interaction styles, from command languages and form-filling through menu selection to direct manipulation~\citep{shneiderman_direct_1981}, each trading expressiveness against the guidance it provides to users. This is especially relevant in \acrshort{task} where users express analytical intent over data whose vocabulary and structure they may not know.

Natural language text constitutes the dominant interaction means in current systems and benchmarks, reflecting the broader emergence of conversational interfaces following the proliferation of chatbots. Text aligns naturally with open-ended insight needs, accommodating varied user goals without requiring familiarity with system-specific syntax. However, natural language is also well-documented as an ambiguous medium for data interaction~\citep{affolter_comparative_2019}, where unconstrained text provides no guidance about what information the system requires~\citep{chopra_conversational_2023}, leaving users unaware of what they must specify to avoid unresolvable underspecification. Complementary interaction means address these limitations by scaffolding specification to make the space of choices explicit, and by enabling direct manipulation at appropriate abstraction levels to correct misaligned resolutions as they emerge. 

\paragraph{Specification Interfaces.}
Specification interfaces impose a structure on the input channel so that users are guided towards providing the relevant information the system requires, adding a layer of guidance beyond free-text interaction. They address users' knowledge gap about what information the system expects by directing input into categories that signal what must be specified, something a blank text box does not do and that has been observed to slow users down and lead to underspecified requests~\citep{chopra_conversational_2023}. They can also surface valid options among which users can select rather than delegating decisions to systems.

Specification interfaces can be organized around the components of the insight need $i_d, i_m, i_o$. To target the data need $i_d$, specification interfaces may prompt for temporal scope, entity types, or domain boundaries, using the selected values to constrain retrieval over the corpus. For the methodological need $i_m$, specification interfaces can enumerate the options for analytical operations or modeling choices, freeing users from having to articulate the corresponding terminology. For the output need $i_o$, interfaces may expose modality choices such as tables, charts, or textual summaries, making the form of presentation a user-controlled rather than system-assumed dimension. \citet{guo_investigating_2024} provide an empirical demonstration of this decomposition, with a prototype that separates fields for dataset descriptions, domain knowledge injection, and selection among system-generated execution plans, finding that users benefit from this explicit separation.

A specification interface can provide static or dynamic scaffolding. Static interfaces fix the structure of the input channel at design time, which makes the system's expectations predictable but assumes that the same components are relevant across all insight needs. In contrast, dynamic interfaces synthesize the scaffolding from analytical context, presenting controls only at points where they serve a concrete underspecification. In natural language interaction with data this has been realized through follow-up questions~\citep{chen_dango_2025}, ambiguity widgets~\citep{gao_datatone_2015, setlur_eviza_2016} that bind controls to specific ambiguous phrases (e.g., a distance slider for ``near Paris''), and through type-aware widgets that adapt to the attribute being referenced (e.g., calendars for temporal attributes or maps for geospatial ones)~\citep{setlur_sneak_2020}. The continuous adaptation of the specification interface throughout the analytical process aligns with the partial observability in \acrshort{task} where the relevant gaps may not be known before retrieval and analytical composition begin. Infrastructure for such dynamic specification is re-emerging in \acrshort{llm}-based agents through toolkits such as MCP UI\footnote{\url{https://mcpui.dev/}}, though design principles for when and how to invoke structured prompts within \acrshort{task} systems remain underdeveloped.


\paragraph{Direct Intervention.} Direct intervention allows users to inspect and act on realizations while or after the system constructs them. It carries the principles of direct manipulation over to \acrshort{task}, namely continuous representation of the objects of interest, physical actions rather than complex syntax, and rapid, incremental, and reversible operations with immediately visible effects~\citep{shneiderman_direct_1981}. Applied to realizations, these principles require interfaces that make the components $r=(O, E, \beta)$ visible and editable, and that propagate user edits back into $r$ to regenerate the dependent operations. Users can thereby correct misaligned choices that become apparent only in light of intermediate results.

Such interfaces differ in the control they afford and the expertise they demand. At one end, users can be exposed to raw code, which gives full control but requires programming experience and imposes significant cognitive load even for proficient users confronted with unfamiliar implementations~\citep{xie_waitgpt_2024}. At the other end, intervention purely through follow-up natural language is accessible to non-programmers but requires users to articulate misalignments precisely and forces the system to ground these descriptions back into specific components of $r$, introducing miscommunication risks. Intermediate representations balance these extremes. Examples include editable natural language explanations of code~\citep{chen_dango_2025, tian_interactive_2023} and editable assumptions and requirements~\citep{kazemitabaar_improving_2024, freund_flowco_2025}. Similarly, visual graph representations expose the operations $(O, E)$ as nodes that can be inspected and parameterized without modifying code directly~\citep{xie_waitgpt_2024, freund_flowco_2025}, while editable analytical plans expose the methodological structure $i_m$ as ordered lists of steps that can be edited, reordered, or marked as optional~\citep{kazemitabaar_improving_2024, tian_interactive_2023}.

Direct interventions on the operations $(O, E)$ can enable users to adjust thresholds, change column selections, modify aggregation functions, or alter the composition of analytical steps. Complementing these, interventions on the data binding $\beta$ enable users to exclude irrelevant tables, request additional data matching specific criteria, or adjust relationships between retrieved tables. \citet{setlur_olio_2023} demonstrate the effectiveness of steering retrieval through visual previews and filtering support by presenting candidate data sources with metadata tooltips summarizing available attributes and value distributions, while filtering widgets enable refinement by properties such as temporal range or data type.

Interaction modes that involve the user throughout the analytical process presuppose some form of direct intervention. Step-wise, system-initiated clarification, and collaborative modes are defined by users acting on the realization as it emerges rather than merely responding to its final output, which requires representations of the emerging realization that users can inspect and act upon.

\paragraph{Combining Interaction Means.}
The means described above are not mutually exclusive. Research on data analysis interfaces has shown that combining natural language with direct manipulation can be more effective than either means alone, as natural language handles open-ended expressions of intent while direct manipulation handles precise, reference-dependent operations such as selecting data points or adjusting parameters~\citep{srinivasan_interweaving_2021, srinivasan_orko_2018}. Among reviewed systems, such combinations remain largely unexplored, as systems mostly rely on natural language as their sole interaction means.

\subsection{Interpretability}
\label{sec:user_facing:interpretability}
Making the realization $r = (O, E, \beta)$ that a system produces accessible and interpretable enables users to verify the analytical process and its alignment with their intent. To allow end-to-end verification, users must be able to confirm that the expressed insight need $\hat{i}$ is correctly interpreted, that the data binding $\beta$ retrieved appropriate tables from the corpus, and that the operations $O$ and their composition $E$ are analytically sound. Empirical studies into verification behavior confirm that users combine assessing the applied operations with inspecting the data they operate on, switching between these levels throughout the verification process~\citep{gu_how_2024, xie_waitgpt_2024}, which indicates that interpretability must address all three components of the insight need to avoid verification gaps.

Interpretability aims to foster warranted trust by allowing users to distinguish sound from unsound realizations, instead of merely projecting transparency by exposing more information~\citep{parasuraman_complacency_2010}. Superficial transparency can induce unwarranted reliance on flawed realizations~\citep{parasuraman_complacency_2010}, which is consequential when outputs inform decisions or are otherwise relied upon.

The conceptual foundations for surfacing (partial) realizations have been established in the study of \textit{data provenance} and \textit{analytical provenance}. \textit{Data provenance} tracks the origin and transformation history of analytical results~\citep{buneman_why_2001,cheney_provenance_2009,woodruff_supporting_1997}, providing a foundation for interpretability by making explicit how outputs derive from the data through sequences of operations. \textit{Analytical provenance}, established in visual analytics~\citep{gotz_characterizing_2009}, captures the trajectory of the analytical process itself, including what operations were considered and applied, making the agent's decision path explicit. Mapped to realizations, data provenance concerns $\beta$ and the data flow induced by $E$, and analytical provenance addresses the operations $O$ and their composition $E$. Beyond surfacing the realization itself, systems can make the interpretation process that derives $r$ from the user input $\hat{i}$ explicit. Disclosing the interpretation, which happens, for instance, through selective resolution, allows users to detect misalignments between the system's reading of their intent and the underlying need $i$ before inspecting the analytical derivation~\citep{xie_waitgpt_2024, kazemitabaar_improving_2024, freund_flowco_2025}.

Not just the \textit{contents} of the interpretation but also its \textit{form} matters because it determines who can effectively verify realizations. There is a trade-off between the fidelity with which the underlying realization is presented and the accessibility to users. Raw code preserves full fidelity but requires programming proficiency to read and imposes cognitive load even on proficient users~\citep{xie_waitgpt_2024}. Visual graph representations of operations~\citep{xie_waitgpt_2024} abstract the realization into nodes and edges that can be interrogated individually without exposing implementation details. Natural language decompositions of generated code into explanations~\citep{narechania_diy_2021, tian_sqlucid_2024, tian_interactive_2023, menon_fisql_2025, nguyen_interpretable_2025} offer another form, providing a human-readable intermediate representation between the raw derivation and the user. \citet{tian_sqlucid_2024} extend these natural language decompositions by aligning each explanation step with the relevant database elements and displaying intermediate results after each step. Empirical evidence indicates that such step-by-step descriptions paired with sample data enable non-experts to locate errors in the analysis~\citep{narechania_diy_2021}. An effective representation remains accessible enough to allow verification while staying structured enough to preserve the analytical content of the realization. Beyond representations of the operations, inspectable intermediate data states~\citep{gu_how_2024, kazemitabaar_improving_2024} address data provenance by allowing users to observe how operations affect the data as the realization is executed, which provides evidence for assessing whether $\beta$ and $E$ align with the intended interpretation.

While interpretations can be surfaced post-hoc, allowing users to inspect a completed realization and decide whether to trust its output, they can also be disclosed concurrently with the analytical process as it unfolds. Such concurrent interpretability is a precondition for the output-driven, step-wise, and collaborative interaction modes discussed in Section~\ref{sec:user_facing:interaction_modes}.

Sufficient interpretability is required to avoid verification gaps, but presenting excessive information risks overwhelming users~\citep{kazemitabaar_improving_2024}. \citet{guo_investigating_2024} find that users particularly seek explanations when outputs do not match their expectations, suggesting that interpretability needs are non-uniform across the verification process. Effective interpretability should therefore address the trade-off between comprehensiveness and cognitive load. \textit{Progressive disclosure}~\citep{shneiderman_eyes_1996}, in which users access a summary-level view of the realization by default and navigate to finer levels of detail on demand, offers a possible design paradigm, which has been extensively explored for visual analytics drill-down interfaces. The appropriate default abstraction and the requirements for drill-down depend on user expertise and on the complexity of the analytical task, neither of which the reviewed works study systematically.

\section{Evaluations in OpenTI}
\label{sec:eval}
In this section we understand assessing the gap between a system's theoretical capabilities and its practical utility as the primary objective of evaluation. This requires systematically quantifying a system's strengths and weaknesses through a robust methodology that mirrors the complexities of real-world data interaction.

Formally, an evaluation benchmark $B = (T, \mathcal{E})$ consists of a corpus of tables $T$ and a set of $n$ evaluation instances $\mathcal{E} = \{e_1,\ldots, e_n\}$. Each evaluation instance $e = (\bar{i}, v)$ comprises an input $\bar{i}$, representing the expression of an insight need $\hat{i}$, and a validation mechanism $v: \mathcal{R} \to [0,1]$ that scores the quality of a system's realization. In most current benchmarks, $\bar{i}$ is a single natural language utterance and $v$ reduces to comparing the output against a predetermined reference output. However, more sophisticated instantiations can capture interactive input protocols and nuanced validation criteria that assess not only the final output but also the analytical process by which it was derived.

\begin{fullpagetable}
    \fullpagecaption{\small Overview of analyzed benchmarks. \textbf{Words} denotes the average number of words in an input. Insight Types are classified following the taxonomy in Section~\ref{sec:problem:spectrum}: \raisebox{-0.6ex}{\pill[okabe_1]{Look}} - Lookup, \raisebox{-0.6ex}{\pill[okabe_1]{Agg}} - Aggregation, \raisebox{-0.6ex}{\pill[okabe_1]{Char}} - Characterization, \raisebox{-0.6ex}{\pill[okabe_1]{Asso}} - Association, \raisebox{-0.6ex}{\pill[okabe_1]{Grp}} - Grouping, \raisebox{-0.6ex}{\pill[okabe_3]{SI}} - Statistical Inference, \raisebox{-0.6ex}{\pill[okabe_3]{Pred}} - Prediction, \raisebox{-0.6ex}{\pill[okabe_5]{Int}} - Interventional, \raisebox{-0.6ex}{\pill[okabe_5]{CF}} - Counterfactual. A bordered box \raisebox{-0.6ex}{\borderpill[okabe_1]{Grp}}/\raisebox{-0.6ex}{\borderpill[okabe_3]{Pred}} denotes instances that require an artifact (e.g., a prediction model) instead of being directed at an insight. \textbf{Tbl/Inst.} details the mean number of tables required per instance, where \singleSettingIcon~indicates that only a single relevant table is provided, \multiSettingIcon~indicates that only the set of relevant tables is provided, \setSettingIcon~indicates that a limited set of relevant and irrelevant tables is provided, and \openSettingIcon~indicates an open setting. ``-'' indicates that the data is not available or the analysis is not applicable to the benchmark.
    }
    \scriptsize
    \setlength{\tabcolsep}{4pt}
    \centering
    \resizebox{\linewidth}{!}{%
        \renewcommand{\arraystretch}{1.6}
\begin{tabular}{p{3cm} p{1cm} p{3cm} p{3.6cm} p{1cm} p{1.25cm} p{1cm} p{3.7cm} 
p{6cm} p{1.65cm}}
\toprule

& \multicolumn{3}{c}{\textbf{Corpus}} & \multicolumn{4}{c}{\textbf{Inputs}} & \multicolumn{2}{c}{\textbf{Validation Functions}} \\
\cmidrule(lr){2-4} \cmidrule(lr){5-8} \cmidrule(lr){9-10} 
\textbf{Name} & \textbf{\#Tab} & \textbf{Source} & \textbf{Supplement Data} & \textbf{\#Inst.} & \textbf{Tbl/Inst.} & \textbf{Words} & \textbf{Insight Types} & \textbf{Function} & \textbf{Output} \\

\midrule \arrayrulecolor{lightline}

    Spider~\citeyear{yu_spider_2018} & 1,056 & University Courses, DatabaseAnswers & Relations & 11,840 & \setSettingIcon~1.55 & 12.2 & \rpill[okabe_1]{Look} \rpill[okabe_1]{Agg} & $\out$: Exact matching\newline $r$: Structural matching & Table, Value \\ \hline
    OTT-QA~\citeyear{chen_open_2020} & 8,891 & Wikipedia & Text Passages, Title & 4,372 & \openSettingIcon~1 & 19.9 & \rpill[okabe_1]{Look} \rpill[okabe_1]{Agg} & $\out$: Exact matching, Similarity matching & Value \\ \hline
    NQ-Tables~\citeyear{herzig_open_2021} & 169,898 & Wikipedia &  & 11,628 & \openSettingIcon~1 & 8.9 & \rpill[okabe_1]{Look} \rpill[okabe_1]{Agg} & $\out$: Exact matching, Similarity matching & Value \\ \hline
    FeTaQA~\citeyear{nan_fetaqa_2022} & 10,330 & Wikipedia & Text Passages, Title & 10,330 & \singleSettingIcon~1 & 13.2 & \rpill[okabe_1]{Look} \rpill[okabe_1]{Agg} & $\out$: Similarity matching, Human evaluation & Text \\ \hline

    ARCADE~\citeyear{yin_natural_2023} & 106 & Kaggle, GitHub &  & 1,078 & \singleSettingIcon~1 & - & \rpill[okabe_1]{Look} \rpill[okabe_1]{Agg} \rpill[okabe_1]{Char} & $\out$: Similarity matching & Value, Table \\ \hline
    Open-WikiTable~\citeyear{kweon_openwikitable_2023} & 24,680 & Wikipedia & Page Title & 67,023 & \openSettingIcon~1 & 18.1 & \rpill[okabe_1]{Look} \rpill[okabe_1]{Agg} & $\out$: Exact matching & Value \\ \hline
    BIRD~\citeyear{li_can_2024} & 611 & Kaggle, ML Data Repository, Manual & Relations, Column\newline Description & 10,962 & \setSettingIcon~2 & 14.1 & \rpill[okabe_1]{Look} \rpill[okabe_1]{Agg} & $\out$: Exact matching & Table, Value \\ \hline
    Archer~\citeyear{zheng_archer_2024} & 68 & Spider & Relations, Text Passage & 1,042 & \setSettingIcon~2.17 & 26.2 & \rpill[okabe_1]{Look} \rpill[okabe_1]{Agg} \rpill[okabe_5]{CF} & $\out$: Exact matching & Table, Value \\ \hline
    BLADE~\citeyear{gu_blade_2024} & 14 & Scientific Works & Dataset/Column Description & 188 & \singleSettingIcon~1 & 68.0 & \rpill[okabe_1]{Asso} \rpill[okabe_3]{SI} & $r$: Component validation, Structural matching & Value \\ \hline
    DA-Code~\citeyear{huang_dacode_2024} & 1,733 & GitHub, Kaggle, Web & Files & 500 & \multiSettingIcon~3.47 & 40.6 & \rpill[okabe_1]{Look} \rpill[okabe_1]{Agg} \rpill[okabe_1]{Char} \rpill[okabe_1]{Asso} \rborderpill[okabe_1]{Grp} \rpill[okabe_3]{SI} \rborderpill[okabe_3]{Pred} & $\out$: Exact matching & Text, Value, Table, Chart \\ \hline
    InfiAgent-DABench~\citeyear{hu_infiagentdabench_2024} & 67 & GitHub &  & 257 & \singleSettingIcon~1 & 27.4 & \rpill[okabe_1]{Agg} \rpill[okabe_1]{Char} \rpill[okabe_1]{Asso} \rpill[okabe_3]{SI} \rborderpill[okabe_3]{Pred} & $\out$: Exact matching & Value \\ \hline
    DA-Dataset~\citeyear{xu_dagent_2025} & 689 & Financial, BIRD & Relations & 735 & \setSettingIcon~4.5 & - & \rpill[okabe_1]{Agg} \rpill[okabe_1]{Asso} \rpill[okabe_1]{Char} & $\out$: LLM-as-judge & Text \\ \hline
    Spider 2.0~\citeyear{lei_Spider20_2025} & 418 & BigQuery, Snowflake Marketplace, Web & Relations & 632 & \setSettingIcon~3.89 & 54.0 & \rpill[okabe_1]{Look} \rpill[okabe_1]{Agg} \rpill[okabe_1]{Asso} & $\out$: Exact matching, Programmatic validation & Text, Table \\ \hline
    TableBench~\citeyear{wu_tablebench_2025} & 886 & Wikipedia &  & 886 & \singleSettingIcon~1 & 20.2 & \rpill[okabe_1]{Look} \rpill[okabe_1]{Agg} \rpill[okabe_1]{Asso} \rpill[okabe_3]{Pred} & $\out$ (text): Similarity matching \newline $\out$ (chart): Programmatic validation & Text, Chart \\ \hline
    DABstep~\citeyear{egg_dabstep_2025} & 3 & Financial & Files & 450 & \setSettingIcon~- & 22.4 & \rpill[okabe_1]{Look} \rpill[okabe_1]{Agg} \rpill[okabe_5]{CF} & $\out$ (value, list): Exact matching\newline $\out$ (text): Similarity matching & Text, Value, List \\ \hline
    MT-RAIG~\citeyear{seo_mtraig_2025} & 19,563 & Spider, Open-WikiTable &  & 18,532 & \openSettingIcon~2.88 & 36.5 & \rpill[okabe_1]{Look} \rpill[okabe_1]{Agg} \rpill[okabe_1]{Char} \rpill[okabe_1]{Asso} & $\out$: LLM-as-judge & Text \\ \hline
    DAComp - DA~\citeyear{lei_dacomp_2025} & 418 & Curated web DBs & Relations & 100 & \setSettingIcon~4.18 & 90.1 & \rpill[okabe_1]{Look} \rpill[okabe_1]{Agg} \rpill[okabe_1]{Char} \rpill[okabe_1]{Asso}  & $\out$: LLM-as-judge & Text + Chart \\ \hline
    MultiTableQA~\citeyear{zou_rag_2025} & 59,307 & Wikipedia &  & 23,785 & \openSettingIcon~2.39 & 14.8 & \rpill[okabe_1]{Look} \rpill[okabe_1]{Agg} & $\out$: Exact matching, Similarity matching & Value \\ \hline
    KramaBench~\citeyear{lai_kramabench_2026} & 1,636 & Scientific Works & Files & 104 & \setSettingIcon~3.33 & 33.8 & \rpill[okabe_1]{Look} \rpill[okabe_1]{Agg} \rpill[okabe_1]{Asso} \rpill[okabe_3]{Pred} & $\out$: Exact matching, Similarity matching\newline $r$: Component validation & Text, Value, List \\ \hline
    ConDABench~\citeyear{duttaConDABenchInteractiveEvaluation2025} & 1,855 & TidyTuesday, Scientific Works, Kaggle & & 1,420 & \multiSettingIcon~1.31 & 13.8 & \rpill[okabe_1]{Look} \rpill[okabe_1]{Agg} \rpill[okabe_1]{Char} \rpill[okabe_1]{Asso} \rpill[okabe_3]{Pred} & $\out$: LLM-as-judge & Text, Chart \\ \hline
    TACO~\citeyear{dengTACOBenchmarkOpenDomain} & 13,004 & Open Data Portals & Relations & 13,000 & \openSettingIcon~- & - & \rpill[okabe_1]{Look} \rpill[okabe_1]{Agg} & $\out$: Exact matching & Table, Value \\ \hline
    TOPBench~\citeyear{ji_topbench_2026} & 35 & Kaggle & Column Description, Value Distribution & 779 & \singleSettingIcon~1 & 108.8 & \rpill[okabe_3]{Pred} & $\out$: Exact matching, Similarity matching, LLM-as-judge & Text, Table\\ \hline
    CausalReasoning\newline Benchmark~\citeyear{sawarni_causalreasoningbenchmark_2026} & 132 & Scientific Works & Column Description, Study Context & 173 & \singleSettingIcon~1 & 32.2 & \rpill[okabe_1]{Agg} \rpill[okabe_3]{SI} \rpill[okabe_5]{Int} & $\out$: Similarity matching,\newline $r$: Component validation & Text, Value\\
\arrayrulecolor{black}
\bottomrule
\end{tabular}

    }
    \label{tab:datasets}
\end{fullpagetable}

This section first provides an overview of the surveyed benchmarks (Table~\ref{tab:datasets}), and then analyzes them along their components. We systematically analyze the benchmarks that we find in our literature review and select a subset of benchmarks relevant for \acrshort{task} that we present in the overview in Table~\ref{tab:datasets}. The methodology for the survey and the selection criteria are detailed in Appendix~\ref{s:appendix:review_method}. Section~\ref{sec:eval:corpora} examines the corpora underlying the benchmarks, Section~\ref{sec:eval:input} analyzes inputs, and Section~\ref{sec:eval:validation} investigates validation mechanisms. Each section lays out key desiderata for the respective component before analyzing how benchmarks address them.

Table~\ref{tab:datasets} gives an overview of the analyzed benchmarks, presenting the characteristics of their corpora, inputs, and validation mechanisms. The benchmarks span a wide range of scope and complexity, from single-table question answering over web tables to multi-step analytical workloads over relational and file-based corpora.

\subsection{Corpora}
\label{sec:eval:corpora}

The corpus $T$ in a benchmark $B$ defines the data environment on which systems are evaluated. In current benchmarks, corpora range from collections of thousands of web-extracted tables to smaller sets of analytical datasets sourced from repositories like Kaggle or GitHub. These corpora differ substantially in their size, the complexity of their tables, and the degree to which they reflect real-world data environments.

\subsubsection{Desiderata}
\label{sec:eval:corpora:desiderata}

To simulate an evaluation environment representative of real-world data environments, $T$ should mirror the complexity and scale of real-world data sources. Thus, a corpus suitable for evaluating \acrshort{task} systems should satisfy the following desiderata:

\begin{itemize}
    \item \textbf{Realism and Scale.} Many academic datasets are built on tables extracted from web pages, which are typically small~\citep{nan_fetaqa_2022, chen_open_2020}. In contrast, representative organizational or scientific tables often have significantly more columns and rows~\citep{li_discovering_2017, hulsebos_gittables_2023} and exhibit different distributions of semantic column types~\citep{hulsebos_gittables_2023}. A benchmark should therefore feature a large-scale corpus~\citep{lai_kramabench_2026, mitsopoulou_analysis_2025} containing numerous, sizable tables from diverse domains to properly test a system's ability to operate in a realistic open setting.
    \item \textbf{Messiness.} Real-world data is rarely pristine. A realistic benchmark should include tables characterized by a degree of messiness, such as missing values, inconsistent formatting, extraneous information, or varying schema quality~\citep{lai_kramabench_2026, mitsopoulou_analysis_2025, duttaConDABenchInteractiveEvaluation2025}. This ensures the evaluation tests not only a system's analytical capabilities but also its essential data preparation and cleaning abilities~\citep{chen_empowering_2025}.
    \item \textbf{Domain Diversity.} In the open setting systems must generalize across heterogeneous sources and topics. A corpus dominated by a narrow domain like sports statistics may overestimate performance by allowing systems to exploit specific world-knowledge or domain-specific patterns~\citep{gan_exploring_2021}. Corpora should span multiple domains to assess whether systems can transfer capabilities across different subject areas and data conventions.
    \item \textbf{Structural Diversity.} Tables have varied structures, including wide versus tall layouts~\citep{dohmen_schemapile_2024}, normalized versus denormalized schemas~\citep{kohita_exploring_2025}, and different levels of metadata availability. A corpus with structural diversity enables testing different integration challenges, particularly for insight needs that span multiple tables~\citep{kohita_exploring_2025}.

\end{itemize}

\subsubsection{Analysis of Corpora}
\label{sec:eval:corpora:analysis}

\begin{figure}[ht]
    \centering
    \includegraphics[width=\linewidth]{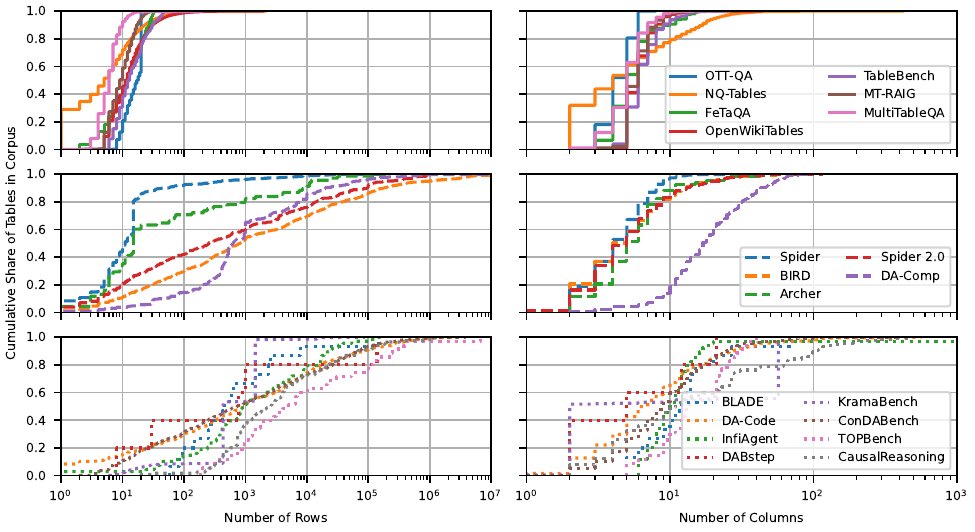}
    \caption{Empirical cumulative distributions of per-table row and column counts for the corpora in Table~\ref{tab:datasets}. Steep curves at low counts mark corpora of small tables, while flatter curves indicate diversity in row and column counts.}
    \label{fig:datasets:statistics}
\end{figure}

Current benchmark corpora exhibit a clear trade-off between the number of tables in the associated corpus $|T|$ and the structural complexity of the tables in these corpora. This becomes apparent when considering the \textit{\#Tab} and \textit{Source} columns of Table~\ref{tab:datasets} together with the distribution of row and column counts across individual tables, depicted in Figure~\ref{fig:datasets:statistics}. Corpora derived from Wikipedia web tables, such as NQ-Tables~\citep{herzig_open_2021}, OTT-QA~\citep{chen_open_2020}, FeTaQA~\citep{nan_fetaqa_2022}, Open-WikiTable~\citep{kweon_openwikitable_2023}, and MultiTableQA~\citep{zou_rag_2025}, offer large scale, often containing tens of thousands of tables. However, the individual tables are predominantly small and clean, with the overwhelming majority featuring fewer than ten columns and rarely exceeding a few hundred rows, intended for presentation rather than analysis. They thereby provide scale at the expense of structural realism and diversity. Additionally, it can be assumed that the overwhelming majority of large language models have been exposed to Wikipedia contents during training~\citep{yu_kola_2024}, potentially leading to data contamination and a better contextual understanding of Wikipedia contents~\citep{ni_survey_2025}, which can have downstream impacts on the measured performance and the domain generalization capabilities beyond Wikipedia tables.

Corpora sourced from relational databases and analytical repositories exhibit more realistic table characteristics at the cost of scale. Relational corpora such as Spider~\citep{yu_spider_2018} and BIRD~\citep{li_can_2024} extend to far wider row distributions, with individual tables reaching hundreds of thousands of rows, while data-analysis corpora drawn from Kaggle, GitHub, and scientific sources, such as DA-Code~\citep{huang_dacode_2024}, InfiAgent-DABench~\citep{hu_infiagentdabench_2024}, and KramaBench~\citep{lai_kramabench_2026}, exhibit the widest spread in both rows and columns, more closely reflecting the complexity of tables used in actual analytical workflows. However, these corpora are substantially smaller, containing hundreds rather than tens of thousands of tables, which limits their utility for evaluations in an open setting for table discovery and retrieval. KramaBench~\citep{lai_kramabench_2026} is a notable exception in coupling raw, realistic data with a data-lake organization spanning multiple formats, though it too remains modest in its number of tables.

Beyond the tables themselves, benchmarks contain diverse additional data, reported under \textit{Supplement Data} in Table~\ref{tab:datasets}. Some establish their corpus as relational data~\citep{yu_spider_2018, li_can_2024, lei_Spider20_2025, dengTACOBenchmarkOpenDomain, xu_dagent_2025, wu_mmqa_2025}, combining tables with titles and relationships in the form of foreign-key relationships. Next to such relations, many benchmarks provide further contextual metadata associated with tables, like contextual descriptions~\citep{nan_fetaqa_2022, chen_open_2020} or documentation files~\citep{huang_dacode_2024, lai_kramabench_2026, egg_dabstep_2025}. Such contextual information may be used to bridge the semantic gap between tabular data and user inputs to aid functional capabilities like interpretation, data retrieval~\citep{gomm_metadata_2025}, and analytical composition and execution~\citep{egg_dabstep_2025}.

The characteristics of the corpora largely follow from where they are sourced rather than from deliberate design. Realism and scale stand in direct tension, as most large-scale corpora are sourced from Wikipedia, yielding tables that do not resemble those in organizational corpora. Exceptions to this are KramaBench~\citep{lai_kramabench_2026}, which pairs raw scientific data with a data-lake organization spanning multiple formats, but at a modest number of tables, and TACO~\citep{dengTACOBenchmarkOpenDomain}, which has a large corpus of 13,000 tables from municipal and federal open data portals but parses and cleans the raw tables into normalized relational databases. Messiness is thus rarely treated as a design property. Synthetically constructed corpora~\citep{yu_spider_2018, egg_dabstep_2025, lei_dacomp_2025} are clean by construction, and corpora assembled from existing data are commonly filtered and standardized during curation, leaving only a few benchmarks that deliberately retain dirty data~\citep{li_can_2024, huang_dacode_2024, duttaConDABenchInteractiveEvaluation2025, lai_kramabench_2026}. The domains a corpus covers are likewise a byproduct of its source. Corpora sourced from Wikipedia or open data portals offer a broad range of domains, whereas the domain coverage is narrow in benchmarks targeting one specific analytical setting~\citep{egg_dabstep_2025, xu_dagent_2025, lei_dacomp_2025}, so that the corpora spanning domains broadly are the same web table corpora with the least structurally realistic tables.

\subsection{Inputs}
\label{sec:eval:input}
The input $\bar{i}$ defines part of the environment for evaluated \acrshort{task} systems, comprising the user inputs that are observable. Following the framework established in Section~\ref{sec:problem}, $\hat{i}$ is the expressed form of the user's latent insight need $i$ communicated in user-system interaction, comprising data, methodological, and output components. The input $\bar{i}$ defines the protocol under which $\hat{i}$ is emulated in the evaluation. Since potential interaction modes range from single inputs to mixed-lead collaborative interactions, requirements on what $\bar{i}$ needs to capture and how interaction should be emulated vary significantly between interaction modes. Benchmarks are usually proposed targeting specific interaction modes. In single-input settings, $\bar{i}$ can be a standalone natural language utterance; in interactive settings, it may comprise a sequence of exchanges or a protocol for responding to system-initiated clarification. Benchmarks vary considerably in the analytical complexity of insight needs, the degree to which inputs are suitable for evaluations under an open premise, and whether they support interaction beyond single utterances.

\subsubsection{Desiderata}
\label{sec:eval:input:desiderata}

Inputs suitable for evaluating \acrshort{task} should satisfy the following desiderata:

\begin{itemize}

    \item \textbf{Data-Independence.} In the open setting, users formulate insight needs without knowledge of specific data structures or contents~\citep{voorhees_trec_2001}. Authentic inputs should thus be \textit{data-independent}, formulated from the user's conceptual perspective, without using privileged knowledge of the underlying corpus a user does not have. Data-privileged inputs, by contrast, reference structural elements such as column headers (e.g., ``\textit{last\_name}''), specific values not in the public domain (e.g., ``\textit{the airplane that took off at 2026-05-08T11:48:11+00:00}''), or data containers themselves (e.g., ``\textit{the user table}''). Such references provide unrealistic signals linking queries to specific data structures, undermining the premise of an open corpus~\citep{gomm_are_2025, suhr_exploring_2020, lee_kaggledbqa_2021}.

    \item \textbf{Sufficient Specification.} As argued in Section~\ref{sec:problem:conceptual_foundations:insight_needs}, insight needs may exhibit a degree of latency where users do not exhaustively articulate them but provide what they judge necessary and rely on systems to resolve the rest. What counts as sufficient specification therefore depends on the validation mechanism used to score a realization. When paired with a validation mechanism that admits only a single reference output or realization, inputs must fully determine the required data and methodology, since alternative interpretations would produce divergent but plausibly correct responses that the validation cannot accommodate. Validation mechanisms that accept multiple valid realizations allow inputs to leave ambiguities in corresponding aspects. Misalignment between the degree of ambiguity in inputs and the validation mechanism can conflate accuracy with interpretation capabilities, yielding unreliable signals~\citep{bhaskar_benchmarking_2023, saparina_disambiguate_2025, gomm_are_2025}.

    \item \textbf{Analytical Complexity.} Real-world analytical workloads often require multi-step reasoning, involving sequences of operations that build upon intermediate results~\citep{gu_blade_2024, wu_tablebench_2025, lai_kramabench_2026}. Inputs should reflect this complexity to test whether systems can plan and execute extended analytical procedures.

    \item \textbf{Multi-Table Sources.} Real-world questions often require integrating information from multiple tables. Therefore, a significant share of the evaluation instances should require retrieving and integrating information from multiple tables~\citep{lai_kramabench_2026, wu_mmqa_2025, akhtar_tanq_2025}.
\end{itemize}

\subsubsection{Input Protocols}

Depending on how inputs are provided, evaluation protocols fall into two categories: \textit{static} and \textit{user simulation}. Static protocols fix all inputs upfront in the benchmark itself, while user simulation protocols define the mechanisms by which inputs that depend on the analytical state or system elicitation are determined.

In their simplest and most common form, static protocols define a single standalone utterance per evaluation instance to emulate single-input interactions, which allows tight control over the input and a consistent, replicable evaluation environment. Similarly, for step-wise interaction a multi-turn sequence of inputs can be fixed upfront~\citep{yu_cosql_2019, yin_natural_2023, liu_tabcqa_2023, li_longtablebench_2025}. While such sequences emulate a multi-step process, the predetermined path does not adapt to the intermediate analytical state a system exposes at each step, trading the realism of a genuinely contingent interaction for replicability.

User simulation protocols instead generate inputs dynamically, emulating a user who provides additional inputs proactively based on observations of the analytical state and/or reactively in response to systems requesting inputs. User simulation has a long lineage in information access research~\citep{balog_user_2024}. It generally centers on a model of the user~\citep{balog_user_2024}, which can be based on rule-based and agenda-based simulators~\citep{schatzmann_agendabased_2007} that generate inputs by reacting to system outputs according to an explicit representation of user goals, or an end-to-end neural architecture~\citep{gur_user_2018, lin_domainindependent_2021} that learns the behavior. LLMs have been proposed as end-to-end user models~\citep{sekulicEvaluatingMixedinitiativeConversational2022, sekulic_reliable_2024, shao_collaborative_2026}. Empirical assessment finds that such models correlate well with human judgments overall~\citep{douSimulatorArenaAreUser2025}, but cautions against possible goal misalignments, particularly in multi-turn contexts~\citep{zhang_recent_2020, mehri_goal_2026}, and against unrealistic response patterns~\citep{seshadriLostSimulationLLMSimulated2026}, against which scaffolding techniques have been proposed~\citep{mehri_goal_2026}.

\subsubsection{Analysis of Inputs}
\label{sec:eval:input:analysis}
Current benchmarks overwhelmingly adopt static input protocols for single-input interaction, whereas a smaller set fixes multi-turn inputs upfront to emulate step-wise interaction~\citep{yu_cosql_2019, yin_natural_2023, li_are_2025, liu_tabcqa_2023, li_longtablebench_2025}. User simulation remains rare, applied only in ConDABench~\citep{duttaConDABenchInteractiveEvaluation2025}, BIRD-INTERACT~\citep{huo_birdinteract_2025}, and COTA~\citep{li_are_2025}. They all employ an \acrshort{llm} to simulate the user but differ in how they align it with the underlying insight need $i$. COTA~\citep{li_are_2025} grounds the user simulator in an extensive description of the insight need, decomposed into a set of sub-questions, as well as a sample of the table contents. BIRD-INTERACT~\citep{huo_birdinteract_2025} and ConDABench~\citep{duttaConDABenchInteractiveEvaluation2025} provide the schema of the data and directly expose the desired analytical operations as SQL and Python code to the LLM, while prompting it to use the code to inform interactions but not to share it, risking leakage of information the system did not ask for. 

As shown in Table~\ref{tab:datasets}, the benchmarks focus mostly on descriptive insights, with little coverage of inferential and causal insights. Lookups and aggregations have the broadest coverage, and many benchmarks do not go beyond these two types. Some benchmarks remain restricted to simple $\min$ and $\max$ aggregations that do not require calculations~\citep{chen_open_2020, nan_fetaqa_2022, herzig_open_2021}, allowing direct inference over the full table to derive answers from the raw data. In comparison, most benchmarks that expand to more complex aggregations, characterization, and association, which require more complex numerical operations, are rooted in data analysis workflows~\citep{huang_dacode_2024, hu_infiagentdabench_2024, lei_Spider20_2025, wu_tablebench_2025, gu_blade_2024} and are mostly absent from the question answering and text-to-SQL benchmarks. At the inferential tier, statistical inference is covered as an explicit analytical objective in BLADE~\citep{gu_blade_2024}, DA-Code~\citep{huang_dacode_2024}, InfiAgent-DABench~\citep{hu_infiagentdabench_2024}, and CausalReasoningBenchmark~\citep{sawarni_causalreasoningbenchmark_2026}. Prediction is covered more widely but to different extents. DA-Code~\citep{huang_dacode_2024} and InfiAgent-DABench~\citep{hu_infiagentdabench_2024} contain inputs for constructing predictive models without seeking a specific insight, whereas TableBench~\citep{wu_tablebench_2025}, KramaBench~\citep{lai_kramabench_2026}, and ConDABench~\citep{duttaConDABenchInteractiveEvaluation2025} target simple forecasts through regression. TOPBench~\citep{ji_topbench_2026} is the first dataset that centers on predictive insights, verbalizing complex predictive needs that require building and applying models to extract the insights. Causal insight needs have been covered on a shallow level as counterfactual insights in Archer~\citep{zheng_archer_2024} and DABstep~\citep{egg_dabstep_2025}, which explore counterfactual scenarios that are resolved under the application of provided static rules. Only CausalReasoningBenchmark~\citep{sawarni_causalreasoningbenchmark_2026} targets genuine interventional analysis, requiring formal observational identification strategies such as instrumental variables, regression discontinuity, and difference-in-differences alongside statistical estimation.

Beyond the covered insight needs, input definitions across benchmarks vary widely in the data requirements they impose. Many benchmarks, particularly those focused on question answering, simplify the problem setting by requiring evidence from a single table only, abstracting away the real-world complexities of data integration that define the full scope of \acrshort{task}. The \textit{Tbl/Inst.} column of Table~\ref{tab:datasets} showcases this, with 9 of the 23 benchmarks assuming that either a single relevant table or a set of relevant tables is directly provided to a system, while only 6 target an open setting that necessitates retrieval. The remainder restrict selection to a bounded set, mostly a single database. Similarly, the complexity of the insight needs expressed in the input definitions is unevenly distributed across benchmarks. We report the average words per input definition as a proxy for the complexity of insight needs in the \textit{Words} column of Table~\ref{tab:datasets}, observing a large range from an average of 8.9 words for inputs in NQ-Tables~\citep{herzig_open_2021} to 108.8 words in TOPBench~\citep{ji_topbench_2026}, whose intent-rich queries embed the analytical scenario in a narrative description. However, while longer inputs generally allow for more detailed instructions, word count alone fails to capture the latency in expressions of insight needs, which allows insight needs to be compressed into shorter forms under greater latent assumptions and delegation. DABstep~\citep{egg_dabstep_2025} and Spider~2.0~\citep{lei_Spider20_2025} are deliberately designed so that the expressed insight needs require multiple analytical steps that build on intermediate results, whereas most other benchmarks can be served by one-shot code generation.

Next to these structural properties, evaluating \acrshort{task} systems in an open setting requires that realistic inputs are formulated independently of the exact data that serves as evidence while providing sufficient specification of the methodological and data need in the expression of the insight need. We assess the degree to which these desiderata are violated following the LLM-judge setup proposed by \citet{gomm_are_2025}, with results shown in Figure~\ref{fig:datasets:ambiguity_privilege}, detailed in Appendix~\ref{s:appendix:input_classification}. Benchmarks vary significantly in the data privilege found in inputs. The analysis reveals high data privilege primarily in data-analysis benchmarks like InfiAgent-DABench (70.0\%), DABstep (78.4\%), and TOPBench (85.0\%), whose inputs reference specific columns and files, limiting the degree to which evaluations in open settings capture realistic interactions. On the other end of the spectrum, NQ-Tables (0.0\%) and FeTaQA (0.4\%) have negligible levels of data privilege. In addition, all benchmarks show high levels of insufficient specification of inputs. All benchmarks but NQ-Tables (31.6\%) and FeTaQA (47.4\%) contain more than 50\% of inputs that are insufficiently specified. While validation mechanisms based on similarity or LLM judgment may absorb some of this ambiguity, the dominance of exact-matching validation in surveyed benchmarks (Section~\ref{sec:eval:validation:output_level}) means that the bulk of insufficiently specified inputs are paired with validation that admits only one valid interpretation, conflating execution accuracy with the ability of systems to guess the intended interpretation.

\begin{figure}[th]
    \centering
    \includegraphics[width=\linewidth]{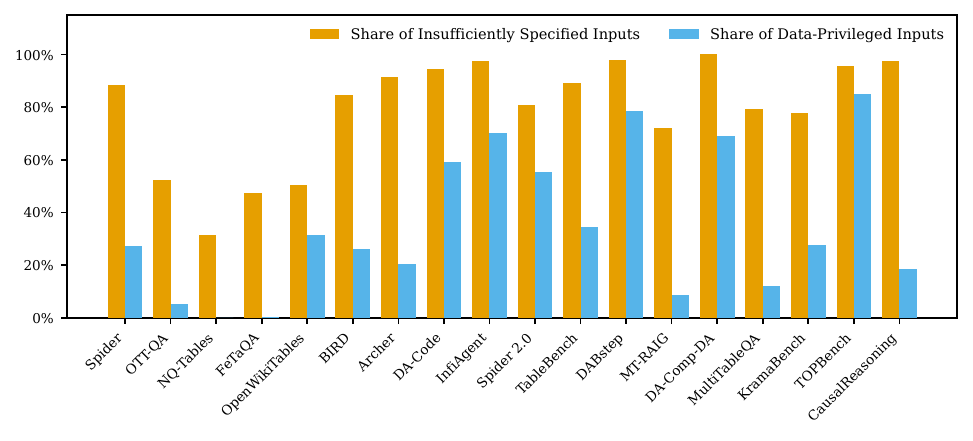}
    \caption{Share of inputs that are insufficiently specified (i.e., admitting multiple interpretations) and that exhibit data privileges (i.e., leakage of privileged information from the data into the input) across benchmarks.}
    \label{fig:datasets:ambiguity_privilege}
\end{figure}

Overall, inputs across the benchmarks remain limited for evaluating \acrshort{task} systems in an open setting. Since most benchmarks are not designed for an open setting, many of their inputs exhibit high levels of data dependence and insufficient specification. The analytical complexity is bounded, with the most widespread focus on lookup and aggregation types, whereas inferential and causal insight types remain only sparsely covered. In contrast, many benchmarks include inputs that require combining evidence from multiple tables to address the insight need, though most do so in a data setting where a small set of tables is provided and the set of relevant tables does not have to be selected from a large corpus.

\subsection{Validation Mechanisms}
\label{sec:eval:validation}

The validation mechanism $v: \mathcal{R} \to [0,1]$ determines how evaluation assesses the quality of a system's realization against the user's insight need. Ideally, validation would assess to which degree the complete realization $r = (O, E, \beta)$, consisting of the operations performed, their composition, and the data to which they are bound, constitutes a sound derivation of the analytical knowledge that addresses the user's insight need. In practice, the validation target ranges from the terminal output $\out(r)$ produced by the realization to components of the realization $r$ itself.

Connecting this to the problem formulation (Section~\ref{sec:problem:conceptual_foundations:problem_formulation}), $v$ acts as an operational stand-in for the validity criterion captured by $R_{\text{valid}}(i)$ and the utility function $u_i$. Since the underlying insight need is fixed by the benchmark's creators, the validation mechanism encodes their judgment of which realizations satisfy the need and how well. An ideal validation mechanism would take the form $v(r) = \mathbf{1}[r \in R_{\text{valid}}(i)] \cdot g(u_i(r))$, where $g$ maps utility monotonically into $(0, 1]$, with its exact form immaterial since the objective in Equation~\ref{eq:objective} depends only on the ordering that $u_i$ induces. Realizations that fail to address the insight need score zero, while valid realizations are scored by how well they serve it. In practice, validation in existing benchmarks primarily approximates the validity indicator, leaving the quality of realizations and outputs largely unjudged.

We separate validation functions that operate on the output level (Section~\ref{sec:eval:validation:output_level}) from those that operate on the realization level (Section~\ref{sec:eval:validation:realization_level}), and consider combined functions (Section~\ref{sec:eval:validation:combining}). \textit{Output-level} validation compares the final output $\out(r)$ against reference outputs or success criteria, without assessing the validity of the analytical process itself. In contrast, \textit{realization-level} validation examines components of $r$ directly, assessing the analytical process itself rather than only its results. Table~\ref{tab:validation_functions} provides an overview of the validation functions we discuss in this section. We focus on functions that score the end-to-end validity of an analysis or its output, setting aside metrics that isolate a single functional capability, such as the recall of the retrieval capability~\citep{herzig_open_2021, zou_rag_2025}.

\begin{table}[ht]
    \centering
    \caption{Overview of validation functions in reviewed benchmarks grouped by whether they validate outputs or the realizations themselves. Formalizations correspond to the specific functions employed by the benchmarks.}
    \label{tab:validation_functions}
    \footnotesize

\setlength{\tabcolsep}{4pt}
\renewcommand{\arraystretch}{1.2}

\newcommand{\vfspan}{\dimexpr 7.0cm+7.5cm+2\tabcolsep\relax}
\newcommand{\vfhead}[2]{\multicolumn{2}{@{\hspace{5pt}}>{\raggedright\arraybackslash}p{\vfspan}@{}}{\textbf{#1} #2}}
\newcommand{\vfsep}{\addlinespace[5pt]\arrayrulecolor{gray!45}\cmidrule[0.5pt](l{5pt}r{0pt}){2-3}\arrayrulecolor{black}\addlinespace[4pt]}

\begin{tabular}{@{} c l p{8.8cm} @{}}
\toprule
& \textbf{Validation Function} & \textbf{Benchmarks} \\
\midrule
\multirow{5}{*}[-11em]{\rotatebox[origin=c]{90}{\textit{Output-level}}}
 & \multicolumn{2}{p{15.4cm}}{\cellcolor{black!7} \textbf{Exact matching} matches the output, or sub-output $\out_j(r)$, against a reference output $\out^*$ under an equivalence function $\mathbf{1}[\cdot]$.} \\
 & $v(r) = \mathbf{1}[\out(r) = \out^*]$ & \scriptsize Spider~\citeyear{yu_spider_2018}, OTT-QA~\citeyear{chen_open_2020}, NQ-Tables~\citeyear{herzig_open_2021}, Open-WikiTable~\citeyear{kweon_openwikitable_2023}, BIRD~\citeyear{li_can_2024}, Archer~\citeyear{zheng_archer_2024}, Spider 2.0~\citeyear{lei_Spider20_2025}, DA-Code~\citeyear{huang_dacode_2024}, MultiTableQA~\citeyear{zou_rag_2025}, TACO~\citeyear{dengTACOBenchmarkOpenDomain}, TOPBench~\citeyear{ji_topbench_2026}, KramaBench~\citeyear{lai_kramabench_2026} \\
 & $v(r) = \mathbf{1}[\out_j(r) = \out_j^* \forall j \in 1, \dots, n]$ & \scriptsize InfiAgent-DABench~\citeyear{hu_infiagentdabench_2024} \\
 \vfsep
 & \multicolumn{2}{p{15.4cm}}{\cellcolor{black!7} \textbf{Similarity matching} measures the similarity of the output and a reference with a similarity measure $\mathrm{sim}$.} \\
 & $v(r) = \mathrm{sim}(\out(r), \out^*)$ & \scriptsize FeTaQA~\citeyear{nan_fetaqa_2022}, ARCADE~\citeyear{yin_natural_2023}, TableBench~\citeyear{wu_tablebench_2025}, KramaBench~\citeyear{lai_kramabench_2026}, TOPBench~\citeyear{ji_topbench_2026}, CausalReasoningBenchmark~\citeyear{sawarni_causalreasoningbenchmark_2026}, NQ-Tables~\citeyear{herzig_open_2021}, OTT-QA~\citeyear{chen_open_2020}, MultiTableQA~\citeyear{zou_rag_2025} \\
 & $v(r) = \max\big(\mathbf{1}[\out(r) = \out^*], \mathrm{sim}(\out(r), \out^*)\big)$ & \scriptsize DABstep~\citeyear{egg_dabstep_2025} \\
 \vfsep
 & \multicolumn{2}{p{15.4cm}}{\cellcolor{black!7} \textbf{LLM-as-judge} prompts a language model to score the output against the reference following instructions $\mathrm{instr}$ or a set of criteria $C$.} \\
 & $v(r) = LLM(\out(r), \out^*, \mathrm{instr})$ & \scriptsize DA-Dataset~\citeyear{xu_dagent_2025}, ConDABench~\citeyear{duttaConDABenchInteractiveEvaluation2025}, TOPBench~\citeyear{ji_topbench_2026} \\
 & $v(r) = \frac{1}{|C|}\sum_{c \in C}LLM(c, \out(r))$ & \scriptsize MT-RAIG~\citeyear{seo_mtraig_2025}, DAComp~\citeyear{lei_dacomp_2025} \\
 \vfsep
 & \multicolumn{2}{p{15.4cm}}{\cellcolor{black!7} \textbf{Programmatic validation} checks the output against a set of per-instance programmatic test cases $\mathrm{tests}$.} \\
 & $v(r) = \mathbf{1}[\mathrm{check}(\out(r), \tau) \forall \tau \in \mathrm{tests}]$ & \scriptsize Spider 2.0~\citeyear{lei_Spider20_2025}, TableBench~\citeyear{wu_tablebench_2025} \\
 \vfsep
 & \multicolumn{2}{p{15.4cm}}{\cellcolor{black!7} \textbf{Human evaluation} asks annotators to judge the output against a set of criteria $C$.} \\
 & $v(r) = \frac{1}{|C|} \sum_{c \in C} \mathrm{human}(\out(r), c)$ & \scriptsize FeTaQA~\citeyear{nan_fetaqa_2022}\\
\midrule
\multirow{4}{*}[-3.5em]{\rotatebox[origin=c]{90}{\textit{Realization-level}}}
 & \multicolumn{2}{p{15.4cm}}{\cellcolor{black!7} \textbf{Program comparison} compares the code $p(\cdot)$ of the realization with that of a reference realization.} \\
 & $v(r) = \mathrm{sim}(p(r), p(r^*))$ & \textcolor{gray}{\textit{none among the benchmarks in Table~\ref{tab:datasets}}} \\
 \vfsep
 & \multicolumn{2}{p{15.4cm}}{\cellcolor{black!7} \textbf{Structural matching} compares the operations of the realization $r=(O,E,\beta)$ and their composition against a reference realization $r^* = (O^*, E^*, \beta^*)$ directly.} \\
 & $v(r) = \frac{1}{|O^*|} \sum_{o \in O}\mathbf{1}[o \in O^*]$ & \scriptsize Spider~\citeyear{yu_spider_2018}, BLADE~\citeyear{gu_blade_2024}\\
 \vfsep
 & \multicolumn{2}{p{15.4cm}}{\cellcolor{black!7} \textbf{Component validation} checks the realization against a set of per-instance criteria $C$.} \\
 & $v(r) = \frac{1}{|C|}\sum_{c \in C}\mathrm{check}(r, c)$ & \scriptsize BLADE~\citeyear{gu_blade_2024}, KramaBench~\citeyear{lai_kramabench_2026} \\
 & $v(r) = \mathbf{1}[\mathrm{check}(r, c) \forall c \in C]$ & \scriptsize CausalReasoningBenchmark~\citeyear{sawarni_causalreasoningbenchmark_2026} \\
\bottomrule
\end{tabular}
\end{table}

\subsubsection{Desiderata}
\label{sec:eval:validation:desiderata}

Validation mechanisms suitable for evaluating \acrshort{task} should satisfy the following desiderata:

\begin{itemize}[itemsep=1pt]
    \item \textbf{Scalability and reproducibility.} Human evaluation offers flexibility in assessing complex, open-ended outputs but is costly, slow, and difficult to reproduce consistently. Instead, a practical validation mechanism should be automated to ensure reproducibility, scalability, and objectivity, requiring minimal human intervention~\citep{gu_blade_2024}.
    
    \item \textbf{Multi-modal coverage.} Analytical results can be presented in different modalities like natural language, numerical results, tables, and visualizations. The validation mechanism should thus be able to handle the modalities relevant to the evaluated insight needs~\citep{zhang_datascibench_2025, huang_dacode_2024}.
    
    \item \textbf{Robustness and faithfulness.} Metrics should reliably quantify what they claim to measure, providing signal rather than noise. This is particularly challenging for open-ended analytical outputs, where traditional text similarity metrics such as BLEU, ROUGE, and BERTScore have been found to correlate poorly with actual response quality and to not reliably separate factually accurate responses from those that are fluent yet factually incorrect~\citep{wolff_how_2025, wang_revisiting_2024}. Validation mechanisms must be calibrated to the specific characteristics of analytical outputs to yield meaningful performance estimates.

    \item \textbf{Acceptance of valid variation.} A single insight need may admit multiple valid realizations producing different but equally valid outputs~\citep{zeng_airepr_2025}, whether through alternative analytical methods, different but appropriate data selections, or varied presentation choices. Validation mechanisms should accommodate this variation rather than penalizing correct responses that diverge from a single gold standard~\citep{gu_blade_2024, zeng_airepr_2025}. This requires either curating sets of valid reference outputs or employing validation approaches capable of assessing correctness beyond exact matching.
\end{itemize}

\subsubsection{Output-Level Validation}
\label{sec:eval:validation:output_level}
Output-level validation assesses the final output $\out(r)$ produced by executing a realization $r$ against one or more reference outputs or a set of success criteria. This disregards the underlying analytical procedure captured in $r$, which simplifies automated evaluation but makes it impossible to distinguish correct reasoning from coincidentally correct answers~\citep{liu_are_2024}.

\textbf{Exact matching.} Exact matching compares the output against a desired reference output under an equivalence function. Since validity is only confirmed if the system's output exactly matches the reference output, the output must follow the exact form of the desired output. Different approaches have been explored to constrain the output space. Multiple-choice questions represent the most restrictive approach, limiting valid outputs to a predefined set of options~\citep{lu_dynamic_2022, li_are_2025}. This makes complex analytical questions easily verifiable, but the few choices may allow systems to take shortcuts and guess instead of deriving the correct output~\citep{balepur_which_2025}. Format-prompting similarly constrains outputs by requiring responses to follow specific templates, such as \texttt{\{@answer[value]\}}, which can be parsed and verified automatically~\citep{hu_infiagentdabench_2024}. These constraints simplify validation at the cost of limiting the expressiveness of responses and potentially leaking information about expected answers. Another approach is to parse the outputs into a standardized format before matching. For instance, \citet{huang_dacode_2024} parse plotting scripts to extract numerical data and plotting parameters into numpy and JSON formats for comparison. Exact matching also often requires some cleaning of outputs to account for permutations in lists, different formatting (e.g., ``16€'' vs. ``16.00''), and further insignificant variations~\citep{egg_dabstep_2025}. Exact matching is the most common validation function used across the analyzed benchmarks. Many benchmarks from tabular question answering are designed to allow only a single value that exactly matches the output~\citep{pasupat_compositional_2015, lu_dynamic_2022, zhang_crtqa_2023, herzig_open_2021, kweon_openwikitable_2023, chen_open_2020, zhao_multihiertt_2022, zou_rag_2025, contalbo_griqa_2025, katsis_aitqa_2022}, whereas benchmarks rooted in text-to-SQL generation often employ execution accuracy~\citep{yu_spider_2018, li_can_2024, wang_retqa_2025, kumar_booksql_2024, chen_beaver_2025}, defining an equivalence function that compares the execution results of SQL statements with expected results, sometimes in a column- or row-order invariant manner. Execution accuracy cannot differentiate between semantically distinct queries that incidentally yield the same result~\citep{klopfenstein_spotit_2026}.
Exact matching extends to insight needs that admit multiple valid outputs by crediting a match against any one of them, though this extension is not employed in any analyzed benchmark. To make exact matching more fine-grained, \citet{hu_infiagentdabench_2024} propose decomposing the output into multiple components, matching each against its own reference value and aggregating the results into a score.

\textbf{Similarity-based matching.} Similarity-based matching relaxes the exact-equivalence requirement by applying a similarity measure to compare a system's output to a reference output. 
Different similarity measures have been proposed for individual output modalities, ranging from numeric tolerances that absorb floating-point differences~\citep{egg_dabstep_2025}, through fuzzy matching of tables that tolerates additional rows and columns~\citep{yin_natural_2023} and image similarity for rendered charts~\citep{li_are_2025}, to lexical and semantic overlap measures such as BLEU~\citep{papineni_bleu_2002}, ROUGE~\citep{lin_rouge_2004}, and BERTScore~\citep{zhang_bertscore_2019} for free-form text~\citep{nan_fetaqa_2022, wu_tablebench_2025}. However, these similarity metrics for free-form text have been found to correlate poorly with human judgments of free-form outputs~\citep{wolff_how_2025, wang_revisiting_2024}. For predictive insight needs, similarity-based matching can be applied to the model used for making the prediction, for instance by calculating the relative performance gap to a well-performing baseline model~\citep{huang_dacode_2024, jing_dsbench_2024}. Similarity-based matching generalizes to sets of reference outputs analogously to exact matching, crediting the best-matching reference.

\textbf{LLM-as-Judge.} To handle genuinely open-ended outputs, recent benchmarks increasingly employ large language models as evaluators, prompted with the output, a reference output, and instructions that specify the evaluation criteria~\citep{gu_blade_2024,chen_scienceagentbench_2025,xu_dagent_2025,wolff_how_2025,lai_ds1000_2023, lei_dacomp_2025, sahu_insightbench_2024, duttaConDABenchInteractiveEvaluation2025, jing_dsbench_2024}. The flexibility of these instructions enables specifying conditions that accept semantically equivalent outputs, multiple valid realizations, or fine-grained success criteria, offering flexibility beyond what surface metrics afford. Judges are structured in different ways to sharpen this signal, decomposing the evaluation criteria into fine-grained rubrics combined into a weighted score~\citep{lei_dacomp_2025}, augmenting the instructions with instance-specific context and criteria~\citep{kim_flex_2025}, or decomposing the output itself into atomic claims verified individually against the tables~\citep{seo_mtraig_2025}. TOPBench~\citep{ji_topbench_2026} further guards against hallucination-induced judgment errors through a multi-step approach that directly matches quotes from the result to expected outputs. A single judge can also span several output modalities at once to account for different ways of presenting the analytical knowledge~\citep{duttaConDABenchInteractiveEvaluation2025}. Rather than scoring outputs in isolation, judges are also prompted to compare two outputs against each other, either by expressing a preference between two analyses~\citep{wu_daco_2024a} or by rating one relative to the other on aspects such as readability, analytical depth, and visualization~\citep{lei_dacomp_2025}. While offering greater flexibility, LLM judges may also introduce biases, cost, complexity, and reproducibility concerns~\citep{egg_dabstep_2025}. While \citet{wolff_how_2025} find a high accuracy of LLM-judges for simple, descriptive insight needs, \citet{hu_infiagentdabench_2024} only find a 67\% consistency between human experts and LLM-judges for more complex insight needs beyond simple description, underscoring the need to validate the alignment of LLM judges with human judgment for the specific setting in which they are applied~\citep{zhu_tableeval_2025, wang_revisiting_2024}.

\textbf{Programmatic validation.} Instead of relying on predefined desired outputs, programmatic validation introduces flexible, task-specific success criteria in the form of test cases capturing conditions the output must satisfy~\citep{chen_scienceagentbench_2025}. These test cases are usually expressed in code and can check aspects like verifying that a value falls within acceptable bounds~\citep{chen_scienceagentbench_2025}, or that a derived output exhibits specific features~\citep{lei_Spider20_2025, wu_tablebench_2025}, or that the output matches any of multiple patterns. Programmatic validation provides unambiguous and reproducible assessment but requires outputs amenable to automated testing and test cases specified per evaluation instance.

\textbf{Human evaluation.} Instead of automated scoring functions, human annotators are employed to rate outputs. Human evaluation can be applied to judge the outputs of individual evaluation instances against explicit criteria~\citep{nan_fetaqa_2022, yu_cosql_2019} or for indicating a pairwise preference for one analysis over another~\citep{wu_daco_2024a, lei_dacomp_2025}.
Human evaluation makes the evaluation more flexible and can adapt to instance-specific particularities or faults outside anticipated criteria, but it introduces cost in terms of scalability and reproducibility, confining human evaluation to relatively small samples~\citep{nan_fetaqa_2022, yu_cosql_2019, zhang_crtqa_2023, wu_daco_2024a} or to a fallback where automated matching is brittle, as demonstrated in CRT-QA~\citep{zhang_crtqa_2023}, where it absorbs formatting errors that exact matching would otherwise penalize.

Output-level validation treats the realization as a black box, meaning systems can arrive at correct outputs through flawed analysis or bypass data evidence entirely~\citep{liu_are_2024}, and partially correct analyses receive no credit. Realization-level validation addresses these gaps by inspecting components of $r$ directly.

\subsubsection{Realization-Level Validation}
\label{sec:eval:validation:realization_level}

Instead of narrowly focusing on the final output, realization-level validation assesses the analytical process captured in $r$ itself. It thereby allows distinguishing correct from incorrect analysis, regardless of the result, and allows more fine-grained scoring that can give partial credit to partially correct analysis. The central challenge in realization-level validation lies in surfacing components of $r$ in a form that allows automated checking, either by requiring systems to declare specific artifacts alongside their output or by extracting them from the executed realization.

\textbf{Program comparison.} Similar to similarity-based matching at the output-level, program comparison directly applies a similarity function to compare the operations and their composition captured in $r$ against those in a reference realization $r^*$, where both $r$ and $r^*$ are represented as code programs. The similarity measure can be algorithmic, such as CodeBLEU~\citep{renCodeBLEUMethodAutomatic2020}, or learned, such as CodeBERTScore~\citep{zhou_codebertscore_2023}. Learned measures can be more permissive of syntactic variation by operating on learned representations of code~\citep{zhou_codebertscore_2023}. Among surveyed works, algorithmic n-gram measures such as ROUGE and BLEU~\citep{wu_mmqa_2025, kumar_booksql_2024} are applied to compare generated with reference SQL, while COTA~\citep{li_are_2025} applies a learned embedding measure to general analysis code. Program comparison is constrained by its reliance on predefined reference realizations, which conflicts with valid variations in structuring and composing realizations where even semantically equivalent programs can diverge substantially in surface form~\citep{pinnaRedefiningTexttoSQLMetrics2025}. It further requires that the realization is fully expressed as a code artifact, limiting the applicability to systems that emit such programs.

\textbf{Structural matching.} Rather than comparing realizations as textual programs, structural matching operates directly on the abstraction of realizations into operations and their composition, measuring the similarity with a reference realization at that level. In text-to-SQL, structural matching has a long lineage in the form of component matching, which decomposes both the predicted and reference SQL into clauses and compares them, either as an exact set match~\citep{yu_spider_2018, lee_kaggledbqa_2021} or with partial credit over parsed query trees~\citep{kumar_booksql_2024}, though such comparisons are tied to the fixed clause structure of SQL. BLADE~\citep{gu_blade_2024} extends structural matching beyond SQL to general data transformations represented in a graph, with nodes corresponding to discrete data transformations (e.g., filter, groupby, derive,\ldots) and edges expressing data flows between them. It introduces two similarity measures. \textit{Value-based matching} executes the transformations and compares intermediate column values at corresponding nodes, matching sequences of operations that produce equivalent data regardless of how they are expressed. \textit{Graph isomorphism matching} compares the topology and node labels of the graph, accepting structurally equivalent transformations that may produce different values. These similarity measures capture complementary aspects, with value-based matching allowing syntactic diversity given identical results and graph matching tolerating differences in values under a consistent analytical structure. Generally, structural matching provides more flexibility than program comparison, but requires the realizations to be surfaced in a form that supports the abstraction. BLADE~\citep{gu_blade_2024} addresses this by employing an LLM to convert transformation code into transform units, which introduces an additional source of error and reliability concerns.

\textbf{Component validation.} Component validation decomposes the realization into a set of components and validates each individually against a set of per-instance requirements, aggregating the results into a composite score. Aggregation is commonly an average over the requirements, which allows partial credit, though CausalReasoningBenchmark~\citep{sawarni_causalreasoningbenchmark_2026} instead credits a realization only if it satisfies every requirement. Unlike program comparison and structural matching, the realization is not compared to a reference realization, but checks are applied to determine whether specific analytical artifacts are present and correct. Component validation approaches vary in what components they extract from the realization and how they perform checks on top of them. Components may be defined as stages of a predefined analytical pipeline~\citep{chen_scienceagentbench_2025}, or task-specific key functionalities that a correct analysis must contain~\citep{lai_kramabench_2026}. This decomposition can also surface how errors propagate through dependent operations by aggregating individual judgments at different steps of the analysis~\citep{lei_dacomp_2025}. In contrast, \citet{gu_blade_2024} require agents to declare specific analytical decisions (independent, dependent, and control variables) alongside their code, matching these declarations against a reference through LLM-judged semantic equivalence. Statistical models are similarly extracted from code into natural-language specifications and matched semantically. CausalReasoningBenchmark~\citep{sawarni_causalreasoningbenchmark_2026} similarly compares a set of conditions within the realizations with annotations on the desired realization. Through the decomposition of realizations into components, component validation yields fine-grained validation scores that not only indicate the end-to-end performance but also surface detailed information on error cases. Depending on the design of the decomposition and checks, component validation can support different valid realizations~\citep{gu_blade_2024}. Component validation inherits the reliability concerns of LLM-as-judge approaches. It further requires expensive curation of per-instance component sets, which must be validated to ensure completeness and correctness. The cost of curation scales with the complexity of the tasks and the number of components per task, which has so far limited the scale of benchmarks.

\subsubsection{Combining Validation Signals}
\label{sec:eval:validation:combining}

Validation mechanisms are not mutually exclusive. A benchmark may combine signals from different targets or different mechanisms to capture a broader or more robust validation signal than any single function~\citep{pinnaRedefiningTexttoSQLMetrics2025}, or to support different output modalities with appropriate functions~\citep{wu_tablebench_2025, egg_dabstep_2025}. Functions can either be aggregated into a single, mixed-target composite score, or deployed separately to collect distinct, multi-criteria validation signals.

\textit{Mixed-target composites} apply and aggregate the results of different validation scores into a single score, acting as an ensemble of validation functions. Output- and realization-level functions can be combined to capture complementary failure modes, where outputs may be correct despite a divergent realization or the realization may be sound despite differences in outputs~\citep{pinnaRedefiningTexttoSQLMetrics2025}. A composite score simplifies comparisons to a single score at the cost of diagnostic granularity, as a mid-range score does not identify which component failed.

\textit{Multi-criteria reporting} maintains separate scores from distinct validation functions rather than collapsing them into one and is commonly employed among the surveyed benchmarks~\citep{lai_kramabench_2026, ji_topbench_2026, gu_blade_2024}. This enables a fine-grained attribution of performance while complicating cross-system comparisons, and can deliberately isolate distinct failure modes by scoring realization-level and output-level signals apart to separate errors in the analysis from errors in its execution~\citep{sawarni_causalreasoningbenchmark_2026}.

Assessed against the desiderata, scalability and reproducibility are met through automatic evaluation protocols that do not rely on human judgment, although the shift towards \acrshort{llm}-based judging reintroduces variance over directly calculated metrics. Coverage of output modalities is narrow, as many benchmarks simplify evaluation by only assessing raw output values or tables, while most others are restricted to text outputs. Only DAComp~\citep{lei_dacomp_2025} evaluates multi-modal outputs that combine text and charts, leaving outputs that combine different modalities mostly uncovered. Robustness and faithfulness are handled by confining outputs to specific formats and by replacing flawed similarity metrics with \acrshort{llm}-based assessments, the faithfulness and robustness of which varies by the specific setup and evaluation design.
Overall, validation remains concentrated at the output level. Mechanisms that inspect the realization appear in a minority of benchmarks and are concentrated in text-to-SQL component matching, with genuine assessment of open-ended analytical processes confined to a few works such as BLADE~\citep{gu_blade_2024}, KramaBench~\citep{lai_kramabench_2026}, and CausalReasoningBenchmark~\citep{sawarni_causalreasoningbenchmark_2026}. Support for multiple valid variations is similarly shallow, as most mechanisms that admit more than one valid realization do so only through output equivalence, where execution accuracy credits results that coincide with the reference, whereas crediting genuinely divergent yet valid realizations or outputs is restricted to component- and rubric-based approaches~\citep{gu_blade_2024, lei_dacomp_2025, lai_kramabench_2026}. Validation across benchmarks thus predominantly approximates the validity criterion $\mathbf{1}[r \in R_{\text{valid}}(i)]$ while the actual utility $u_i(r)$ remains largely unmeasured. Even where mechanisms produce intermediate scores, these mostly arise from aggregating validity checks over components of the output or realization, reflecting partial validity. The utility of different valid realizations remains largely unaddressed, only partially surfacing in rubric-based judges that score aspects of data selection, methodological rigor, and the effectiveness of the presentation explicitly~\citep{lei_dacomp_2025, lai_kramabench_2026}, leaving two systems producing valid realizations of substantially different utility largely indistinguishable under current benchmarks.

\section{A Research Agenda for OpenTI}
\label{sec:research_agenda}
The preceding sections surface a landscape of related but fragmented work making progress towards \acrshort{task}. Yet, none of the reviewed systems addresses \acrshort{task} end-to-end, and significant gaps remain before such systems can deliver practical value. This section puts forward a research agenda towards closing these gaps and advancing \acrshort{task} towards real-world applications.

Successfully implementing (data) systems and democratizing access to insights requires that systems are both reliable and trustworthy~\citep{shneiderman_humancentered_2020}, and that they provide accessible means for extracting insights~\citep{jagadish_big_2014}. The gaps that separate current work from these ends come from simplifying assumptions in current methods, missing integration between capabilities, evaluation infrastructure that lags behind the systems we wish to build, and a limited perspective on the role of humans in the insight extraction process. We organize the research agenda following these requirements. To holistically address \acrshort{task} end-to-end, systems must reliably retrieve relevant tables from the corpora available to them and address complex analytical needs that reach beyond descriptive insights into inferential and causal ones (Section~\ref{sec:research_agenda:needs}). Further, we argue that \acrshort{task} systems need to be trustworthy so that results can be relied upon and that new user interaction paradigms are required to align the analysis with the actual insight need a user has (Section~\ref{sec:research_agenda:trust}). Lastly, we take a step back to rethink how systems are built and evaluated, arguing that realizing \acrshort{task} requires new paradigms for both (Section~\ref{sec:research_agenda:foundations}).

\subsection{Addressing Data and Methodological Needs}
\label{sec:research_agenda:needs}
The insight needs a system can satisfy are bounded by the data it can locate, corresponding to the data need $i_d$, and by the analysis it can perform, corresponding to the methodological need $i_m$. 

\subsubsection{Retrieving Compatible and Relevant Data at Scale}
\label{sec:research_agenda:retrieval}

Locating sufficient tables from a large corpus is a prerequisite for any downstream operation. As discussed in Section~\ref{sec:orchestration:capabilities:retrieval}, existing retrieval methods have made progress on alignment between user inputs and individual tables, yet several challenges remain unaddressed.

\textbf{What makes data relevant?} Current retrieval methods predominantly operationalize relevance as semantic similarity between textual representations of data needs and tables, while accounting for structural compatibility (Section~\ref{sec:orchestration:capabilities:retrieval:representations}). However, relevance and the fit of tables to the specific task are multidimensional. In particular, when multiple sets of tables satisfy semantic and structural criteria, the realizations they enable may exhibit different utility $u_i$. Data quality literature has established such multidimensional relevance frameworks for tabular data~\citep{wang_accuracy_1996, batini_methodologies_2009}, highlighting aspects like completeness, consistency, and timeliness. Future work should thus define and operationalize quality-aware relevance criteria and integrate them with retrieval.

\textbf{How to design indexes that support the discovery of relevant data?}
Retrieval methods predominantly build indexes as flat collections of independent table, column, or row representations that are scored against the input (Section~\ref{sec:orchestration:capabilities:retrieval:granularity}). Such indexes treat each entry in isolation and remain agnostic to relationships between tables, data quality criteria, and the analytical use the retrieved data will serve. Instead, future work should explore indexes that are aware of the data they are serving, designed for surfacing relevant data rather than just similar data. 

\textbf{How can retrieval scale to heterogeneous real-world corpora?} Real-world data environments such as organizational data lakes exhibit large numbers of tables, heterogeneous schemas, inconsistent naming conventions, varying quality, and potentially overlapping content~\citep{nargesian_data_2019}. Maintaining retrieval performance requires methods that are robust to the messiness and ambiguity in such environments. Scaling also amplifies the challenge of distinguishing more from less relevant data. Progress on this is limited by evaluation, since current benchmarks have corpora that trade scale against structural realism (Section~\ref{sec:eval}), leaving methods largely untested against the heterogeneity and ambiguity of the environments they are meant to operate in.

\subsubsection{Addressing Inferential and Causal Insights}
\label{sec:research_agenda:scope}

Insights span descriptive, inferential, and causal tiers (Section~\ref{sec:problem:spectrum}), yet reviewed systems and evaluations remain mainly concerned with descriptive insights. Inferential and causal insights require deriving knowledge that reaches beyond the observed data, justified against assumptions about the process that produced the data. Current models show a limited capacity to do so across inferential and causal insights. LLMs mainly rely on causal relationships learned from world knowledge~\citep{kiciman_causal_2024} but show limitations in reliably inferring them from evidence~\citep{jin_can_2023, liu_are_2024}. While recent reasoning models narrow this gap, they remain susceptible to bias in instructions~\citep{fu_correlation_2026}. They also show limitations in building predictive models, with reported tendencies to overfit validation signals~\citep{chan_mlebench_2025} and difficulties in identifying methodological errors~\citep{kumar_navigating_2025}. Evaluation is covered unevenly and mostly on a surface level. Within inferential and causal types, prediction is covered most widely, though mostly as a modeling task rather than focused on deriving an insight (Section~\ref{sec:eval}), with TOPBench~\citep{ji_topbench_2026} the first to center on predictive insights and CausalReasoningBenchmark~\citep{sawarni_causalreasoningbenchmark_2026} the first to target interventional analysis, while statistical inference appears mostly as one objective among many~\citep{gu_blade_2024, huang_dacode_2024} and counterfactuals remain limited to simplified rule-based instances~\citep{egg_dabstep_2025, zheng_archer_2024}.

\textbf{How can systems establish and verify the data and analysis assumptions?}
Since the validity of inferential and causal analysis relies on assumptions about the process that generated the data, systems should make these explicit and verify them against the data where possible, and surface the rest to the user to judge. The assumptions differ by tier, as prediction rests on the sample being representative of the targets and free of leakage, statistical inference on the distributional and sampling conditions of the applied test, and causal estimation on identification conditions such as the absence of unobserved confounding. Benchmarks like BLADE~\citep{gu_blade_2024} take a first step in this direction by requiring systems to declare analytical decisions such as control variables, yet none of the surveyed systems establishes or verifies such assumptions. Treating assumptions as first-class components of the analytical process could also extend interpretability from the provenance of results to their analytical validity (Section~\ref{sec:research_agenda:interaction}). Progress on this is also required to enable realization-level validation of inferential and causal results, which should establish whether a predictive model generalizes beyond its training sample rather than reproducing a stored value, and whether an inferential or causal estimate rests on justified assumptions and appropriate methodology.

\textbf{How to select a methodology that matches the insight need and the data?} Inferential and causal needs require choosing among methods whose validity depends on properties of the selected data. Predictive needs require selecting a model and features, while interventional needs demand formal identification strategies such as instrumental variables, regression discontinuity, and difference-in-differences, depending on the structural conditions~\citep{sawarni_causalreasoningbenchmark_2026, wang_causalcopilot_2025}. In \acrshort{task}, however, the appropriate methodology depends on retrieval and the data binding $\beta$ it surfaces, that is, whether the corpus supplies the features a model needs or a variable that serves as a valid instrument. Future work should reason jointly over the methodological need $i_m$ and the data need $i_d$, retrieving data that enables valid methods rather than fitting methods to whatever data was retrieved.

\subsection{Making Systems Trustworthy and Cooperative}
\label{sec:research_agenda:trust}
Beyond addressing the full spectrum of insight needs end-to-end, \acrshort{task} systems should do so in a trustworthy manner and ensure that they address the actual need a user has. We thus argue that they need to produce robust realizations whose results users can rely upon, and that systems should be designed for cooperative interactions with users to align realizations with the users' needs.

\subsubsection{Ensuring Robustness of Results and Realizations}
\label{sec:research_agenda:robustness}
\acrshort{task} systems are intended to provide factual information grounded in data to inform decisions. To make outputs trustworthy for real-world decision-making, systems need to either produce reliable results or transparently communicate their limitations. Automation bias creates the risk of unwarranted reliance on system outputs when the underlying realization is flawed~\citep{parasuraman_complacency_2010}, for instance, building on incomplete data, inappropriate methods, or compounding errors across operations.

\textbf{How to detect and handle unanswerable insight needs?} In practice, not all insight needs can be resolved~\citep{dorschner_more_2026} since corpora may lack sufficient data, inputs may be underspecified beyond what conventional or selective resolution can address, or the request may be logically ill-formed. Work on unanswerability is limited and largely confined to text-to-SQL settings~\citep{li_are_2025, ren_latentrefusal_2026, wang_know_2023}, where unanswerability is typically reduced to schema-level coverage gaps. The open setting introduces additional challenges of insufficient data quality, insufficient system capabilities to solve a task (e.g., a system that can only interact with data through SQL is limited to the operations supported by the specific SQL dialect), and data that supports a result but lacks statistical validity, for instance, when samples are too small or confounders undermine the intended interpretation. Future research should develop taxonomies of unanswerability in \acrshort{task} and equip systems with the capacity to detect, categorize, and communicate such unanswerability accordingly.

\textbf{How can the uncertainty of realizations and outputs be quantified?} The reliability of produced realizations may vary substantially. A realization chains multiple operations, each introducing potential uncertainties and errors, for instance from imprecise retrieval and noisy data to inappropriate analytical choices. Errors may compound across the realization, such that individually reasonable steps may yield unreliable results~\citep{zhao_uncertainty_2025}. None of the surveyed systems provides systematic means to quantify this uncertainty, presenting outputs without confidence assessment or caveats about analytical limitations. 
Future work should propose methods that can quantify and propagate uncertainties, assess whether data and methodology support the conclusions, and surface these assessments to users~\citep{kirchhof_position_2025}.

\textbf{How to detect when realizations fail to serve the insight need?}
A realization can be methodologically sound yet fail to deliver the insight the user actually needs. Users may express their insight need $\hat{i}$ in a way that requests analytical knowledge that does not itself fully satisfy the underlying insight need $i$. For instance, a user interested in understanding the effect of one variable on another may ask for the correlation between them, yet actually require an analysis that establishes the direction and significance of the effect while accounting for confounders to satisfy their underlying insight need. This divergence may be driven by insufficient methodological knowledge, as users articulate high-level analytical goals without identifying the concrete analytical operations that would satisfy them~\citep{grammel_how_2010, amar_lowlevel_2005}, or by cognitive biases that influence how needs are framed, for instance, in users who seek analytical knowledge that confirms a prior belief~\citep{lehner_confirmation_2008, cook_human_2008}, where a system that faithfully answers may introduce or reinforce misinterpretations rather than contributing factual insights. Systems should thus recognize when an accurate response to $\hat{i}$ would not serve the underlying need, communicate this divergence, and propose better-suited analyses, requiring reasoning about the latent need beyond its surface expression.

\subsubsection{Enabling Cooperative Human-System Interaction}
\label{sec:research_agenda:interaction}
Section~\ref{sec:user_facing} establishes insight extraction as a cooperative process in which users deliberately delegate unspecified aspects of their insight need to the system and both parties carry obligations of revealment. Research into analytical systems has traditionally placed the burden of alignment on the user, who is expected to form an accurate mental model of what the system can do and how it interprets inputs. Following the obligation for mutual revealment formulated by \citet{radlinski_theoretical_2017} and the division of labor argued for by \citet{gomm_are_2025}, we instead regard this alignment as a mutual responsibility in which the system maintains and refines a model of the user and the context their insight needs arise in, just as the user builds a model of the system's data and capabilities. Progressing towards such mutual alignment requires work on interaction modes, interpretability mechanisms, and contextual knowledge.

\textbf{What interaction modes best serve users?}
Complex analytical workloads may require iterative exchange, as argued in Section~\ref{sec:user_facing:interaction_modes}. Yet, system-initiated and collaborative interaction modes remain largely conceptual. Future research should investigate which interaction modes are most effective for different combinations of task complexity and user expertise, and develop systems that support richer interaction protocols. This requires extensive user studies and novel approaches to validate the accuracy of systems under these interaction protocols.

\textbf{How to make realizations accessible and interpretable to users?}
Users verify system outputs by assessing both the operations performed and the data involved~\citep{gu_how_2024, xie_waitgpt_2024}. Making the realization $r$ explicit, interpretable, and actionable for intervention is essential to delegate control to users, fostering trust and enabling users to judge the reliability of realizations~\citep{shneiderman_humancentered_2020}. We argue for future work drawing on data provenance and lineage research~\citep{buneman_why_2001, cheney_provenance_2009}, surfacing data flows and operations at appropriate levels of abstraction. This requires balancing comprehensiveness with cognitive effort~\citep{kazemitabaar_improving_2024} and developing representations that serve both verification and direct intervention.

\textbf{How to adapt insight extraction to its context?} Tabular data mostly exists in organizational settings where analytical tasks are deeply contextualized, with domain-specific vocabulary, conventions, implicit constraints, policies, and regulations~\citep{cao_evolution_2007, egg_dabstep_2025} shaping how insight needs should be interpreted. Operating within this context reduces the degree to which insight needs must be explicitly specified, as shared understanding resolves ambiguities that would otherwise require specification. Instead of burdening users with accounting for the system's missing contextual knowledge through redundant specification, which users embedded in a domain may not even recognize as necessary since much of this context is tacit~\citep{polanyi_tacit_1967}, we argue that systems should acquire and maintain accurate models of organizational and domain context, facilitating efficient interactions by reducing the requirements on input specification. Beyond the organizational level, systems should also adapt to the specific user since different users bring distinct expertise, expectations, and analytical perspectives~\citep{zhang_how_2020}, implying that similar insight needs may require different realizations depending on who expresses them. Similarly, users may regard their interaction histories as shared context, under which they express their insight needs~\citep{yin_natural_2023}. Such contextual knowledge could be informed by organizational documentation, domain ontologies, interaction histories, or user profiles.

\textbf{How to determine appropriate output presentations?}
The output need $i_o$ is frequently left unspecified by users, yet the choice of output modality, like textual summaries, tables, charts, or a combination thereof, may substantially affect the utility of a realization. Surveyed systems either rely on explicit output specification in the input~\citep{hu_infiagentdabench_2024, huang_dacode_2024} or default to a fixed modality. Future work should explore inferring appropriate output presentations from the analytical context, the nature of the results, and user expectations, making the output need a first-class component of the realization process.

\subsection{Rethinking How Systems Are Built and Evaluated}
\label{sec:research_agenda:foundations}
The preceding directions ask what systems should be able to do. We now shift the focus to how the paradigms that current systems inherit from the narrow task framings they are built and benchmarked for need to adapt to support this, spanning how analytical work is orchestrated, what models carry it out, and how systems are evaluated.

\subsubsection{Dynamically Orchestrating Analytical Workloads}
\label{sec:research_agenda:orchestration}
For simple insight needs, the analytical process can often be determined from the input alone. For instance, a lookup usually only requires locating the right table and extracting a value. However, as the complexity of insight needs increases, the realization that satisfies them becomes increasingly harder to specify upfront. Intermediate results may reveal that an initially chosen analytical method is unsuitable for the data, that required variables are distributed across tables not yet retrieved, or that the initial plan has a flaw. User inputs throughout processing also require mechanisms to adjust the analytical process. While reactive systems demonstrate initial capabilities in this direction~\citep{zhang_reactable_2024, wang_chainoftable_2024}, their adaptivity remains largely confined to selecting among predefined operations, with data bindings $\beta$ fixed upfront and limited capacity for backtracking and revising the analytical strategy. Advancing \acrshort{task} towards complex workloads requires orchestration that can dynamically adapt all components of a realization $r=(O, E, \beta)$ in response to intermediate results and interactions.

\textbf{How to acquire data during the analytical process?} Most existing systems operating in an open setting separate retrieving data from the downstream usage of that data in a two-stage design~\citep{kong_opentab_2023, chen_open_2020}, requiring all relevant tables to be identified before any data operations take place. Yet complex insight needs may reveal data requirements only as the analysis unfolds, for instance, when intermediate results require linking tables or additional reference data not initially anticipated. While research into dynamic retrieval in retrieval augmented generation~\citep{jiang_active_2023, asai_selfrag_2024} highlights the potential of retrieving data when it becomes relevant, initial steps by \citet{wang_aop_2025} in applying this paradigm to \acrshort{task} fall short of systematically representing and utilizing intermediate analytical context for retrieval, or of addressing data compatibility with intermediate results.

\textbf{How can user input be integrated during processing?} 
As discussed in Section~\ref{sec:user_facing:interaction_modes}, interaction modes beyond single-input remain underexplored. From an orchestration perspective, this requires systems to recognize the risk of producing misaligned realizations when proceeding with an interpretation that is uncertain due to ambiguity or data challenges, and to collect user input instead~\citep{gomm_are_2025, saparina_reasoning_2025}. This demands mechanisms to estimate confidence in interpretations and to incorporate intermediate user inputs into the ongoing workflow.

\textbf{How should orchestration adapt to the task?}
The different types of insights (Section~\ref{sec:problem:spectrum}) have fundamentally different analytical requirements. A lookup may be best served by direct retrieval and a single SQL query, while a prediction may require integrating data from different tables, pre-processing it, building, validating, and applying a model. Current systems apply uniform orchestration strategies regardless of task characteristics. Future work should explore how to route insight needs to appropriate orchestration strategies based on the complexity and requirements of the analytical workload, potentially drawing on task classification or complexity estimation as an initial step.

\subsubsection{Superseding and Complementing Language Models}
\label{sec:research_agenda:models}

Surveyed systems rely almost exclusively on language models for orchestration, reasoning, and code generation. Yet language models are architecturally mismatched with core aspects of insight extraction. Tables are two-dimensional structures that exhibit column-order invariance~\citep{cong_observatory_2023}, contain data of various types, and stand in complex relationships to each other. Linearization into text does not capture these attributes, leading to limited performance in understanding tabular data, particularly as table size increases~\citep{wolff_how_2025}. Beyond tables, the unfolding analytical process, comprising partial results, transformed tables, and the evolving directed graph of operations, is inherently non-sequential, yet must be communicated to language models through linear context windows.

\textbf{Which architectures fit tabular data and the analytical process?} Initial efforts towards table-native models through modified training and architectures~\citep{herzig_tapas_2020, li_tablegpt_2024, su_tablegpt2_2024} have shown limited improvements over general-purpose language models on downstream tasks. Large improvements have been made in recent in-context learning for regression and classification on tabular data, so-called Tabular Foundation Models (TFMs)~\citep{hollmann_tabpfn_2023, qu_tabicl_2025, hollmann_accurate_2025}. Yet, these advancements have only transferred to insight extraction insofar as TFMs have been proposed as tools for language model agents~\citep{cong_unlocking_2025}. Future research should thus explore architectures and methods that better fit the structure of the analytical process and the data it is applied to.

\textbf{To what degree can single models cover the analytical process?} The functional capabilities examined in Section~\ref{sec:orchestration:capabilities} span locating, understanding, and transforming tabular data as well as orchestrating, reasoning within, and compiling outputs from the analytical process. It remains an open question to what degree these skills can be effectively covered by single models, and where specialization or composition of dedicated components yields more capable systems.

\subsubsection{Designing Realistic and Faithful Evaluations}
\label{sec:research_agenda:evaluations}
Evaluation methodology must faithfully assess the systems we aim to build. As shown in the analysis in Section~\ref{sec:eval}, current benchmarks exhibit corpora that trade off scale against structural realism, input specifications suffer from ambiguities and data privileges, tasks remain concentrated on descriptive insights with limited coverage of inferential and causal types, and validation mechanisms predominantly rely on output-level exact matching. These limitations mean that benchmark performance may not reflect system capabilities under realistic conditions. Addressing these gaps requires not only more representative datasets but also methodological advances in how system behavior is assessed.

\textbf{How to design robust realization-level verification?} Benchmarking systems on their outputs alone risks conflating correct realizations with coincidentally correct answers, which has been shown to happen when systems bypass data evidence and instead guess a response~\citep{liu_are_2024}. Moreover, for open-ended analytical tasks where multiple valid realizations exist~\citep{gomm_are_2025, saparina_disambiguate_2025}, output-level validation against singular gold answers penalizes legitimate alternative interpretations. Realization-level verification, which explicitly examines operations $O$, their composition $E$, and data bindings $\beta$, offers more fine-grained diagnostic signal and enables partial credit for correct substeps~\citep{chen_scienceagentbench_2025}. However, this requires systems to expose realizations in standardized forms and demands validation methodology capable of assessing analytical soundness.

\textbf{How to evaluate interactive systems and multimodal outputs?} Current evaluations mostly assume a standalone input scored against a fixed reference output (Section~\ref{sec:eval}), an assumption that breaks on both sides as \acrshort{task} advances. On the input side, the cooperative interaction modes argued for in Section~\ref{sec:research_agenda:interaction} replace the standalone utterance with a composition of inputs provided at different times, where the system's state at each step shapes the subsequent user input. This interdependence between system and input complicates both the design of evaluation instances and the attribution of performance. Initial benchmarks emulate such interaction through \acrshort{llm}-based user simulation~\citep{duttaConDABenchInteractiveEvaluation2025, huo_birdinteract_2025, li_are_2025}, yet user models remain prone to goal misalignment and unrealistic response patterns~\citep{seshadriLostSimulationLLMSimulated2026, mehri_goal_2026}, calling for methods that faithfully emulate user behavior. On the output side, realizations increasingly produce results that combine text, tables, and visualizations, demanding validation mechanisms that assess these modalities jointly rather than in isolation~\citep{duttaConDABenchInteractiveEvaluation2025}.

\section{Conclusion}
\label{sec:conclusion}
Most of the knowledge that makes tabular data valuable does not reside in its cells but has to be discovered and derived, imposing barriers to locating and analyzing the relevant data. In this work, we posit \acrfull{task} as the task of making this knowledge accessible to those who need it, treating the problem end-to-end, from the expressed need to presenting the extracted knowledge. We develop \acrshort{task} as an emerging field centered on insights rather than the surface-level answers that narrower framings such as table question answering, text-to-SQL, and data analysis largely optimize for. By formulating the problem around the insight needs that users hold, the realizations that describe how analytical knowledge is extracted from tabular data, and a notion of their utility that reaches beyond literal accuracy, we establish a basis on which we conceptualize systems that address \acrshort{task} as agents that navigate a combinatorially vast realization space under partial observability. On this basis, we recast the human side of the task under a framing of cooperative interaction and formalize the evaluation of such systems. Interleaving this conceptualization with a structured survey lets us take stock of where the field stands. We find that many of the pieces \acrshort{task} requires already exist across communities but remain fragmented and, where present, are mostly assembled around narrow answer-centered targets. Capabilities such as data retrieval, output contextualization, cooperative interaction, and inferential and causal analysis remain thin, and evaluation methodology lags behind the systems we ultimately wish to build.

We intend the conceptualization and vocabulary developed here to give the communities working towards these ends a common ground from which to advance, making it possible to understand scattered efforts as facets of one problem and to transfer results across the conventions that have kept them apart. Realizing \acrshort{task} will require assembling these facets into systems that operate end-to-end, yet we believe it is paramount that such systems are designed and evaluated around the insights they produce rather than accuracy alone. A result that is correct on its face can still mislead when it is interpreted outside the context that gives it meaning, as when an output accurately answers the question a user posed while missing the context for the insight they actually seek. The goal we consider most important to keep in view is therefore that \acrshort{task} systems leave people better informed, producing results that are robust and, just as crucially, contextualized so that users arrive at factual, data-driven conclusions rather than confidently mistaken ones. Realizing this makes the cooperation between user and system a first-class concern in system design rather than a peripheral one. Pursued this way, \acrshort{task} aims to make the knowledge held in tables as accessible as the knowledge we already retrieve from documents, democratizing access to data-driven insights in a trustworthy manner.

\section{Acknowledgements}
This work was supported by the Dutch Research Council (NWO, AiNed Fellowship grant NGF.1607.22.045) and a grant from SAP. We also gratefully acknowledge support through compute credit grants from Google and OpenAI.

\bibliography{references_compact}
\bibliographystyle{tmlr}

\newpage

\appendix
\addtocontents{toc}{\protect\setcounter{tocdepth}{1}}

\section{Overview of Notation}
\label{s:appendix:notation}

\begin{longtable}{@{} l p{13cm} @{}}
\small
\label{tab:notation} \\
\toprule
\textbf{Symbol} & \textbf{Description} \\
\midrule
\endfirsthead

\multicolumn{2}{l}{\textit{Table~\ref{tab:notation} continued}} \\
\toprule
\textbf{Symbol} & \textbf{Description} \\
\midrule
\endhead

\midrule
\multicolumn{2}{r}{\textit{continued on next page}} \\
\endfoot

\bottomrule
\endlastfoot

\multicolumn{2}{@{}l}{\textbf{Insights and Insight Needs (Section~\ref{sec:problem:conceptual_foundations})}} \\
$i$ & An insight need, a person's requirement for analytical knowledge which, when interpreted within their contextual understanding, constitutes an insight. (Definition~\ref{def:insight_need})\\
$i_d$ & The data need, specifying what data entities, temporal scope, and domain boundaries are relevant to the insight need. \\
$i_m$ & The methodological need, specifying what analytical operations and procedures should be applied. \\
$i_o$ & The output need, specifying how results should be presented. \\
$\hat{i}$ & The expression of an insight need $i$ communicated by a user, often only a partial approximation of $i$. \\
$\hat{i}^{s}$ & A system's representation of the insight need, constructed from $\hat{i}$ to approximate $i$. \\
\addlinespace
\multicolumn{2}{@{}l}{\textbf{Tabular Data and Corpus (Section~\ref{sec:problem:conceptual_foundations:open_tix})}} \\
$t=(H,D,M)$ & A table, consisting of a header $H$, a data matrix $D$, and a metadata set $M$. \\
$T=\{t_1,\ldots,t_n\}$ & A corpus of $n$ tables over which insight needs are expressed. \\
$\mathcal{T}$ & The space of possible tables, where a corpus $T \subset \mathcal{T}$. \\
\addlinespace
\multicolumn{2}{@{}l}{\textbf{Realizations (Section~\ref{sec:problem:conceptual_foundations:realizations})}} \\
$r=(O,E,\beta)$ & A realization, a directed acyclic graph of fully parameterized operations $O$ connected by dataflow edges $E$ with a data binding $\beta$ (Definition~\ref{def:realization}). \\
$\mathcal{R}$ & The space of all possible realizations, with $r \in \mathcal{R}$. \\
$O=\{o_1,\ldots,o_m\}$ & The set of fully parameterized operations applied in a realization's derivation. \\
$o$ & A single operation, a discrete functional unit ranging from granular primitives to high-level analytical functions. \\
$E \subset O \times O$ & The set of directed edges connecting operations in a realization, defining the flow of data between operations. \\
$\beta \subset \mathcal{T} \times O$ & The data binding connecting specific tables from $T$ to source operations in a realization. \\
$\out(r)$ & The analytical knowledge produced as output by executing realization $r$. \\
\addlinespace
\multicolumn{2}{@{}l}{\textbf{Utility and Problem Formulation (Section~\ref{sec:problem:conceptual_foundations:problem_formulation})}} \\
$u_i$ & The realization utility function for insight need $i$, $u_i: \mathcal{R} \to \mathbb{R}$, scoring how well a realization satisfies $i$ (Definition~\ref{def:utility}). \\
$R_{\text{valid}}(i)$ & The set of realizations valid for insight need $i$, i.e., realizations whose data, operations, and output address $i$. \\
$R_{\text{feasible}}(T)$ & The set of realizations feasible given corpus $T$, i.e., where bound tables are in $T$. \\
$R_{\text{cand}}(i,T)$ & The set of candidate realizations, $R_{\text{valid}}(i) \cap R_{\text{feasible}}(T)$, that both address the insight need and are grounded in available data. \\
$r^*$ & The candidate realization that maximizes utility (Equation~\ref{eq:objective}); reused in Section~\ref{sec:eval:validation:realization_level} to denote a benchmark's reference realization. \\
\addlinespace
\multicolumn{2}{@{}l}{\textbf{Evaluation (Section~\ref{sec:eval})}} \\
$B=(T,\mathcal{E})$ & An evaluation benchmark, comprising a corpus of tables $T$ and a set of evaluation instances $\mathcal{E}$. \\
$\mathcal{E}=\{e_1,\ldots,e_n\}$ & The set of $n$ evaluation instances in a benchmark. \\
$e=(\bar{i},v)$ & An evaluation instance, comprising an input $\bar{i}$ and a validation mechanism $v$. \\
$\bar{i}$ & The benchmark input, representing the expression of an insight need $\hat{i}$ under the benchmark's input protocol. \\
$v$ & The validation mechanism scoring a realization, $v: \mathcal{R} \to [0,1]$, acting as an operational stand-in for $R_{\text{valid}}(i)$ and $u_i$. \\

\end{longtable}

\section{Comparison with Related Work}
\label{s:appendix:related_work}

Several existing works survey and systematize areas related to \acrshort{task}, yet each addresses only a part of its problem space from within the conventions of particular research communities. A larger group of works exists that systematize systems that perform complex tasks over unseen information without targeting tabular data, spanning open-domain question answering~\citep{zhang_survey_2023}, retrieval augmented generation~\citep{zhao_retrieval_2024}, and \acrshort{llm}-agents~\citep{kapoor_ai_2024, sumers_cognitive_2024}. More closely related, other works target tabular data but organize contributions under the labels of the communities they originate from, such as text-to-SQL~\citep{liu_survey_2025}, table question answering~\citep{zhou_table_2026}, the application of \acrshort{llm}s to tabular tasks~\citep{fang_large_2024, lu_large_2025}, and agents for data science and data analysis~\citep{chen_large_2025, tang_llm_2025, tian_realworld_2026}. Table~\ref{tab:survey_comparison} compares these works against the dimensions foregrounded in our framing of \acrshort{task}, namely the coverage of the end-to-end scope, the treatment of an open setting in which the corpus is unknown when the need is expressed, an insight-centered rather than answer-centered framing, a formal definition of the problem, and the treatment of user-facing concerns.

\begin{table}[ht]
    \centering
    \caption{Comparison of this paper with other works that survey related areas. $-$ indicates aspects that are not addressed, $\circ$ ones that are partially addressed or acknowledged for future work, and $\bullet$ signifies that the work foregrounds the aspect as a central part of its contribution. For the end-to-end scope coverage, we indicate the explicit treatment of parts of the end-to-end scope using the abbreviations UI for \textit{User Interaction}, R for \textit{Table Retrieval}, DI for \textit{Data Integration}, DA for \textit{Data Analysis}, and O for \textit{Output Presentation}.}
    \small
    \resizebox{\linewidth}{!}{%
        \begin{tabular}{l p{2.6cm} p{2.4cm} p{3.7cm} p{1.3cm} p{1.5cm} p{1.7cm} p{1.9cm}}
\toprule
\textbf{Work} & \textbf{Focus} & \textbf{Research Communities} & \textbf{End-to-End Scope Coverage} & \textbf{Open Setting} & \textbf{Insight-\newline Centered} & \textbf{Formal Definition} & \textbf{User-facing Concerns} \\
\midrule
\citet{liu_survey_2025} & Text-to-SQL\newline & DB, NLP & \pipeline{1}{0}{0}{1}{0} & $-$ & $-$ & $\circ$ & $-$ \\ \addlinespace
\citet{fang_large_2024} & LLMs for tabular tasks & NLP, DB & \pipeline{0}{0}{0}{1}{1} & $-$ & $-$ & $\circ$ & $-$ \\ \addlinespace
\citet{lu_large_2025} & LLMs for tabular tasks & NLP, DB & \pipeline{0}{0}{0}{1}{1} & $-$ & $\circ$ & $\circ$ & $\circ$ \\ \addlinespace
\citet{zhou_table_2026} & Table Question Answering & NLP, IR & \pipeline{0}{1}{0}{1}{0} & $\circ$ & $\circ$ & $-$ & $\circ$ \\ \addlinespace
\citet{tian_realworld_2026} & LLM-Agents for tabular tasks & NLP, DB & \pipeline{0}{1}{1}{1}{0} & $\circ$ & $\circ$ & $\circ$ & $\circ$ \\ \addlinespace
\citet{chen_large_2025} & LLM-Agents for tabular tasks & NLP, ML/AI & \pipeline{0}{1}{1}{1}{1} & $-$ & $-$ & $-$ & $\circ$ \\ \addlinespace
\citet{tang_llm_2025} & LLM-Agents for data science & DB, NLP, ML/AI & \pipeline{0}{1}{1}{1}{1} & $\circ$ & $-$ & $-$ & $\circ$  \\ \addlinespace
\midrule
This paper & Open Tabular\newline Insight Extraction & DB, HCI, IR, ML/AI, NLP  & \pipeline{1}{1}{1}{1}{1} & $\bullet$ & $\bullet$ & $\bullet$ & $\bullet$ \\ \addlinespace
\bottomrule
\end{tabular}%
    }
    \label{tab:survey_comparison}
\end{table}

Compared to this paper, these works cover only part of the path from an expressed need to a presented insight, mostly the analysis, while assuming that the relevant tables are supplied rather than discovered from an unknown corpus. The setting they address is thus a closed one, and where openness appears it does so as one task dimension among many rather than the premise of the problem~\citep{zhou_table_2026, tang_llm_2025}. None of these works takes an insight-centered perspective, treating the tabular tasks as targeting surface-level answer correctness, not as the search for the insights that serve the user. Formalization, where offered, stays confined to isolated subtasks such as translating a question into SQL~\citep{liu_survey_2025}, and the user-facing side of the problem, from interaction to personalization and interpretability, is raised as an open challenge more than it is developed~\citep{lu_large_2025, tian_realworld_2026}. Instead, our framing of \acrshort{task} treats these dimensions as one problem, spanning the full path, taking the open setting and the insight need as its starting point, grounding both in a formal definition, and placing user-facing concerns as a central aspect.

\section{Literature Review Protocol}
\label{s:appendix:review_method}

We collect the works that provide the evidence base for this paper following a structured, multi-stage protocol organized into identification, screening, and selection stages, inspired by the PRISMA guidelines for reporting systematic reviews~\citep{page_prisma_2021}. We start with a broad keyword search over a curated set of venues (Section~\ref{s:appendix:review_method:identification}) and proceed through manual screening against inclusion criteria (Section~\ref{s:appendix:review_method:screening}), yielding a set of relevant works that informs this paper throughout. From this set we extract the systems and benchmarks that we analyze in Sections~\ref{sec:orchestration} and~\ref{sec:eval} (Section~\ref{s:appendix:review_method:overview}). The sets obtained through this protocol further provide the basis for the analysis of benchmark inputs in Appendix~\ref{s:appendix:input_classification} and for the analysis of citation behavior in Appendix~\ref{s:appendix:citation_analysis}. Figure~\ref{fig:survey_flow} summarizes the resulting funnel.

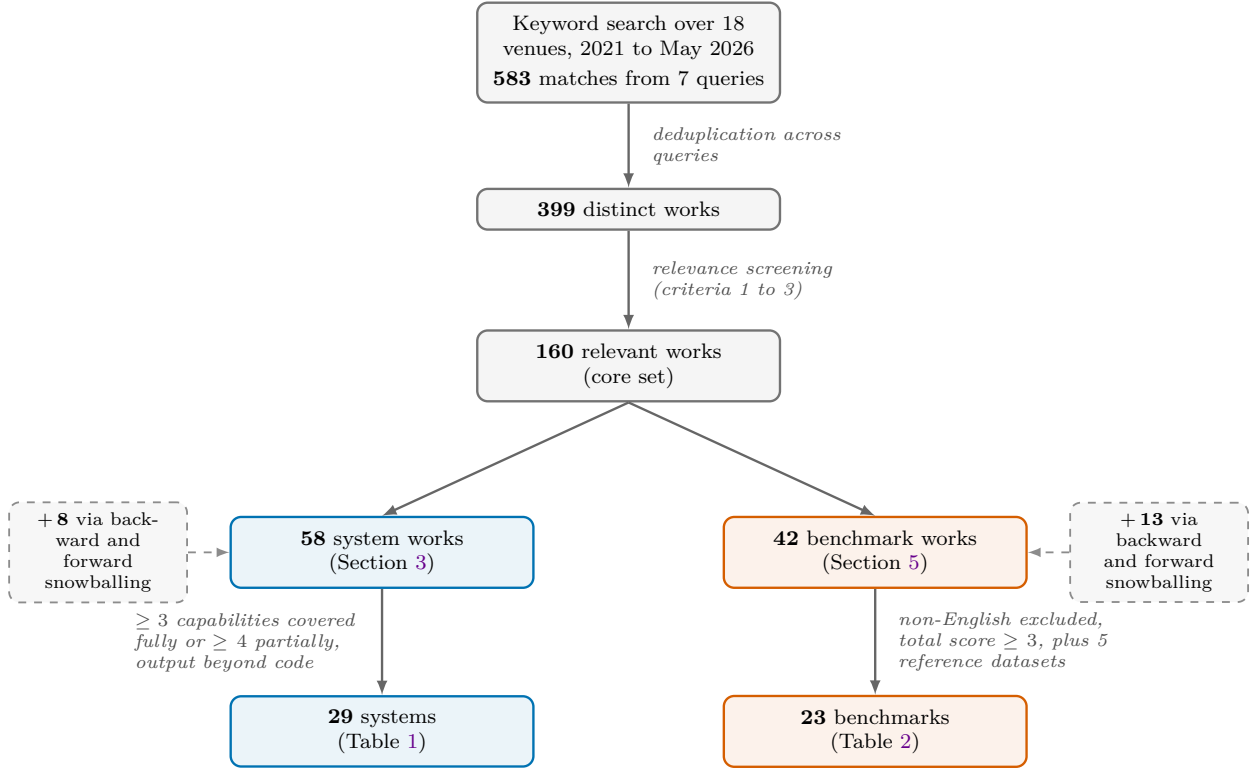
\begin{figure}[ht]
    \centering
    \resizebox{\linewidth}{!}{%
        \begin{tikzpicture}[
    font=\sffamily,
    >=latex,
    stage/.style={draw, rounded corners=4pt, line width=0.9pt, align=center, inner sep=5pt, text width=3.7cm, font=\footnotesize},
    shared/.style={stage, draw=black!55, fill=black!4},
    methods/.style={stage, draw=okabe_5, fill=okabe_5!8},
    benches/.style={stage, draw=okabe_6, fill=okabe_6!8},
    snow/.style={draw, rounded corners=4pt, line width=0.7pt, dashed, align=center, inner sep=4pt, text width=2.1cm, font=\scriptsize, draw=black!45, fill=black!3},
    note/.style={font=\scriptsize\itshape, text=black!70, align=left, text width=3.0cm},
    flow/.style={->, line width=0.9pt, draw=black!60},
    add/.style={->, line width=0.7pt, draw=black!45, dashed},
]

\node[shared] (n1) at (0,0) {Keyword search over 18 venues, 2021 to May 2026\\[2pt]\textbf{583} matches from 7 queries};
\node[shared] (n2) at (0,-2.1) {\textbf{399} distinct works};
\node[shared] (n3) at (0,-4.2) {\textbf{160} relevant works\\(core set)};

\node[methods] (m1) at (-3.3,-6.7) {\textbf{58} system works\\(Section~\ref{sec:orchestration})};
\node[benches] (b1) at (3.3,-6.7) {\textbf{42} benchmark works\\(Section~\ref{sec:eval})};
\node[methods] (m2) at (-3.3,-9.1) {\textbf{29} systems\\(Table~\ref{tab:functional_capabilities_overview})};
\node[benches] (b2) at (3.3,-9.1) {\textbf{23} benchmarks\\(Table~\ref{tab:datasets})};

\node[snow] (sm) at (-7.1,-6.7) {$+\,\mathbf{8}$ via backward and forward snowballing};
\node[snow] (sb) at (7.1,-6.7) {$+\,\mathbf{13}$ via backward and forward snowballing};

\draw[flow] (n1) -- node[note, anchor=west, xshift=5pt] {deduplication across queries} (n2);
\draw[flow] (n2) -- node[note, anchor=west, xshift=5pt] {relevance screening (criteria 1 to 3)} (n3);

\draw[flow] (n3.south) -- (m1.north);
\draw[flow] (n3.south) -- (b1.north);

\draw[add] (sm.east) -- (m1.west);
\draw[add] (sb.west) -- (b1.east);

\draw[flow] (m1) -- node[note, anchor=east, xshift=-5pt] {$\geq 3$ capabilities covered fully or $\geq 4$ partially, output beyond code} (m2);
\draw[flow] (b1) -- node[note, anchor=west, xshift=5pt] {non-English excluded, total score $\geq 3$, plus 5 reference datasets} (b2);

\end{tikzpicture}%
    }
    \caption{Overview of the literature review funnel. A keyword search over 18 venues yields 399 distinct works after deduplication, which relevance screening reduces to a core set of 160 works. From this set, extended through backward and forward snowballing, we extract the works surveyed in Sections~\ref{sec:orchestration} and~\ref{sec:eval} and apply the overview-table criteria that select the 29 systems in Table~\ref{tab:functional_capabilities_overview} and the 23 benchmarks in Table~\ref{tab:datasets}.}
    \label{fig:survey_flow}
\end{figure}

\subsection{Identification}
\label{s:appendix:review_method:identification}

We consider works published between 2021 and May 2026 at the venues most relevant to \acrshort{task} across machine learning, information retrieval, human-computer interaction, database research, and natural language processing. We begin in 2021, the point from which work relevant to \acrshort{task} accelerates markedly (Figure~\ref{fig:stats:papers_by_year}). The venue set is not exhaustive but deliberately targets the leading venues of the communities we identify in Section~\ref{sec:introduction} as contributing towards \acrshort{task}, selecting the top venues of each to capture the most influential and representative work. To these we add the TRL workshops, which sit precisely at the intersection of tabular data and machine learning that \acrshort{task} occupies and thereby concentrate directly relevant work that the general-purpose venues disperse. The search spans the following 18 venues.

\begin{itemize}
    \item \textbf{Machine Learning.} NeurIPS, ICML, ICLR, JMLR, TMLR
    \item \textbf{Information Retrieval.} SIGIR, ECIR
    \item \textbf{Human-Computer Interaction.} CHI, UIST
    \item \textbf{Database Research.} VLDB, SIGMOD, EDBT
    \item \textbf{Natural Language Processing.} ACL, EMNLP, NAACL, EACL
    \item \textbf{Tabular.} TRL Workshop@\{NeurIPS/ACL\}, AI for Tabular Data Workshop@EurIPS
\end{itemize}

We extract the titles and abstracts of all papers published at these venues and organize them in a database. We then run seven keyword searches with diverse queries, mirroring the fragmentation of vocabulary across these communities. The queries target different labels under which communities frame facets of \acrshort{task}, spanning data analysis agents, table understanding and reasoning, table retrieval, table question answering, and text-to-SQL. Table~\ref{tab:survey_search_terms} reports the queries together with the number of matches each returns and the number of works from it judged relevant during the screening described below. Deduplicating the matches across queries yields a pool of 399 distinct works that enters screening.

\begin{table}[ht]
    \centering
    \caption{Keyword queries run over the titles and abstracts, with the number of matches returned and, of these, the number of works judged relevant to the survey during screening (Section~\ref{s:appendix:review_method:screening}).}
    \small
    \begin{tabular}{p{0.66\linewidth} r r}
\toprule
\textbf{Query} & \textbf{Matches} & \textbf{Retained} \\
\midrule
\texttt{"data analysis" AND ("agent" OR "automated" OR "LLM")} & 91 & 25 \\ \addlinespace
\texttt{"data science" AND "table" AND ("agent" OR "automated" OR "LLM")} & 13 & 6 \\ \addlinespace
\texttt{"open-domain" AND "table"} & 81 & 15 \\ \addlinespace
\texttt{"table understanding" OR "table reasoning"} & 110 & 36 \\ \addlinespace
\texttt{"table retrieval"} & 20 & 11 \\ \addlinespace
\texttt{"table question answering" OR "TableQA" OR "(TQA)"} & 62 & 39 \\ \addlinespace
\texttt{"text-to-sql"} & 206 & 54 \\
\bottomrule
\end{tabular}

    \label{tab:survey_search_terms}
\end{table}

\subsection{Screening and Selection}
\label{s:appendix:review_method:screening}

We screen the pool of 399 works by assessing each title and abstract for relevance to the subject of the survey. We retain a work if it informs the treatment of \acrshort{task}, which we operationalize through three criteria. A work qualifies if it (1) describes systems, components of systems, or assessments of systems that fall within the end-to-end scope of \acrshort{task}, (2) proposes a benchmark or evaluation that targets directed insight needs over tabular data such as table question answering, text-to-SQL, or data analysis, or (3) concerns user-system interaction relevant to \acrshort{task}. We further retain works that contribute neither a system nor a benchmark but that inform the conceptual treatment of the subject. Deduplicating the relevant works across queries yields a core set of 160 works, which provide the backbone of the literature review and inform its writing throughout.

From this core set we isolate the works that receive detailed treatment in the separate surveys of systems and benchmarks, extracting the works that propose systems spanning the functional capabilities of \acrshort{task} agents and the works that propose benchmarks targeting directed insight needs over tabular data. We extend each of these two sets separately through targeted backward and forward snowballing, adding works with substantial contributions that fall outside the venues, date range, or queries defined above. This adds 8 works to the systems and 13 to the benchmarks, yielding 58 works covering methods and 42 works covering benchmarks that form the basis for the surveys in Sections~\ref{sec:orchestration} and~\ref{sec:eval} respectively.

\subsection{Selection for the Overview Tables}
\label{s:appendix:review_method:overview}

From these works we select a subset for detailed presentation in the overview tables of Section~\ref{sec:orchestration} (Table~\ref{tab:functional_capabilities_overview}) and Section~\ref{sec:eval} (Table~\ref{tab:datasets}), applying criteria aimed at presenting the most relevant works concerning the scope of \acrshort{task}.

For methods, we analyze each system along the five functional capabilities established in Section~\ref{sec:orchestration} and score its coverage of each as fully, partially, or not covered. We include a system if it covers at least three capabilities fully or four at least partially, selecting systems covering a significant extent of the scope of \acrshort{task}. We further require that a system produces some output beyond raw results or code, so that it exercises at least partial freedom in contextualizing and presenting the analytical knowledge it derives. This yields the 29 systems presented in Table~\ref{tab:functional_capabilities_overview}.

For benchmarks, we first exclude non-English datasets, which we cannot fairly assess against English-language ones. We then score the remaining datasets on the degree to which they cover the data need $i_d$, the methodological need $i_m$, and the output need $i_o$, assigning each component a score from 0 to 2 following the rubric in Table~\ref{tab:benchmark_scoring}. We include datasets with a total score of at least 3, ensuring that the presented datasets cover substantial parts of the scope of insight needs, which yields 18 datasets. We supplement these with five datasets that fall below the threshold but are widely used and highly cited, whose inclusion readers would likely expect as points of reference, namely Spider~\citep{yu_spider_2018}, BIRD~\citep{li_can_2024}, OTT-QA~\citep{chen_open_2020}, NQ-Tables~\citep{herzig_open_2021}, and FeTaQA~\citep{nan_fetaqa_2022}. This results in the 23 benchmarks presented in Table~\ref{tab:datasets}.

\begin{table}[ht]
    \centering
    \caption{Rubrics for scoring the degree to which benchmarks cover data need $i_d$, methodological need $i_m$, and output need $i_o$. Insight types are counted following the taxonomy in Section~\ref{sec:problem:spectrum}. A dataset is presented in Table~\ref{tab:datasets} if its scores sum to at least 3.}
    \small
    \begin{tabular}{l l p{3.4cm} p{3.4cm} p{3cm}}
\toprule
\textbf{Need} & \textbf{Scored by} & \textbf{0} & \textbf{1} & \textbf{2} \\
\midrule
$i_d$ & Data setting & Only the relevant tables provided per instance & A set of tables provided per instance & Open setting \\ \addlinespace
$i_m$ & Insight types covered & Shallow descriptive only (lookup, aggregation) & At least three insight types covered & Any inferential or causal type covered \\ \addlinespace
$i_o$ & Output modalities & Single bare modality (value or table) & Two modalities or text synthesis & Outputs include charts \\
\bottomrule
\end{tabular}

    \label{tab:benchmark_scoring}
\end{table}

\subsection{Limitations}
\label{s:appendix:review_method:limitations}

The results and discussion presented in this paper have to be read as claims about the surveyed works, which are limited in scope by the review methodology. The identification stage covers 18 peer-reviewed venues, so works published solely on preprint servers or at venues outside this set enter the review only through snowballing, which may underrepresent the most recent developments in this fast-moving space. Additionally, commercial and organizational systems are excluded, since their internals are not documented in a form that can be assessed against our framework, although several deployed products plausibly attempt parts of \acrshort{task}. Our protocol departs from full PRISMA compliance. Notably, neither the search nor the inclusion criteria were registered ahead of time, so the criteria in Sections~\ref{s:appendix:review_method:screening} and~\ref{s:appendix:review_method:overview} were refined alongside our developing understanding of \acrshort{task} rather than fixed in advance, reflecting that \acrshort{task} is currently emerging from research.

\section{Input Classification Methodology}
\label{s:appendix:input_classification}

The analysis of data privilege and insufficient specification reported in Section~\ref{sec:eval:input:analysis} (Figure~\ref{fig:datasets:ambiguity_privilege}) adopts the LLM-judge setup of \citet{gomm_are_2025} unchanged, applying their classifiers to the benchmarks in Table~\ref{tab:datasets}. We classify the 18 benchmarks whose inputs are self-contained natural language utterances for single-input interaction, excluding five benchmarks for which this does not hold. TACO~\citep{dengTACOBenchmarkOpenDomain} and DA-Dataset~\citep{xu_dagent_2025} are excluded because their inputs are not available to us, ARCADE~\citep{yin_natural_2023} and ConDABench~\citep{duttaConDABenchInteractiveEvaluation2025} because their step-wise inputs are incompatible with the LLM-judge setup, and similarly BLADE~\citep{gu_blade_2024} because its inputs follow a multiple-choice format over code rather than a natural language expression of an insight need. For each remaining benchmark we randomly sample 500 input definitions, or all inputs for benchmarks with fewer instances, and classify each sampled input individually. The full prompts and implementation are available in the repository accompanying \citet{gomm_are_2025}.

Data privilege is assessed along three dimensions: references to structural elements such as column headers, references to non-public values specific to the underlying data, and references to data containers such as files or tables. To stabilize the classification, we apply self-consistency, classifying each input five times and assigning the final label by majority vote requiring at least three identical votes. An input counts as data privileged if at least one dimension is flagged, where for value references both the explicit and the obscure label count as a flag. For this classification we employ \texttt{gpt-5-mini-2025-08-07}, following the original setup by \citet{gomm_are_2025}.

Sufficient specification is assessed by two classifiers covering the five specification dimensions of \citet{gomm_are_2025}, with data specification comprising entity, temporal, and domain specification and procedural specification comprising intent and methodological specification. The per-dimension labels are aggregated into boolean flags following their criteria, under which temporal specification also admits underspecification assuming recency or inapplicability, and domain specification admits underspecification assuming a universal domain. An input is fully specified if all five flags hold and insufficiently specified otherwise. We use \texttt{gpt-5-2025-08-07} as an LLM-judge.

The validity of the LLM-judges is thoroughly established by \citet{gomm_are_2025}. To verify that the setup generalizes to our selection of benchmarks, we replicate their annotation study on a stratified sample of 168 annotations over the 18 classified benchmarks, following their validation protocols. For data privilege, a data-science expert independently annotates each sampled input along three dimensions of data privilege (schema, value, and container privilege), and we report the agreement between these annotations and the judge in Table~\ref{tab:llm_judge_eval:data_privilege}. For sufficient specification, we follow the expert-corrected protocol of \citet{gomm_are_2025}, in which the expert corrects the labels emitted by the judge and we report the agreement between the initial and the corrected labels in Table~\ref{tab:llm_judge_eval:sufficient_specification}. \citet{gomm_are_2025} adopt this correction-based protocol because resolving underlying ambiguities requires broader world knowledge than a single independent annotation pass affords. We report both raw agreement and Cohen's $\kappa$~\citep{cohen_coefficient_1960}, which corrects for the agreement expected by chance.

\begin{table}[ht]
    \centering
    \caption{Agreement between the data privilege judge and an independent expert annotation. We report raw agreement (\%) and Cohen's $\kappa$.}
    \small
    \begin{tabular}{lcc}
    \toprule
    \textbf{Privilege Dimension} & \textbf{Raw} & $\boldsymbol{\kappa}$ \\
    \midrule
    Schema Privilege & 0.881 & 0.606 \\
    Value Privilege & 0.911 & 0.585 \\
    Container Privilege & 0.946 & 0.760 \\
    \addlinespace
    \textbf{Data Privilege (agg.)} & 0.893 & 0.747 \\
    \bottomrule
\end{tabular}
    \label{tab:llm_judge_eval:data_privilege}
\end{table}

\begin{table}[ht]
    \centering
    \caption{Agreement between the sufficient specification classifications and expert-corrected labels. We report raw agreement (\%) and Cohen's $\kappa$.}
    \small
    \begin{tabular}{lcc}
    \toprule
    \textbf{Specification Dimension} & \textbf{Raw} & $\boldsymbol{\kappa}$ \\
    \midrule
    Entities & 0.924 & 0.838 \\
    Temporal & 0.906 & 0.786 \\
    Domain & 0.953 & 0.904 \\
    Intent & 0.971 & 0.861 \\
    Methodology & 0.929 & 0.850 \\
    \addlinespace
    \textbf{Full Specification (agg.)} & 0.953 & 0.869 \\
    \bottomrule
\end{tabular}
    \label{tab:llm_judge_eval:sufficient_specification}
\end{table}

In this annotation study, we find a substantial agreement of $\kappa = 0.747$ on the aggregated data privilege judgments and an almost perfect agreement of $\kappa = 0.869$ on the aggregated judgments of sufficient specification, following the scale of \citet{landis_measurement_1977}. The raw agreement scores are in line with those reported by \citet{gomm_are_2025}, indicating that the judges generalize to the benchmarks we assess.

\section{Citation Analysis}
\label{s:appendix:citation_analysis}

To support the claim that works relevant to \acrshort{task} are organized by the task label under which they frame their contributions instead of the problem they address, and that works converging on similar system designs frequently remain unaware of one another, we analyze how surveyed works cite each other against what label they use to describe the task they target. In this citation analysis we construct a citation graph among the surveyed works, assign each work the task label it uses to frame itself, and measure whether citation is assortative on that label. We analyze the 181 works that we surface in our literature review, including the works identified through forward and backward snowballing.

\subsection{Citation Graph Extraction}
\label{s:appendix:citation_analysis:graph}

We extract the reference list of each work from its PDF with \textsc{grobid}\footnote{\href{https://github.com/grobidOrg/grobid/}{https://github.com/grobidOrg/grobid/}}, yielding 8,473 references across 180 of the 181 works. We match references against the surveyed works by comparing normalized titles. We apply a hierarchical matching procedure, first matching the full title, then matching only the title preceding any subtitle, and then fuzzy matching with a similarity threshold of $0.85$. Since many titles begin with a short system name, we accept a match on the shortened title only when the first authors of both works match. In total, we identify 959 edges between the papers, of which $96\%$ come from an exact title match.

We additionally employ \href{https://www.semanticscholar.org/}{Semantic Scholar} and \href{https://openalex.org/}{OpenAlex} to compare and complement our extraction, which yield 809 (Semantic Scholar) and 109 (OpenAlex) edges between surveyed papers respectively. We add the 40 edges not extracted through \textsc{grobid}, giving a final citation graph with 999 edges.

\subsection{Assigning Task Labels}
\label{s:appendix:citation_analysis:labels}

We assign each work one of three labels, namely \textit{text-to-SQL}, \textit{table question answering}, and \textit{data analysis agents}. These are the three framings under which the surveyed works predominantly present themselves. Works framed around table retrieval are assigned the label of the downstream task they serve, since retrieval is a functional capability rather than a task in the sense of \acrshort{task} (Section~\ref{sec:orchestration:capabilities:retrieval}).

To label how each paper frames the task it addresses we use a multi-step process. First, we automatically assign a label to each paper by extracting how often the label, or a derivation of it, occurs in the title and the abstract of the papers, assigning the label that appears most often, not assigning a label if two labels appear the same number of times or if none appears. We cross-check the assignment by an independent classifier that labels the same titles and abstracts against a written codebook, using \acrshort{llm}-based classification with a majority vote over three samples, using \texttt{Qwen3.5-9B}~\citep{qwen3.5} as \acrshort{llm}. The two procedures agree on $86.2\%$ of the works, corresponding to a substantial agreement of $\kappa = 0.799$. We manually review the 37 works on which they disagree, or for which neither procedure finds a task term, to correct any remaining inconsistencies. This yields 85 table question answering works, 54 text-to-SQL works, and 33 data analysis agent works, while 9 works are not assigned to any task label.

\subsection{Analysis}
\label{s:appendix:citation_analysis:analysis}

We measure assortativity over feasible citations as pairs of works $(A, B)$ for which $A$ could have cited $B$, meaning that $B$ is published before or in the same year as $A$. Restricting to these feasible pairs removes the differing sizes of the label groups and the age distribution of the works as confounds, and yields 18,243 pairs. We compare the citation rate within labels against the rate across labels and test the ratio between the two by permuting the labels across works over $10{,}000$ draws.

\subsubsection{Results}
\label{s:appendix:citation_analysis:analysis:results}

Works cite within their own label in $10.6\%$ of the feasible pairs and across labels in $1.8\%$, a ratio of $6.02$, significant at $p < 0.001$. Looking into the surveyed systems specifically, an increase in the number of functional capabilities shared between two systems raises the citation rate within a label from $5.6\%$ to $18.5\%$ while the rate across labels stays at or below $3.0\%$ (Figure~\ref{fig:fragmentation:capabilities}), so systems converging on similar designs are no more likely to cite each other across a label boundary than systems sharing no capability at all.

 Importantly, the absence of a citation is evidence of non-engagement rather than proof that works are unaware of each other. Nonetheless, these results show that literature engages less rigorously with prior works across task labels than within them. Additionally, around a quarter of the works we label under data analysis agents cite none of the surveyed works at all, so the corresponding row of Figure~\ref{fig:fragmentation:matrix} reflects not just the isolation of that literature but also the reach of our review.

\end{document}